\documentclass[a4paper,twocolumn,showpacs,superscriptaddress,floatfix,nofootinbib,hyperref]{emulateapj}
\usepackage{color,amsfonts,amsmath,amssymb,times}
\usepackage{ulem}

\newcommand{\red}[1]{\textcolor{red}{#1}}

\def\be{\begin{equation}}
\def\ee{\end{equation}} 
\def\msun{M_\odot}
\def\lsim{\mathrel{\rlap{\lower 3pt\hbox{$\sim$}}\raise 2.0pt\hbox{$<$}}}
\def\gsim{\mathrel{\rlap{\lower 3pt\hbox{$\sim$}} \raise 2.0pt\hbox{$>$}}}

\slugcomment{}

\shorttitle{Linking black-hole spins to galaxy kinematics}
\shortauthors{Sesana, Barausse, Dotti and Rossi}

\begin{document}


\title{Linking the spin evolution of massive black holes to galaxy kinematics}

\author{A. Sesana\altaffilmark{1}, E. Barausse\altaffilmark{2,3}, M. Dotti\altaffilmark{4,5}, E. M. Rossi\altaffilmark{6}}

\altaffiltext{1}{Max Planck Institute for Gravitational Physics, Albert Einstein Institute, Am M\"uhlenberg 1, 14476, Golm, Germany\\ Email: {\sf alberto.sesana@aei.mpg.de} }
\altaffiltext{2}{CNRS, UMR 7095, Institut d'Astrophysique de Paris, 98bis Bd Arago, 75014 Paris, France\\ Email: {\sf barausse@iap.fr} }
\altaffiltext{3}{Sorbonne Universit\'es, UPMC Univ Paris 06, UMR 7095, 98bis Bd Arago, 75014 Paris, France}
\altaffiltext{4}{Dipartimento di Fisica G. Occhialini, Universit`a degli Studi di Milano Bicocca, Piazza della Scienza 3, 20126 Milano, Italy\\ Email: {\sf massimo.dotti@mib.infn.it}}
\altaffiltext{5}{INFN, Sezione di Milano-Bicocca, Piazza della Scienza 3, I-20126 Milano, Italy}
\altaffiltext{6}{Leiden Observatory, Leiden University, PO Box 9513 2300 RA Leiden, the Netherlands\\ Email: {\sf emr@strw.leidenuniv.nl}}

\begin{abstract}
We present the results of a semianalytical model that evolves the
masses and spins of massive black holes together with the properties
of their host galaxies along the cosmic history. As a consistency
check, our model broadly reproduces a number of observations, 
e.g. the cosmic star formation
history, the black hole mass and luminosity function and the galaxy
mass function at low redshift, the black hole to bulge mass relation,
and the morphological distribution at low redshift. 
For the first time in a semianalytical investigation,
we relax the simplifying assumptions of perfect coherency or perfect
isotropy of the gas fueling the black holes. The dynamics of gas is
instead linked to the morphological properties of the host galaxies,
resulting in different spin distributions for black holes hosted in
different galaxy types.  We compare our results with the observed
sample of spin measurements obtained through broad K$\alpha$ iron line
fitting. The observational data disfavor both accretion along a fixed
direction and isotropic fueling. Conversely, when the properties of 
the accretion flow are anchored to the kinematics of 
the host galaxy, we obtain a good match between theoretical expectations
and observations. A mixture of coherent accretion and phases of activity 
in which the gas dynamics is similar to that of the stars in bulges 
(i.e., with a significant
velocity dispersion superimposed to a net rotation) best describes the data,
adding further evidence in support to the coevolution of massive black
holes and their hosts.

\end{abstract}

\keywords{black hole physics -- accretion, accretion disks -- galaxies: evolution -- galaxies: kinematics and dynamics -- galaxies: active -- (galaxies:) quasars: supermassive black holes}

\maketitle

\section{Introduction}

Observational evidence for the existence of massive black holes (MBHs) in
the center of galaxies is circumstantial but convincing. Stellar orbits near
the Galactic center point at the presence of a dark and massive object, which is
too heavy and compact to be a cluster of low-luminosity bodies~\citep{darkcluster} or a fermion star~\citep{schoedel}.
Also, near-infrared observations of this object show that it is likely to 
have an event horizon, because if it did not, it would emit thermal radiation
and be more luminous than observed~\citep{horizon1,horizon2}. This disfavors exotic horizon-less alternatives to
General Relativity's black holes, such as e.g. boson stars and gravastars.
Similarly, MBHs are expected to be present in the center of most galaxies (and not just ours), 
as their presence is required to explain quasars (QSOs), active galactic nuclei (AGNs)~\citep{soltan82}
and the cosmic downsizing~\citep{AGN_downsizing1,AGN_downsizing2,AGN_downsizing3}. Indeed, 
observations of nuclear stellar and gas dynamics, reverberation mapping,
spectroscopic single epoch measurements and SED fitting \citep[see
  e.g.][and references therein for a discussion of the different
  methods]{ghez08, gillessen09, kormendy95, miyoshi95, hicks08,
  blandford82, kaspi05, calderone13, castignani13}
corroborate this expectation and allow the mass of MBHs to be estimated. Furthermore, these MBH mass
measurements have in turn resulted into the discovery of the MBH-galaxy
relations \citep[e.g.][]{gebhardt00, ferrarese00, marconi03, haring04,
  gultekin09}, hinting at a symbiotic evolution of the central compact
object and its host galaxy.

MBHs are also expected to possess a spin angular momentum. The combination
of mass and spin completely characterizes these objects 
if they are described by the black-hole solutions of General Relativity~\citep{kerr}\footnote{Black holes in General Relativity 
can also have an electric charge~\citep{kerr_newman}, 
but that is expected to be quickly canceled by the charges in 
the plasma surrounding astrophysical black holes, as well as
by quantum effects such as Schwinger pair production~\citep{Gibbons:1975kk,Hanni:1982}, or vacuum breakdown mechanisms triggering
cascades of electron-positron pairs~\citep{1969ApJ...157..869G,1975ApJ...196...51R,blandford1977}.}. 
If instead the gravity theory describing our Universe is not exactly given by General Relativity, other charges may be needed to describe MBHs, but the spin will still be one of them ~\citep[c.f. ][for a few examples of black holes in alternative gravity theories]{AE_BHs,CS_BHs,Pani_BHs,GB_BHs}. 
Indeed, future spaced-based gravitational-wave interferometers such as
ESA's L3 experiment eLISA~\citep{lisa13}
will measure MBH masses and spins with fantastic accuracy (respectively $\sim 0.1$\% and $\sim 1$\%),
and also put General Relativity to the test by e.g. measuring possible additional black-hole charges.
More importantly from an astrophysical point of view, measurements of MBH 
spins could unveil still unknown links between the MBHs and their galactic hosts. As an example, knowledge of the spins
could constrain the properties of the gas accreting onto the MBHs, e.g.
whether it spirals toward it on a preferential plane, on completely
isotropic directions, or with other more complex dynamics
\citep[][hereinafter D13]{dotti13}. 

Until eLISA or a similar space-based gravitational-wave interferometer is launched, however, 
measuring the spins of MBHs is considerably more difficult than measuring their masses. This is because the spin affects the dynamics
of stars and gas surrounding the MBH only  at distances orders of
magnitudes smaller than the MBH influence radius
\citep[e.g.][]{bardeen75}.
Currently, the most accurate way of
measuring the spins of MBHs is through the spectra of
relativistically broadened K$\alpha$
iron lines \citep[see][for recent reviews on the topic]{reynolds13,
  brenneman13}. Indeed, using current X-ray spectrographs (in
particular XMM and Suzaku), a significant sample of MBHs with spin estimates is
being built up. Also, more stringent constraints on the spin estimates are
becoming available from hard X-ray data from NuSTAR
\citep{risaliti13,marinucci14a,marinucci14b}. Finally, spin measurements with K$\alpha$
iron lines are one of the main
scientific drivers of ESA's upcoming L2 mission 
ATHENA+ \citep{nandra13}. 

This situation calls for a
theoretical framework capable of interpreting available
data for MBH spins and making testable predictions for future measurements.
However, detailed modeling of the MBH spin evolution is still
missing. Since the spin is a vector, any theoretical model should take the
 evolution of its direction into account, because that also affects the
evolution of the 
magnitude. This is particularly clear when one considers MBH mergers, which
are expected in the late stages of galaxy mergers \citep[e.g.][]{begelman80}. 
The final spin, mass and kick velocity of the resulting MBH remnant  critically depend on the orientations of the spins
of the progenitors \citep{spin_formula1,spin_formula2,BKL,mass_formula,kick_RIT,kick_goddard}.

The main driver of the MBH evolution, however, is not expected
to be given by mergers but rather by radiative
efficient gas accretion \citep{soltan82, merloni04, shankar04, shankar13b, wang, li}. In
this case, the spin magnitude evolution depends on whether the gas in the
very central regions of the accretion disk 
is co-rotating or counter-rotating relative to the MBH spin.
This, in turn, depends on
the larger-scale geometry of the fueling process (i.e. if the gas
keeps falling toward the MBH on some preferential plane), as well
as on the possible re-alignment of the MBH-disk system due to the
Bardeen-Petterson effect \citep{bardeen75}. 
The latter is the interplay between the frame dragging from the rotating 
spacetime geometry and the viscous stresses in the disk.
The details of the effect
of the spacetime's geometry on the properties of accretion disks are
still not completely understood \citep[see, e.g.][and references
  therein]{sorathia13a, sorathia13b}. All the studies that follow
the coupled evolution of the magnitude and direction of the MBH spin
have been performed under some simplifying hypotheses: i.e. assuming that
 the disk has isotropic viscosity \citep[see
  the discussion in][]{sorathia13a, sorathia13b}, using order of
magnitude estimates for the effect of the alignment
\citep[e.g.][]{king06}, or under the small-deformation approximation for
the warped disk \citep[e.g.][]{perego09}. 

These simplifying hypotheses allowed for simple recipes for the spin
evolution that have been used in semi-analytical models of structure
formation and evolution \cite[see, e.g.][]{volonteri05, berti08,
  lagos2009, fanidakis11, fanidakis2, mymodel,volonteri_sikora,
  volonteri13}.  
  However, most of these studies~\citep{volonteri05,
  berti08, lagos2009, fanidakis11, fanidakis2} assume that the MBH
fueling is either always coherent (i.e. accretion always occurs
exactly on the equatorial plane), or statistically isotropic
(i.e. the MBH accretes small gas clouds with randomly oriented angular
momenta). These assumptions were partly relaxed
in \cite{mymodel}, \cite{volonteri_sikora} and \cite{volonteri13}, who noted that the
Bardeen-Petterson effect is likely to align the MBH spin with the
angular momentum of the accreting gas in gas-rich nuclear
environments, thus making the accretion flow effectively coherent in
such situations, as opposed to isotropic accretion, which is more
likely in gas-poor nuclear environments.  Even with this improvement,
however, the possible parameter space of the accretion flow is still
limited to two points (coherent accretion in gas-rich environments vs
isotropic one in gas-poor environments).

Recently D13 demonstrated that the spin evolution can 
be significantly different if a more realistic description of the 
accretion flow is adopted, namely one that allows regimes intermediate 
between a perfectly coherent and a perfectly isotropic flow. 
\cite{dubois13} studied the evolution of the spins of a population of MBHs 
taking into consideration the dynamical properties of the fueling gas. They 
extracted the direction of the angular momentum of the accreting matter on the
smallest scales resolved by hydrodynamical simulations (down to $\approx 10$
pc in their highest resolution run). In a similar spirit, in this paper we
couple the semianalytic model of  \citet[][hereinafter B12]{mymodel},  to the spin evolution model
of D13. In particular, we link the degree of coherency/isotropy of the accretion flow
to the dynamical parameters of the gas and stars observed in galactic nuclei.
We stress that, also in our case, most of the observations that we use have a spatial resolution $\gsim 100$ pc,
significantly larger than typical accretion disks
scales, and that the dynamical properties of the gas could change at
smaller unresolved scales \citep[e.g.][]{hopkins12, maio13, dubois14}.
In this sense, therefore, a comparison of the predictions
of our model for the MBH spin evolution with actual spin measurements 
provides a way of testing the dynamics and coherency of
the nuclear gas at yet unresolved scales.
We attempt such a test with the MBH spin measurements from 
K$\alpha$ iron lines available to date.
We note, however, that our model provides a flexible framework to
also interpret future, more accurate measurements from experiments such
as eLISA and ATHENA+.

The paper is organized as follows. In Section \ref{sec:model} we present the details
of our model for the evolution of MBHs in their galactic hosts. Improving on B12,
we adopt a different star-formation law (see Section \ref{sec:SF} and Appendix \ref{ap:SF})
and, more importantly, we update the prescription for the evolution of the MBH spins under accretion, following D13 (see Section \ref{sec:spin}).
 The degree of coherence of the  accretion flow is then linked to the kinematic properties of the host galaxy in Section
 \ref{sec:link}.
We consider in particular three models: one where the degree of coherence is related to the gas kinematics, one that links it to the stellar kinematics,
and a hybrid model where information from both the gas and stellar kinematics is used. In Section \ref{sec:cali}, we calibrate the free parameters of our semianalytical galaxy-formation
model against observations.
 We present our predictions for the evolution of MBH spins in Section \ref{sec:results},
and in Section \ref{sec:obs} we compare them to measurements from
 K$\alpha$ iron line profiles. In Section \ref{sec:conclusions} we draw our conclusions.

\begin{figure}
\includegraphics[width=8.5cm]{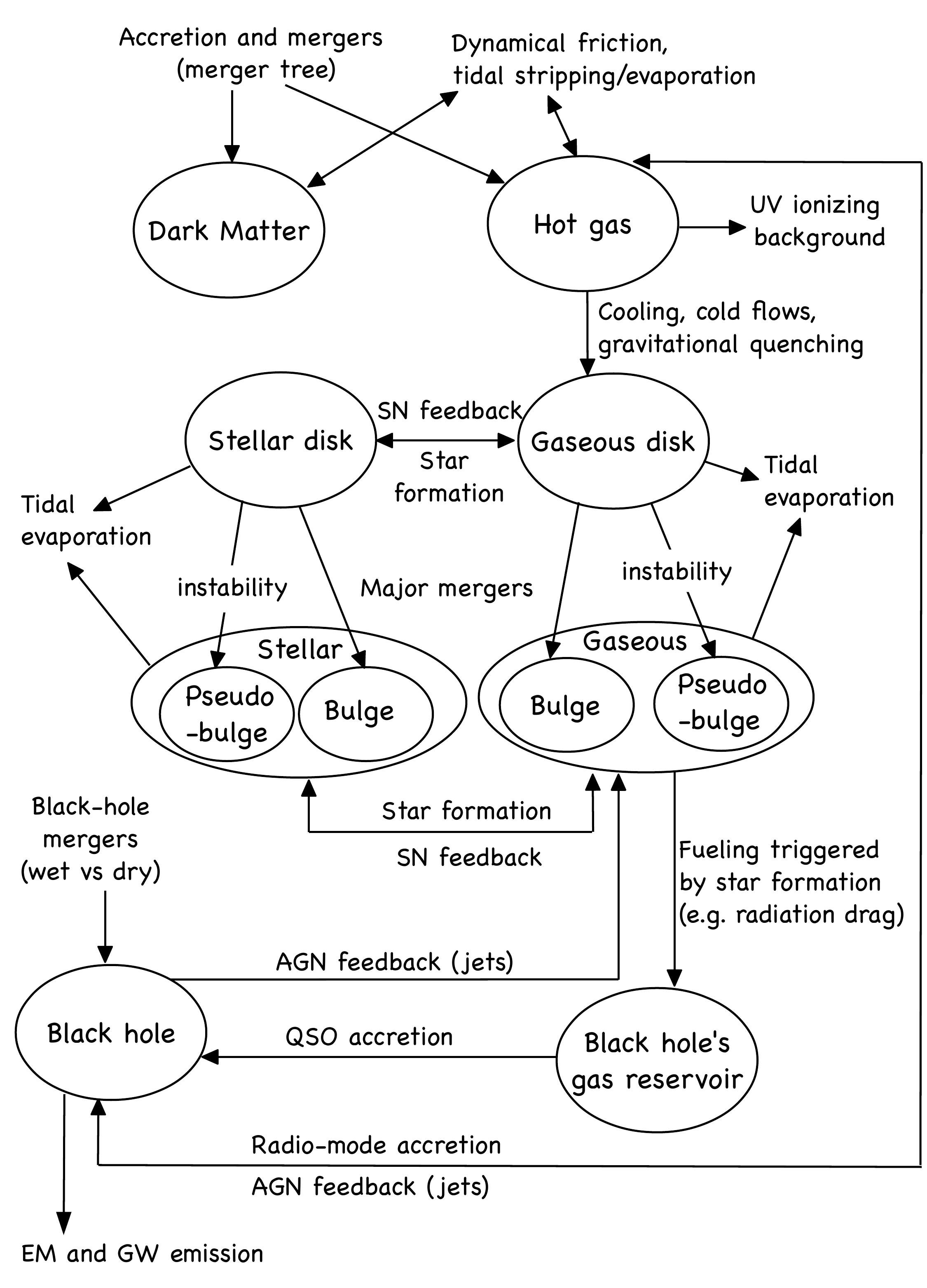}
\caption{\label{fig:model} Schematic representation of our semianalytical model, based on that of B12. Note that
B12 did not use the distinction between ``bulges'' forming from major galactic mergers and
``pseudobulges'' forming from bar-instabilities of disk galaxies (although both formation channels were present).
In this paper, instead, we assume that bulges and pseudobulges have different star-formation laws (c.f. Section \ref{sec:SF})
and, in some realizations of our model, different accretion properties onto the central MBH (c.f. Section \ref{sec:link}).
The other major change from the model of B12 is indeed the prescription for the spin evolution
of MBHs under accretion and mergers, which is schematically represent in Figure~\ref{fig:flow} and described in detail in Section \ref{sec:spin}.}
\end{figure}

\section{The model}
\label{sec:model}

Our model is based on B12, 
who studied the cosmic evolution of MBH masses and spins
using a state-of-the-art semianalytical galaxy-formation model. More specifically,
B12 describes the Dark-Matter evolution by merger trees
produced with the extended Press-Schechter formalism and suitably
modified to reproduce the results of N-body Dark-Matter
simulations~\citep{DMtrees}.  The baryonic content of galaxies is
evolved along the branches of the merger tree and includes several
components: a hot-gas phase, a cold-gas disk, a stellar disk, a
cold-gas bulge, a stellar bulge, a MBH and a reservoir of cold gas
fueling accretion onto the MBH. Gravitational interactions among the
various components are accounted for (although in simplified ways), 
allowing the computation of the density and velocity profiles of the various components inside each galaxy.  Also modeled are a
variety of non-gravitational interactions, which are summarized
schematically in Figure~\ref{fig:model}.  

For more details, we refer the reader to B12. Here, we focus instead on our
improvements to that model: the star formation law (Section \ref{sec:SF}) and the description of the cosmic MBH spin evolution  (Section \ref{sec:spin}).

\subsection{The star formation model}
\label{sec:SF}

The star formation adopted here differs from and improves on the one used in B12. In this section, we describe 
the general framework and main equations, and confine a more detailed presentation in Appendix \ref{ap:SF}.

The first new ingredient is that the star formation is now an explicit function of metallicity, which allows for better modeling of the build-up of stars in different galaxies, at various redshifts and environments (c.f. the star-formation history of Figure~\ref{fig:sfr}, to be compared with Figure 8 of B12). 
The second is a different prescription for {\it classical bulges} and {\it pseudobulges}. The former are the result of major galactic mergers, 
which cause bursts of star formation  \citep{daddi10, genzel10}, while the latter result from bar instabilities in galactic disks and undergo a slower,
disk-like star formation \citep[see, e.g.][]{kormendy13}. In B12, both bar instabilities and major mergers were modeled, but 
star formation in the bulge component was the same irrespective of the bulge's formation mechanism. In practice, our new prescription is based on the work by \cite{krumholz09}, extended to include the low metallicity \cite[less than 1$\%$ solar,][] {krumholz_gnedin, krumholz12, forbes13, kuhlen13} and starburst regimes \citep{daddi10, genzel10}. 

In galactic disks, the star formation density rate is expressed in terms of the fraction $f_{\rm c}$ of the background gas surface density ($\Sigma_{\rm g}$) 
that is converted into stars on a time scale $t_{\rm SF}$,
\be\label{sf1} \dot{\Sigma}_* =  \frac{f_{\rm c} \Sigma_{\rm g}}{t_{\rm SF}}\,.\ee
The metallicity dependence enters explicitly in the fraction  $f_{\rm c}$ (see equation \ref{eq:fh2} in Appendix \ref{ap:SF}): down to 2\% solar, the lower the metallicity, the smaller the amount of gas available for star forming. 
The timescale $t_{\rm SF}$ (see equation \ref{eq:k7} in Appendix  \ref{ap:SF}) depends on the properties of Giant Molecular Clouds, 
which vary with the gas richness of the environment. This timescale is generally longer than the local free fall timescale.
The same star formation law applies to pseudobulges, but it is calculated per unit volume,
\be\label{sf2} \dot{\rho}_* =  \frac{f_{\rm c} \rho_{\rm g}}{t_{\rm SF}}. \ee
Finally, in classical bulges, we describe the burst of star formation as the sudden consumption of gas in a local dynamical timescale,
\be\label{sf3} \dot{\rho}_{*}=  \frac{\rho_{\rm g}}{t_{\rm ff}}\,,\ee
where $t_{\rm ff}=\sqrt{3\pi/(32 G \rho_{\rm g})}$.

The total star formation rates are then performed integrating equations \eqref{sf1}--\eqref{sf3} over the whole disk/pseudobulge/bulge.


\subsection{The spin evolution}
\label{sec:spin}
Two main processes drive the cosmic evolution of MBH spins: {\it gas accretion} from the inter-stellar (ISM) or the inter-galactic (IGM) medium,
and {\it mergers} between MBHs.

\begin{figure*}
\includegraphics[width=1\textwidth]{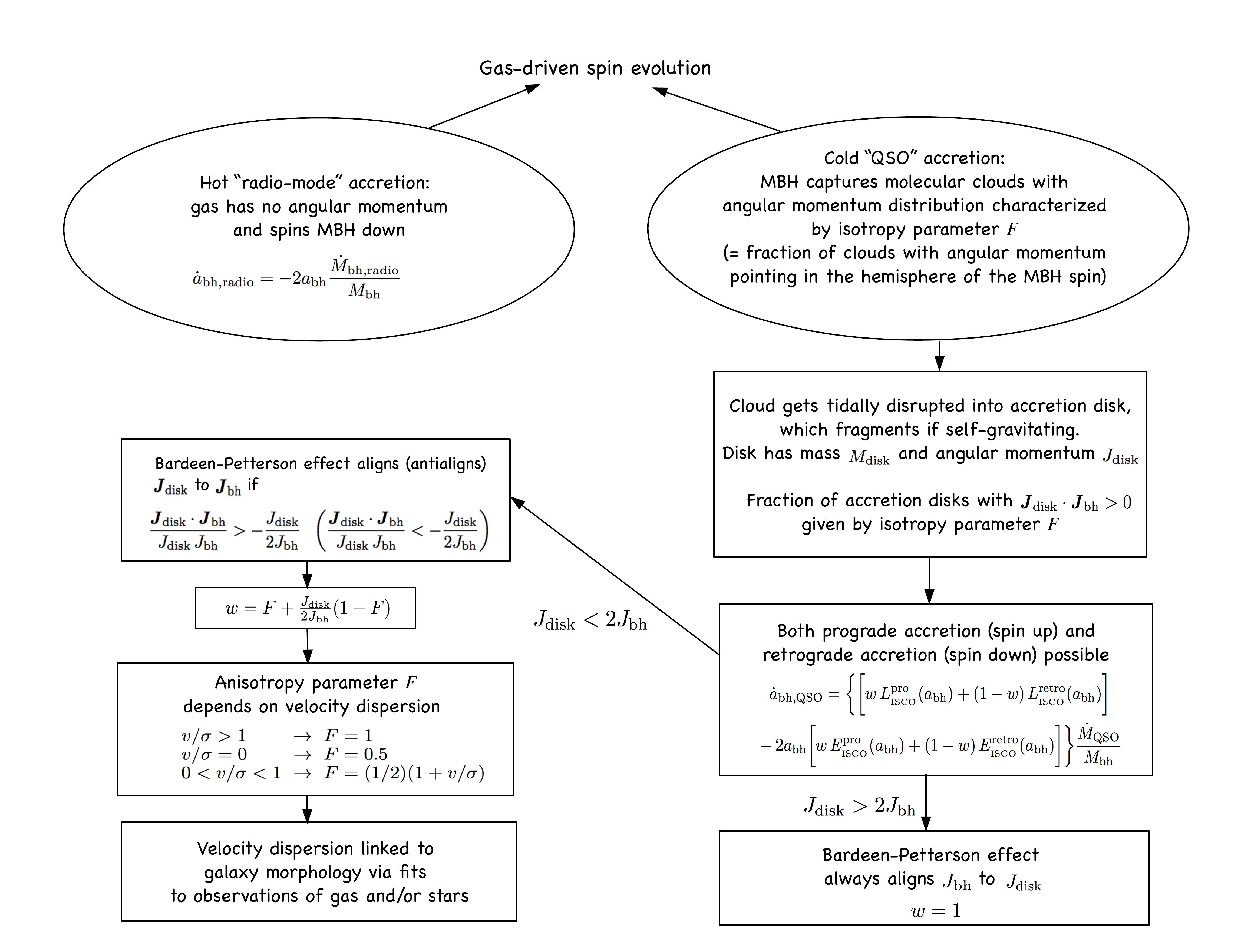}
\caption{Scheme of the possible accretion driven MBH spin evolution channels.}
\label{fig:flow}
\end{figure*}

\subsubsection{Gas accretion}
Accretion is believed to be the main driver of the MBH spin evolution
\citep[][B12]{berti08}, with the possible exception of the most
massive black holes in massive, gas-poor, low-redshift ellipticals
\citep[e.g.,][]{fanidakis11}. 
Because of the spin's vectorial nature, its evolution depends on the angular momentum magnitude and direction 
of the accreting gas.
In our implementation, based on D13,  the circumnuclear gas reservoir surrounding the MBH
is accreted in a sequence of clouds, with a non-trivial angular momentum distribution. Unlike in D13, 
however, this distribution is not arbitrary, but is linked to the galactic properties at larger scales and therefore 
intimately connected to the galaxy morphology (see Section \ref{sec:link}).
 
Following B12, we distinguish two main astrophysical modes of gas accretion, which prompt different spin evolution channels  as illustrated in Figure~\ref{fig:flow}:

$\bullet$ A hot ``radio'' mode. In this regime, the MBH accretes IGM gas  that cools on a timescale longer than its free fall time. 
 The gas is 
  approximately spherically symmetric, so no net angular
  momentum is acquired by the MBH and the spin parameter $a_{\rm bh}$ changes
  (to first approximation) only because of the increase of the MBH mass:
\begin{equation}
 \dot{a}_{\rm bh,radio} = -2 a_{\rm bh} \frac{ \dot{M}_{\rm bh,radio}}{{M}_{\rm bh}}\,,
\end{equation}
where $\dot{M}_{\rm bh,radio}$ is the Bondi accretion rate of the hot IGM component  (see equation (44) in B12).
This mode has a negligible direct impact on the  mass and spin evolution of MBHs,
because it is typically characterized by very low accretion rates, but plays a major role in activating
jets capable of exerting an ``AGN feedback'' on the growth of large cosmic structures (see B12 for more details on
the implementation of AGN feedback).

$\bullet$ A cold ``QSO'' mode. It is associated with the star formation in the bulge,
which drives cold ISM gas into a low-angular momentum
reservoir of mass $M_{\rm res}$, available for accretion onto the MBH.
In practice, we assume that the influx of gas into the reservoir is 
proportional to the bulge star formation rate (see B12; \cite{granato,haiman,2014ApJ...782...69L}).
As a result, in our current framework for the star formation (Section~\ref{sec:SF}), 
major mergers tend to produce larger reservoirs and MBH accretion rates than disk instabilities.
Once the cold gas has settled in the reservoir, we follow \cite{granato} and assume that it accretes onto the MBH with
instantaneous rate
\begin{equation}
\dot{M}_{\rm QSO}=\min(\dot M_{\rm visc}, A_{\rm Edd}\dot{M}_{\rm Edd})\,.
\end{equation}
The viscously driven accretion rate reads
$\dot M_{\rm visc}= k_{\rm accr}\sigma^3 M_{\rm res}/(G M_{\rm bh})$, where\footnote{Note that $k_{\rm accr}$  is
related to the critical Reynolds number ${\cal R}_{\rm crit}$ for the onset of turbulence by $k_{\rm accr}=1/{\cal R}_{\rm crit}$.} $k_{\rm accr}=10^{-3}$, $\sigma =0.65 V_{\rm vir}$ and $V_{\rm vir}$ is the halo's virial velocity  ~\citep{granato,granato_private}.
 The Eddington rate $\dot{M}_{\rm Edd}=L_{\rm Edd}/[\eta(a_{\rm bh}) c^2]$ is derived from the Eddington luminosity $L_{\rm Edd}$, assuming a radiation efficiency $\eta$. 
As shown below [c.f. equation~\eqref{eq:eta}],
$\eta$ can be computed as a function of the MBH spin and the fraction of prograde vs retrograde accretion events.  The free parameter $A_{\rm Edd}\geq1$ sets the
amount of super-Eddington accretion allowed in our model. This
prescription for $\dot{M}_{\rm QSO}$ replaces the simpler recipe of B12, which assumed
$\dot{M}_{\rm QSO}=M_{\rm res}/t_{\rm accr}$, with $t_{\rm accr}$ a free parameter on the order of $5 \times 10^8$ yr. 

{ The main difference with respect to B12 consists in {\it how} the gas in the reservoir is accreted. In B12, depending on the ratio $M_{\rm res}/M_{\rm bh}$, $M_{\rm res}$ is either 
accreted coherently or as a collection of small randomly oriented clouds with vanishing total angular momentum. 
Here, we employ a model that allows for a broader range of configurations for the accreting flow, as we proceed to explain. 
This model is based on results by \cite{perego09} and D13, to which we refer for more details.

Accretion is described as a series of  transient accretion disks, each one formed following the tidal disruption of a cloud. The mass $M_{\rm disk}$ of each individual disk 
is the minimum between
the typical mass of a molecular cloud --- $M_{\rm cloud}=3\times 10^4 M_\odot$ is our fiducial value ---  and  the mass $M_{\rm sg}$ of a disk truncated at the
self-gravity radius, where the Toomre parameter $Q$ reaches a value of 1.} This is because outside this radius, the
accretion disk that forms from the infalling cloud will fragment under its self gravity and be consumed by star
formation. {The explicit expression for $M_{\rm sg}$ is given by (D13)
\begin{equation}\label{eqn:msg}
M_{\rm sg} \approx 2 \times 10^{4} {\alpha}_{0.1}^{-1/45}   \left(\frac{f_{\rm Edd}}{\eta_{0.1}}\right)^{4/45}
\left(\frac{M_{\rm bh}}{10^6 M_\odot}\right)^{34/45}  M_\odot\,,
\end{equation}
 where $f_{\rm Edd}$ is the Eddington ratio, $\eta_{0.1}$ is the disk radiative efficiency normalized to 0.1, 
and $\alpha_{0.1}$ is the disk viscosity parameter in units of 0.1.  Throughout this paper we assume $\alpha_{0.1}=1$. Each accreting lump gets tidally disrupted and settles in a standard thin accretion disk with an associated angular momentum
$J_{\rm disk}$ obtained from its mass, $M_{\rm disk}=\min(M_{\rm cloud},M_{\rm sg})$:
\begin{align}
J_{\rm disk}=\;&\frac{8}{21}\frac{\dot{M}_{\rm bh,QSO} \sqrt{G M_{\rm bh}}}{A_{\nu}}R_{\rm disk}^{7/4}\,\\
A_{\nu}=\; & 9.14 \times 10^{6}\alpha_{0.1}^{4/5}\left(\frac{M_{\rm bh}}{10^6 M_\odot}\right)^{1/20}\nonumber\\&\times\left( \frac{f_{\rm Edd}}{\eta_{0.1}} \right)^{3/10}{\rm cm^{5/4}s^{-1}}\,
\end{align} 
(c.f. D13). The extension of the disk  $R_{\rm disk}$ depends on its mass. If $M_{\rm sg}< M_{\rm cloud}$, it is given by (D13)
\begin{equation}\label{eqn:Rdisksg}
{R_{\rm disk}}\approx  10^5 R_{\rm g}\,  \alpha_{0.1}^{28/45}\left(\frac{M_{\rm bh}}{10^6 M_\odot}\right)^{-52/45}\left ( {f_{\rm Edd}\over \eta_{0.1}}\right )^{-22/45} \,,
\end{equation}
where $R_{\rm g}={2 G M_{\rm bh}}/{c^2}$ is the Schwarzschild radius. If instead $M_{\rm sg}> M_{\rm cloud}$, one has (D13) 
\begin{align}\label{eqn:rdiskcl}
{R_{\rm disk}}\approx\; & 4\times 10^4\, R_{\rm g} \left ({M_{\rm disk}\over 10^4\,\msun}\right )^{4/5}\alpha_{0.1}^{16/25}\nonumber\\&\times
\left(\frac{M_{\rm bh}}{10^6 M_\odot}\right)^{-44/25} 
\left ({f_{\rm Edd}\over \eta_{0.1}}\right )^{-14/25}.
\end{align}
}

{ According to  the initial direction of the lump's angular momentum, the Bardeen-Petterson effect will either align or antialign the disk and the MBH spin directions [c.f. equation (\ref{eq:w})], resulting in prograde or retrograde accretion respectively.
As the total mass in the reservoir is consumed by several lumps accreting onto the MBH, 
the time-averaged spin evolution is determined by the fraction $w$ ($1-w$) of lumps that accrete on prograde (retrograde) orbits:}
\begin{multline}
 \dot{a}_{\rm bh,QSO} = \bigg\{
\bigg[w \, {L}^{\rm pro}_{\rm_{\rm ISCO}}(a_{\rm bh})+(1-w)\,{L}^{\rm retro}_{\rm_{\rm ISCO}}(a_{\rm bh})\bigg]\\-2 a_{\rm bh}
\bigg[w \,{E}^{\rm pro}_{\rm_{\rm ISCO}}(a_{\rm bh}) +(1-w)\,{E}^{\rm retro}_{\rm_{\rm ISCO}}(a_{\rm bh})\bigg]\bigg\}
\frac{ \dot{M}_{\rm QSO}}{{M}_{\rm bh}}\,,\label{adot_chaotic}
\end{multline}
 where $L_{\rm_{\rm ISCO}}^{\rm pro}(a_{\rm bh})$ and $E^{\rm pro}_{\rm_{\rm ISCO}}(a_{\rm
   bh})$ are respectively the specific angular momentum and specific energy
  at the prograde innermost stable circular orbit (ISCO), while
$L_{\rm_{\rm ISCO}}^{\rm retro}(a_{\rm bh})$  and $E^{\rm retro}_{\rm_{\rm
     ISCO}}(a_{\rm bh})$ are the same quantities for the retrograde ISCO.
Likewise, the accretion efficiency is calculated by weighing the prograde and retrograde 
thin-disk efficiencies $\eta_{\rm pro}(a_{\rm bh})$ and $\eta_{\rm retro}(a_{\rm bh})$:
\begin{equation}
\eta(a_{\rm bh}) = w \eta_{\rm pro}(a_{\rm bh})+ (1-w) \eta_{\rm retro}(a_{\rm bh})\,.\label{eq:eta}
\end{equation}

The fraction $w$ of prograde accretion disks is determined by the Bardeen-Petterson effect \citep[][D13]{bardeen75,king05,king06,perego09} and is given by
\begin{equation}
\begin{array}{lcc}
w=1 & {\rm if} & J_{\rm disk}> 2J_{\rm bh},\\
w=F + \frac{J_{\rm disk}}{2 J_{\rm bh}}(1-F)& {\rm if} & J_{\rm disk} < 2J_{\rm bh} \,.
\end{array}\label{eq:w}
\end{equation}
Here,  the ``isotropy parameter'' $F$ measures  the fraction of accretion events with
angular momentum initially pointing in the hemisphere of the MBH spin (i.e. with angular momentum initially tilted by 
an angle $\theta_{\rm out}<\pi/2$ relative to the MBH spin)\footnote{Note that D13 assumes the opposite definition for $F$, which is there defined as
fraction of accretion events with angular momentum initially tilted by 
an angle $\theta_{\rm out}>\pi/2$ relative to the MBH spin.}. 
When $J_{\rm disk}> 2J_{\rm bh}$, 
the MBH spin aligns
with the angular momentum of the outer regions of the accretion disk that forms when the lump of matter falls into the MBH, 
and the alignment happens on a timescale
smaller than the accretion timescale~\citep[][D13]{perego09}.
Most of the accreted gas then has
 angular momentum aligned with the MBH spin, hence $w=1$ and the MBH
spins up. In this limiting case, equation~\eqref{adot_chaotic} tends to equation (38)
of B12. If instead 
$J_{\rm disk}< 2J_{\rm bh}$, \cite{king05} and \cite{king06} showed that on  a similarly short timescale, 
 the inner accretion flow's angular momentum \textit{antialigns}  
with the MBH spin if the angle between the outer disk angular momentum and the MBH spin satisfies $\theta_{\rm out}>\pi/2$ and  $\cos \theta_{\rm out}< -J_{\rm disk}/(2 J_{\rm bh})$.
Assuming that the fraction $1-F$ of accretion events with  $\theta_{\rm out}>\pi/2$ is distributed isotropically, one then obtains the second expression for
$w$ in equation~\eqref{eq:w}.

\begin{figure}
\includegraphics[width=0.45\textwidth]{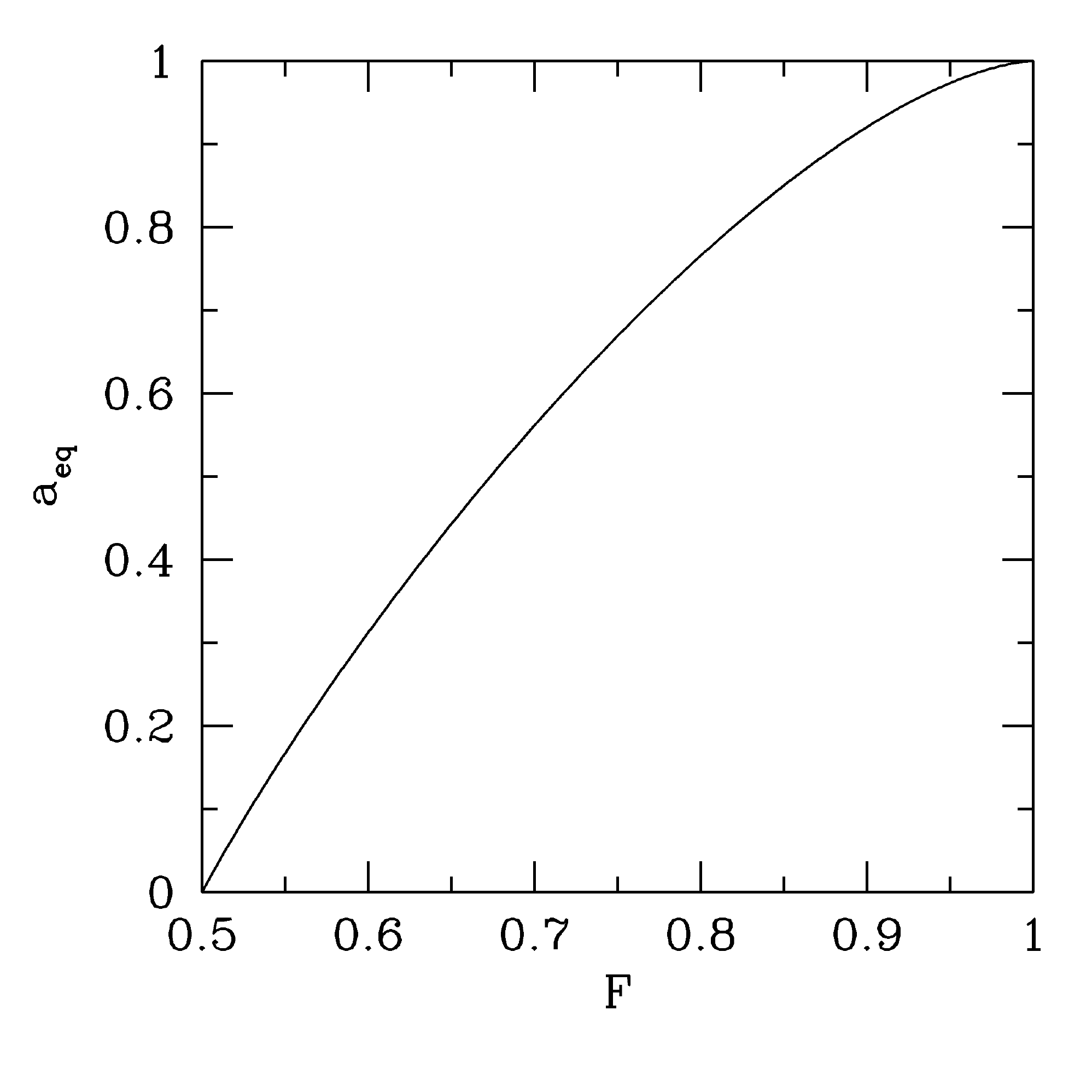}
\caption{Equilibrium spin parameter as a function of the ``isotropy parameter'' $F$. Here we assume that
  each single accretion episode has a very small angular momentum compared to
  the MBH, so that both alignment and anti-alignment can occur (see text for details).}
\label{fig:a_eq}
\end{figure}

Note that in the limit
of $J_{\rm disk}/2J_{\rm bh}\ll 1$ (always valid for very massive
MBHs, $M_{\rm bh} \gsim 10^9 \msun$), 
our model becomes very simple, as it predicts that the MBHs will move toward an
equilibrium spin parameter $a_{\rm eq}$ function of $F$, obtained by  setting $\dot{a}_{\rm bh,QSO}=0$ 
in equation~(\ref{adot_chaotic}), after substituting $w$ with $F$ as given by equation~\eqref{eq:w}. The value
of $a_{\rm eq}$ as a function of $F$ is shown in Figure~\ref{fig:a_eq}.

At this point, however, we should stress that equations~\eqref{adot_chaotic} and \eqref{eq:w} \textit{i)}
only provide an ``averaged'' version of the stochastic scenario
proposed by D13: i.e. our prescription does not follow each
individual accreting cloud, but only approximates the model of D13 over many such clouds, and \textit{ii)} do not take
into account the detailed evolution of the accretion disk during the
(short) alignment process.\footnote{The details of the alignment
  mechanism are, in any case, still subject of debate, see e.g. \cite{sorathia13a, sorathia13b}
  and references therein.}  A more accurate treatment of
the spin evolution addressing these two shortfalls is beyond the scope
of the paper, but we expect the impact of these simplifications on
the spin evolution to be small. Regarding \textit{i)}, the mass
$M_{\rm res}$ of the reservoir available for accretion is
typically much larger than $M_{\rm disk}$, thus
accretion takes place in several ``lumps'' and the stochastic character of the model of D13
averages out. An exception is in the
``tails'' of the accretion process,  when $M_{\rm res}$ is small
before and after a QSO event. In those tails, however, accretion rates and spin changes 
are small due to the scarcity of gas.  We have indeed tested
that replacing $M_{\rm cloud}\to\min(M_{\rm res},M_{\rm cloud})$,
which corresponds to changing our model prescription in the tails of
the accretion process, does not alter our results significantly.  As
for \textit{ii)}, neglecting the short alignment process is
typically justified, as the alignment takes place on timescales
shorter than accretion.  Nevertheless, even a small amount of
retrograde accretion during the alignment can potentially lower the
spin significantly near the $a_{\rm bh}=1$ limit; more details on this point
are given in Appendix \ref{app:sanity}.

\subsubsection{MBH Mergers}
Our prescription for MBH mergers closely follows that of B12, to which we refer for more details.
In summary, we use the phenomenological formulas of \cite{spin_formula2} and \cite{kick_goddard} to predict the final spin and 
recoil velocity of the  MBH resulting from the merger, and check whether the recoil ejects the MBH from the galaxy. 
As for the mass of the MBH produced by the merger, we update the prescription of 
 B12 by adopting the phenomenological formula of \cite{mass_formula}. 
We stress that these formulas produce results in accurate agreement
with fully general-relativistic simulations. 
The only other change from the implementation of B12 regards the
criterion adopted to discriminate ``wet'' (i.e. gas-rich) mergers from ``dry'' (i.e. gas-poor) mergers. This distinction is important because in wet mergers the spins of the two MBHs are expected to align to each other -- we follow here \citet{dotti10} and assume that the MBH spins are almost aligned, to within $\sim 10^\circ$, by the circumbinary disk via the Bardeen Petterson effect --, whereas in dry mergers such alignment does not take place and the spins are isotropically distributed. Improving on  B12,
who compared $M_{\rm res}$ to the mass of the MBH binary to distinguish the two regimes,
here we assume that a merger happens with almost aligned spins if $J_{\rm res}> 2 (S_1 + S_2)$ (where $J_{\rm res}$ is
the angular momentum of the cold gas reservoir, and $S_{1,2}$ are the spins of the two MBHs),
while otherwise we assume that the spins are randomly oriented.
This prescription is preferable to that of  B12, because the Bardeen Petterson effect
is sensitive to the angular momenta of the circumbinary disk and the MBHs, rather than to their masses.

\section{The link between accretion flows and host dynamics}
\label{sec:link}
In the previous section, we described how to physically relate the degree of anisotropy of the accretion flow to the MBH spin changes.
This relation may need to be still understood in the details, but there is a general consensus in the community on the overall physical picture~\citep[][D13]{bardeen75,king05,king06,perego09}.
The outstanding question is instead what the {\it actual}  conditions of the flow at sub-parsec scales are, and what determines these conditions.
This problem is considerably harder to solve because \textit{(i)} there is hardly any direct observational evidence at those scales and
\textit{(ii)} those scales are extremely challenging to explore theoretically with galaxy-formation simulations due to lack of
resolution and the unknown role of dissipative/non-linear processes.

In an attempt to contribute  some insight into the problem, we link the dynamical properties of the host galaxies, {\it which can be easily observed}, to 
the properties of the environment in the immediate vicinity of the MBH. As we will argue later in this paper, {\it existing} MBH spin measurements allow
for testing and placing constraints on this conjectured link.
The basic assumption behind our attempt is that the mechanism fueling the MBH --- either disk instabilities or galaxy mergers --- bridges large and small scales, 
ensuring that the degree of ``disorder'' of each galactic component (gas and stars) 
does not change much from galactic to circumnuclear scales.
At kpc scales, a routinely observed measure of ``disorder'' is the $v/\sigma$ ratio, where $v$ is the bulk rotation velocity of the system and $\sigma$ is its velocity dispersion. 
Therefore, the remaining issue is whether the gas that is brought down to the MBH and {\it gets accreted} has a $v/\sigma$ more similar to that of the surrounding stars in the bulge, or to that of the gas in the galactic disk. In the absence of solid theoretical predictions and observational evidence, we explore a few models, 
further differentiating between elliptical and spiral galaxies. As a consequence, we are able to predict different spin evolution in different 
galaxy types, which allows us to constrain our models (and therefore the physics of the fueling mechanism) by comparing them 
to observed MBH spin distribution in the local universe.

In the following, we first discuss observations of the  $v/\sigma$ ratio in galaxies (Section \ref{sec:measures}). 
These data will guide us in constructing models to assign a $v/\sigma$
ratio to the accreting reservoir, according to the morphology of the host galaxy (Section \ref{sec:implementation}). Finally, in Section \ref{sec:Fvsigma}  we will  describe 
how to geometrically translate a given a value of the  $v/\sigma$ ratio for the reservoir into the isotropy parameter $F$ used
in the previous section to characterize the MBH spin evolution, c.f. equations \eqref{eq:w} and \eqref{adot_chaotic}.

\subsection{Measurements of $v/\sigma$}
\label{sec:measures}
The ratio $v/\sigma$ for the stellar and/or gaseous components of a 
galaxy has been measured for a variety of galaxy samples. We collected these data  
to construct $v/\sigma$ distributions to implement in our galaxy evolution model. In the following we describe these measurements in detail,
discriminating between gas measurements in galactic disks and stellar measurements in bulges.

\subsubsection{Gaseous disks}
Measurements of $v/\sigma$ for the gas component in spiral galaxies have been 
obtained with integral field emission line spectroscopy with sub-arcsec
resolution, tracing the gas from 10 kpc down to sub-kpc scales,
depending on the redshift of the system. More specifically \cite{kassin12} 
discuss explicitly the dependence of $v/\sigma$ on redshift and
stellar mass for disk galaxies at $z<1.2$ using a sample of several hundred
objects. Here, we complement their data with measurement at higher
redshift. \cite{wisnioski11,epinat12,swinbank12} report $v/\sigma$ measurement for
star forming galaxies in the redshift range $1.2<z<1.7$; their data are
combined in a catalog of 30 objects that we take as a representative sample at
$z\approx1.5$. \cite{schreiber2009,law09,newman13} explore the range $1.5<z<2.5$; we
combine their data in a catalog of 56 objects that we take as a representative
sample at $z\approx2$. \cite{gnerucci11} measure $v/\sigma$ of 33 star
forming galaxies at high redshift, in the range $3<z<4.8$. They find a
fraction of rotating systems (defined as systems having $v/\sigma>1$) of
$\sim1/3$, claiming that ``the comparison between the SINS analysis at $z\sim2$
and the AMAZE analysis at $z\sim3.3$ suggests that the fraction of rotating
objects does not evolve within this redshift interval''. Unfortunately they do
not provide a full set of mass and $v/\sigma$ measurements for all the objects
in their sample. For fitting purposes, we then just assume another set of
measurements at $z\sim3.3$ mirroring the ones at $z\approx2$. Each data sample
is divided into galaxy stellar mass ($M_*$) bins, and for each mass bin we
compute the mean $v/\sigma$ and the dispersion around the mean. We therefore
end up with a series of measurements of $v/\sigma$ in the $z$-$M_*$ plane,
which we then fit with a ``Nuker's law'' of the form
\begin{equation}
\frac{v}{\sigma}=a\left(\frac{x}{x_0}\right)^{\alpha}\left[1+\left(\frac{x}{x_0}\right)^\beta\right]^{-\alpha/\beta},
\label{vsusigmagas}
\end{equation}
where $x={\rm log}(1+z)$, and $x_0=b({\rm log}M_*+c)$.

\begin{figure}
\includegraphics[width=\columnwidth]{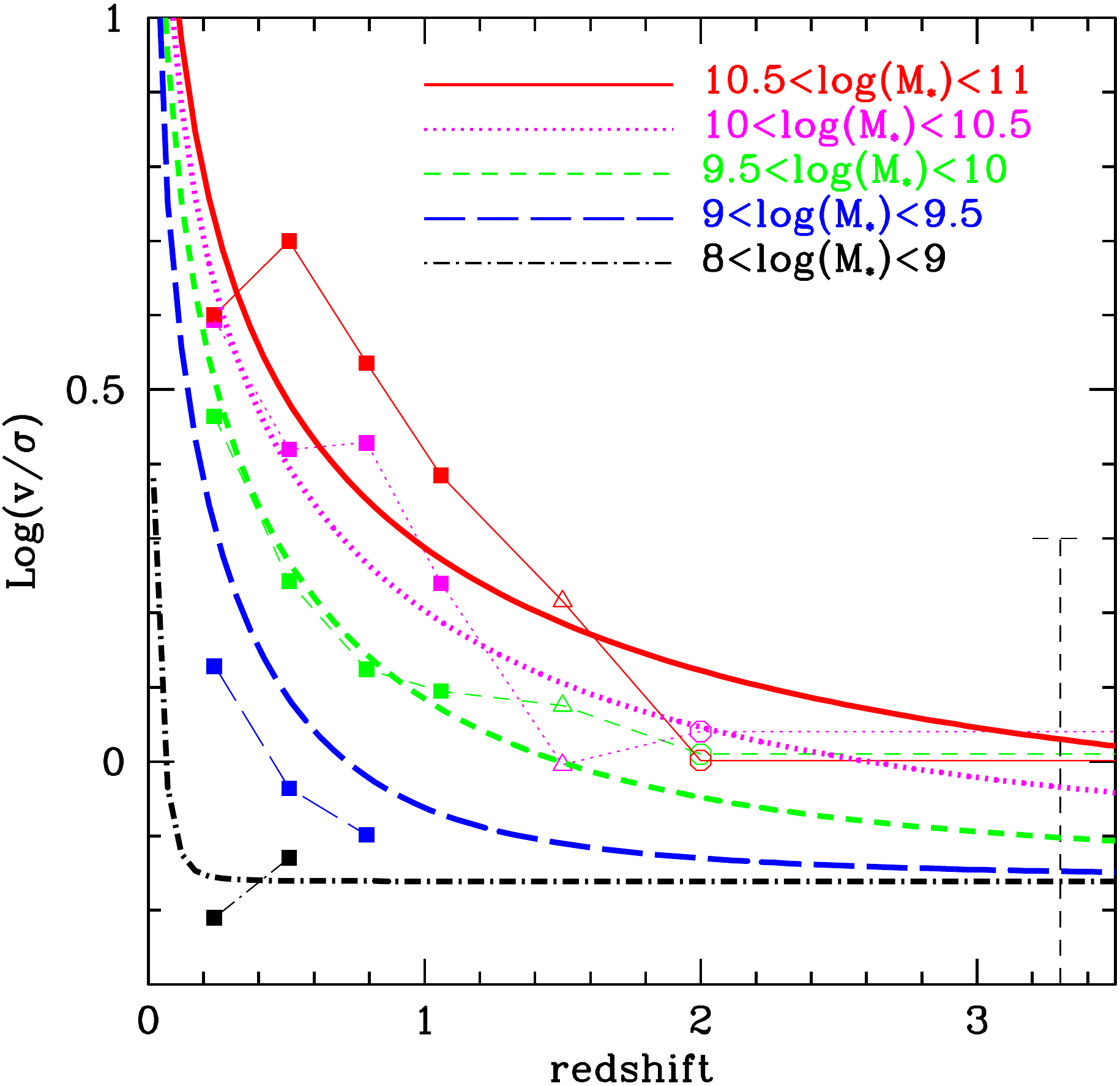}
\caption{$v/\sigma$ of gas on kpc scales in gas-rich galaxies. Points are derived from the literature: filled squares are derived from \cite{kassin12}, whereas open triangles and open circles are derived from the $z\approx1.5$ and $z\approx2$  samples respectively. Data sets has been grouped and binned as described in the main text. The dashed vertical line at $z=3.3$ marks the $v/\sigma$ range reported by \cite{gnerucci11}. Different colors and linestyles correspond to different mass bins, as labeled in the figure. Smooth thick lines represent the best fit to the data given by equation (\ref{vsusigmagas}), evaluated at the middle point of each ${\rm log}M_*$ bin. The average dispersion of the measurement of each data point is $\approx0.34$ dex in ${\rm log}(v/\sigma)$.}
\label{gas}
\vskip 0.5cm
\end{figure}

Equation (\ref{vsusigmagas}) describes a broken power-law where the slope at $x>x_0$ is
forced to zero, and the location of the break, $x_0$, depends on the stellar
mass of the system. The parameter $\beta$ describes the `sharpness' of the
transition between the two power-laws. The best least-square fit to the data
gives the following value for the five parameters: $a=0.6949$; $b=0.4548$;
$c=-8.5255$; $\alpha=-0.8680$; $\beta=3.2126$. 
\footnote{Note that the fitting function \eqref{vsusigmagas} becomes singular for low stellar masses
${\rm log}M_*<c$. For those masses, we simply assume $v/\sigma\approx 1$. We stress that
this simplifying assumption has a negligible impact on our results as such low stellar masses are only frequent at high redshifts
and for galaxies hosting small MBHs with mass $M_{\rm bh}\lesssim 10^6 M_\odot$. For such small MBHs, as will be clear later in this paper,
the Bardeen Petterson effect makes accretion effectively coherent, irrespective of the $v/\sigma$ ratio.} 
Data and best fit are shown in
Figure \ref{gas}. To account for the data dispersion around the best fit, we assume
that the density distribution for $v/\sigma$ is log-normal with
standard deviation of 0.34 dex around equation \eqref{vsusigmagas}. 

\subsubsection{In stellar bulges}
\begin{figure}
\includegraphics[width=\columnwidth]{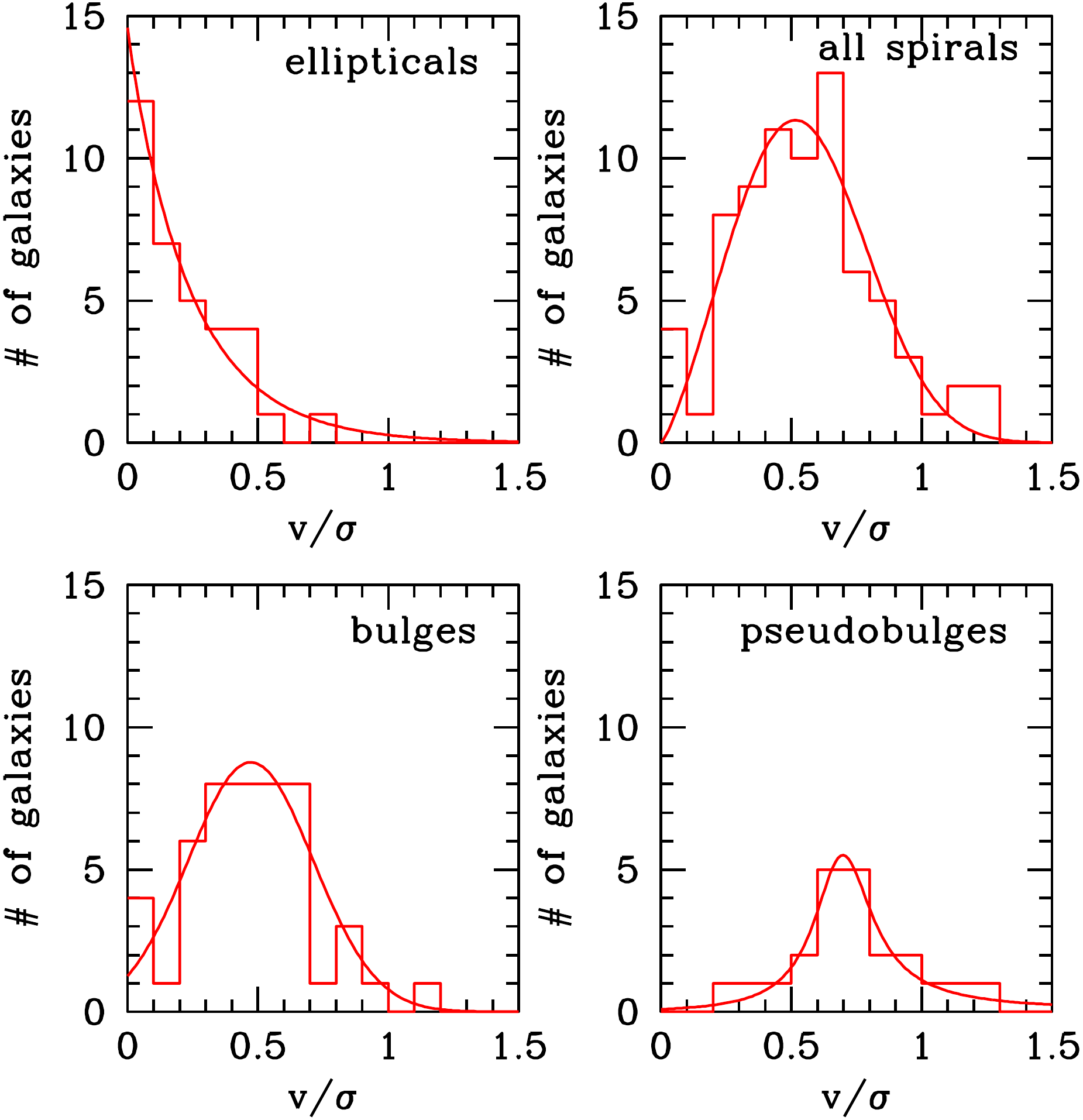}
\caption{Distribution of $v/\sigma$ for different galaxy types (as labeled in the panels) at low redshifts. In each panel, the red histogram is the distribution of observed $v/\sigma$, and the overlaid curve is the analytical fit reported in the main text.}
\label{bulges}
\end{figure}

The kinematics of stellar bulges has been also extensively studied in the literature. The SAURON project \citep{cappellari07} focuses on a sample of 48 ellipticals and lenticular galaxies, completed with 18 objects from the literature, for a total of 66 objects. Kinematic measurements are also available for stellar bulges of spiral galaxies; \cite{fabricius12} report measurements of $v/\sigma$ for bulges in 43 spiral galaxies of all classes, from S0 to Sc, also making a distinction between bulges and ``pseudobulges''. Using the total sample of 109 objects, we note that, in general, a distinction can be made between ellipticals and lenticular/spirals. The former have lower $v/\sigma$, clustering at zero values, whereas the latter have generally higher $v/\sigma$, extending to values greater than 1, and with an average greater than 0.5. Moreover, ``pseudobulge'' spirals tend to rotate faster than classical bulges, as shown in Figure \ref{bulges}. The $v/\sigma$ distribution is analytically fitted with the 
following functions:
\begin{itemize}  
\item ellipticals
\be
f(x)=ae^{bx^c},
\label{ellipticals}
\ee
with $a=14.59, b=-3.98, c=0.96$;
\item bulges and pseudobulges in spirals
\be
f(x)=ax^be^{cx^d},
\label{spirals}
\ee
with $a=42.24, b=1.29, c=-2.95, d=2.81$;
\item bulges in spirals
\be
f(x)=\frac{a}{b}x^be^{-0.5[(x-c)/b]^2},
\label{bulgefit}
\ee
with $a=2.10, b=0.24, c=0.47$;
\item pseudobulges in spirals
\be
f(x)=\frac{a(x-b)}{x^2+cx+d},
\label{pseudobulges}
\ee
with $a=0.083, b=-0.53, c=-1.38, d=0.50$.
\end{itemize}
The distribution of $v/\sigma$ for the different galaxy samples together with the corresponding $f(x)$ fitting formulas given by equations (\ref{ellipticals}-\ref{pseudobulges}) are shown in figure \ref{bulges}. When implemented in our code, we normalize each $f(x)$ so that $\int_0^\infty f(x)dx=1$, to get a probability density function (PDF). The $v/\sigma$ of each individual galaxy is then drawn randomly from the appropriate PDF, according to its morphological type, as described in detail in the next Section. 

\begin{table}
\begin{center}
\begin{tabular}{ccc}
\hline
 & {\rm light seeds} & {\rm heavy seeds} \\
\hline
 $M_{\rm cloud}$ & $3\times 10^4 M_\odot$ & $3\times 10^4 M_\odot$ \\
 $\epsilon_{\rm SN,b}$  &  0.4 & 0.4  \\
 $\epsilon_{\rm SN,d}$  &  0.1 & 0.1  \\
 $f_{\rm jet}$    & 10   &  10\\
 $A_{\rm res}$    & $6\times 10^{-3}$   &  $5.75\times 10^{-3}$  \\
 $A_{\rm Edd}$   & 2.2 & 1\\
 $k_{\rm accr}$   & $10^{-3}$& $10^{-3}$\\
\hline
\end{tabular}
\caption{The calibrated values of the free parameters of the model (see B12 and
Section~\ref{sec:spin} for their meaning). 
\label{table:parameters}}
\end{center}
\end{table}

\subsection{Implementation in our galaxy formation model}
\label{sec:implementation}
Following what is usually done in semianalytic models \cite[see e.g.][]{guo,2013MNRAS.433.2986W,2014MNRAS.437.1576B}, we 
define ellipticals as those galaxies with a bulge to total mass ratio $B/T>0.7$, and spirals the remaining galaxies (i.e. $B/T<0.7$).

Since in {\it ellipticals} (in particular the most massive ones at lower redshifts) cold gas is subdominant, 
we always assume that in these galaxies  the dynamics of the gas feeding the MBH traces that of the stellar population, 
and we thus extract $v/\sigma$ from the observational distribution given by equation (\ref{ellipticals}) for ellipticals. 

Fueling of the MBH in {\it spiral galaxies} might be more subtle. 
We therefore explore three different models that encompass plausible scenarios:

\renewcommand{\labelenumi}{\roman{enumi}}

{\it i) pseudobulge} model. Stars in the bulge are formed out of the same galactic flow that 
simultaneously replenishes the gas reservoir available to accrete on the MBH. 
Therefore, one may conjecture that this common origin should imply a similar velocity dispersion. 
This may happen if the dissipation (through shocks) in the gas of the reservoir acts on a longer timescale than accretion. 
In our model, spiral bulges form either via disk instabilities or via mergers. In the former case, we assume that a pseudobulge forms,
assigning the reservoir a $v/\sigma$ ratio drawn from the distribution given in equation (\ref{pseudobulges}) for pseudobulges hosted in spirals; in the latter case 
we assume that a classical bulge forms, extracting $v/\sigma$ from the observational distribution given in equation (\ref{bulgefit}) for classical bulges hosted in spirals.
The assigned value of $v/\sigma$ is then kept constant during the galaxy's evolution, as long as the morphology does not change. 
In our model, morphology changes can be triggered by  
 disk instabilities or major mergers. Thus, whenever one such event changes the galaxy morphology, we reset the value of $v/\sigma$ according to the PDFs above. 
 In the case of minor mergers, instead, galactic disks are usually not disrupted, preserving the galaxy morphology.
Hence, we simply take the $v/\sigma$ of the resulting galaxy to be the average 
of the $v/\sigma$ of the progenitors, weighed with their respective baryonic
masses. In practice, the resulting $v/\sigma$ is always much closer to that of the 
more massive progenitor. \footnote{We refer the reader to B12 for more detail on the definition and  implementation of minor/major mergers and disk instabilities.}

{\it ii) disk} model. The gas fueling the MBH comes from kpc scales. An alternative model can thus be envisaged where 
the low angular momentum reservoir accreting onto the MBH shares the kinematic properties of the large-scale gaseous disk. 
This is because one may conjecture that gaseous disks at all scales should be affected by major mergers and/or bar instabilities in similar ways, when such events occur.
To apply this scenario, we need to extract, for each spiral galaxy, a $v/\sigma$ ratio from a log-normal distribution with average given by equation (\ref{vsusigmagas}) and standard deviation of 0.34 dex.
However, tracking the cosmic evolution of MBH spins in this model is slightly more complex than in the {\it pseudobulge} model, because the typical $v/\sigma$ 
of the large scale disk in spirals depends explicitly on the stellar mass $M_*$ and on $z$ (c.f. equation \eqref{vsusigmagas}), unlike what happens
for ellipticals (c.f. equation \eqref{ellipticals}). This calls for a different implementation of
the velocity dispersion evolution in ellipticals and in spirals.
More specifically, when a galaxy has $B/T>0.7$, we assign it a $v/\sigma$ ratio from the observational distribution given by equation \eqref{ellipticals}
for ellipticals, and keep it constant until a disk instability or a major merger changes the morphology. When $B/T<0.7$, we assign the galaxy an ``intrinsic dispersion''
$\chi$ relative to the typical value given by  equation (\ref{vsusigmagas}). More precisely, we draw
$\chi$ from a normal distribution
with zero average and standard deviation $0.34$, and then relate the galaxy's $v/\sigma$ to this intrinsic dispersion via
${\rm log}(v/\sigma)={\rm log}(v/\sigma)\vert_{\rm av}(M_*,z)+\chi$, where $(v/\sigma)\vert_{\rm av}(M_*,z)$ is the value of the observational fit given by equation (\ref{vsusigmagas}) for gas in spirals. 
This procedure thus assigns the galaxy a $v/\sigma$
ratio drawn from a log-normal distribution with average given by equation (\ref{vsusigmagas}) and dispersion 0.34 dex.
The intrinsic dispersion $\chi$ is then kept constant along the quiescent galaxy evolution phase, 
while $(v/\sigma)\vert_{\rm av}(M_*,z)$ evolves as $M_*$ and $z$ change. 
When a morphology change (following a major merger or a disk instability) occurs we reset the value of $v/\sigma$ based on the
PDF that is appropriate, as dictated by the new galaxy morphology, 
while at minor mergers we combine the $v/\sigma$ ratios in a weighed average as in the {\it pseudobulge} model.

 {\it iii) hybrid} model. This model is intermediate between the
  {\it disk} and the {\it pseudobulge} models, because
  in the case of classical bulges it relates the MBH fueling in spirals to the stellar bulge
  kinematics, 
  while in the case of pseudobulges the MBH fueling is linked to the kinematics of the large scale gaseous disk. 
  The motivation is a possible different origin for the gas reservoir in the two cases.
  The gas accreting on the MBH in pseudobulges may be more likely to retain a
  certain degree of coherence, because it originated from a bar instability. This residual coherence may be
  larger than in the stellar component, because unlike gas, stars do not dissipate random motions through
  shocks. In the case of classical bulges, we thus
  extract $v/\sigma$ from the observational distribution given in equation
  (\ref{spirals}) for all spirals\footnote{The observational distributions for classical bulges in spirals (equation \eqref{bulgefit}) or
  for pseudobulges in spirals (equation \eqref{pseudobulges}) yield analog results, but the combined
  distribution given by  (\ref{spirals}) has better statistics because based on a larger sample.},
  while for pseudobulges we assume, as in model \textit{ii)}, a log-normal distribution with average given by equation (\ref{vsusigmagas}) and standard deviation of 0.34 dex. 
  The evolution of the $v/\sigma$ ratios for the accreting reservoir is tracked using the (combined) techniques applied in model \textit{i)} and \textit{ii)}.
 
\begin{figure}
\includegraphics[width=8.5cm]{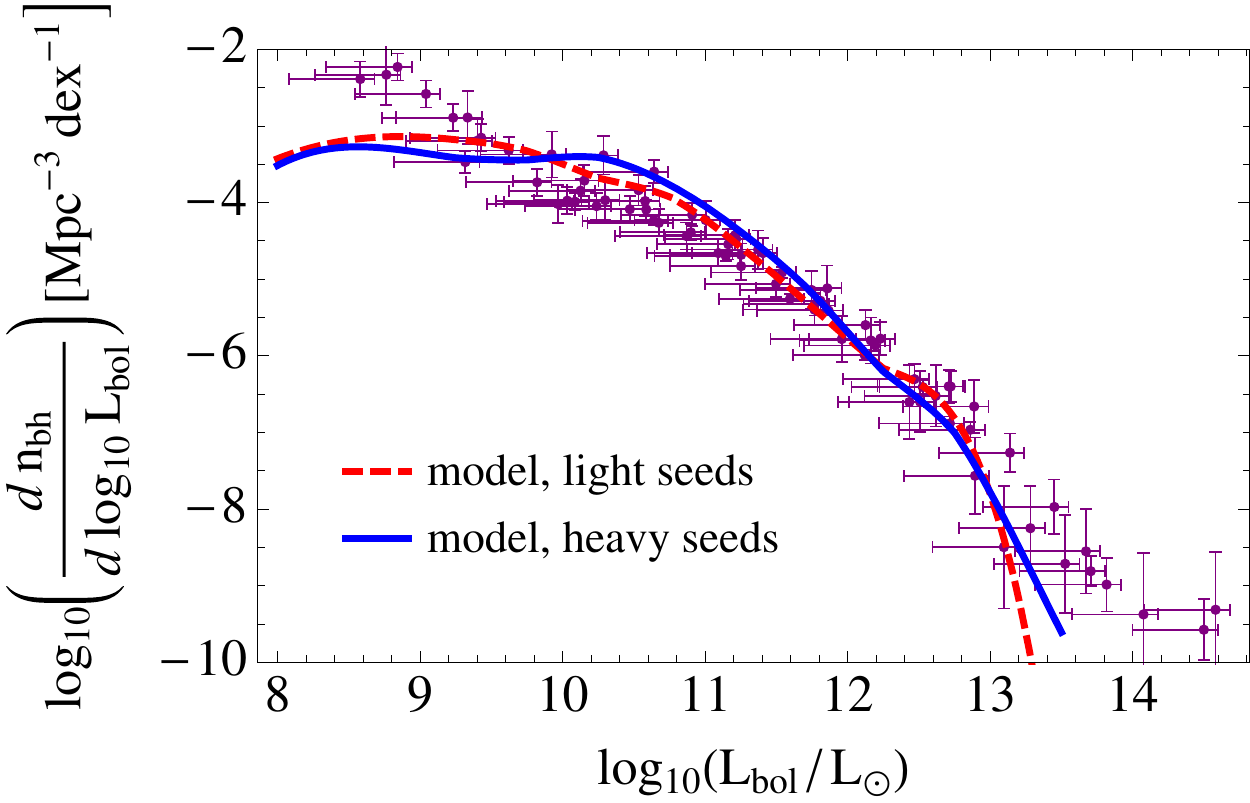}
\caption{
The MBH bolometric luminosity function at $z=0.1$ predicted by our model 
vs the compilation of observational data by \citet{hopkins}. The vertical error bars on the data are from \citet{hopkins},
while for the horizontal ones we have assumed $-0.5$ dex and $+0.1$ dex to account 
for possible overestimation of the bolometric correction \citep{lusso}.
\label{fig:phiL}}
\end{figure}

For the sake of comparison, we also investigate three more idealized models for accretion onto the MBH that are often adopted in the literature, namely
a purely coherent accretion scenario (where accretion always takes place on a prograde accretion disk orthogonal to the MBH spin) and two variants of a purely chaotic accretion scenario
(where the accreting gas has a perfectly isotropic angular momentum distribution). We postpone a detail description of these models and their implementation to Appendix \ref{app:co-cha}.

For each of the accretion models that we consider, we perform two sets of runs seeding MBHs 
either as PopIII remnants at $z\sim 20$ ({\it light seeds}) or as end product of direct collapse at $z\sim 15$  ({\it heavy seeds}); see B12 for more details. 
 Each run consists of $\sim 24000$ merger trees{\footnote{For the coherent/chaotic models we perform shorter 
runs with a factor of 20 less halos.}} in the dark-matter halo mass range $10^{10}$ -- $10^{15} M_\odot$. 
In this paper we focus on MBHs of $M_{\rm bh}>10^{6}\msun$ at 
relatively low $z$, because we aim at a comparison with measured spins in the local Universe. 
Since memory of the seeding process fades away already at high redshift, we did not find any 
significant difference between the light and heavy seed prescriptions at any considered redshift, 
for any considered galaxy subsample. We therefore add up light and heavy seed runs to increase 
the statistics and ``smoothness'' of our theoretical samples to be compared with the observed data (see Section \ref{theosample}). 

\subsection{Linking $v/\sigma$ to the isotropy parameter $F$}
\label{sec:Fvsigma}

When the dispersion of the low angular momentum reservoir available for accretion is known, one can compute
the amount of prograde/retrograde accretion and derive the MBH spin evolution (see Section \ref{sec:spin}).
The relation between the $v/\sigma$ ratio of the accretion flow and the geometrical parameter $F$ introduced earlier
to describe the anisotropy of the angular momentum distribution of the accreting clouds can be derived as follows.
 Suppose that the net
rotation is along the $z$ axis. Each lump of matter falling into the MBH has
an angular momentum with $z$-component
 $L_z= m r\sin\theta (v+\sigma_\varphi)$, where $\theta$ and $\varphi$ are the lump's polar and
azimuthal angles in a spherical coordinate system, and $\sigma_\varphi$ is 
the projection of  $\boldsymbol{\sigma}$ on the tangential direction $\boldsymbol{e}_\varphi$.
Let us now write $\sigma_\varphi = \sigma \cos\alpha$, where $\alpha$ is the angle between  $\boldsymbol{\sigma}$
and $\boldsymbol{e}_\varphi$. If $\boldsymbol{\sigma}$ is isotropically distributed, 
$\cos\alpha$ has a uniform probability distribution, i.e. d$P/$d$\cos\alpha=1/2$.
If $v/\sigma>1$, $L_z$ is always positive and $F=1$. If $0\leq v/\sigma<1$,
the probability of having $L_z>0$ (or $L_z<0$)
is proportional to $1+v/\sigma$ (or $1-v/\sigma$), hence $F=(1+v/\sigma)/2$.
In summary, we assume
\begin{equation}\label{eq:vsigma}
\begin{array}{lll}
v/\sigma>1 &\rightarrow &F=1\,;\\
v/\sigma=0 &\rightarrow &F=0.5\,;\\
0<v/\sigma<1 &\rightarrow &F=(1+v/\sigma)/2\,.
\end{array}
\end{equation}

{ One subtlety should be noted about equation \eqref{eq:vsigma} and its derivation. When using 
the parameter $F$ resulting from this equation into equation \eqref{eq:w}, one
is implicitly identifying 
the reference system used above to derive equation \eqref{eq:vsigma}
(i.e. one where the $z$ axis is along the total angular momentum of
the gas reservoir, out of which the accreting lumps are drawn) 
with the reference system used to derive equation \eqref{eq:w} (i.e.
one where the polar angle 
$\theta_{\rm out}$, is defined 
relative to a $z$ axis parallel to the MBH spin).
Therefore,
we are implicitly assuming that the MBH spin is exactly aligned with the average angular
  momentum of the clouds, already when the {\it first} cloud hits the MBH.
This may not be the case under many
  circumstances, but the impact of this
assumption on our predictions for the MBH spin \textit{magnitudes}
is expected to be small, for the following two reasons:
\textit{(i)} When the MBH is light enough ($M_{\rm bh}\lsim 10^7 \msun$) to be in the $J_{\rm
    disk}/2J_{\rm bh}>1$ regime, the MBH spin realigns with each cloud angular momentum 
on a timescale shorter than the accretion time. Therefore, assuming the MBH spin to be always aligned with the clouds average rotation axis does not affect significantly the evolution of the spin \textit{magnitude};
 \textit{(ii)} The high mass regime of our model is exact for the
  two limiting cases $F=0.5$ and $F=1$, and also preserves the correct
  monotonic trend for $0.5<F<1$ (i.e. the higher $F$, the higher the MBH spin magnitude). 
 It would indeed be possible to estimate the detailed effect of the initial MBH orientation
for  $0.5<F<1$ by comparing Figure~\ref{fig:a_eq} with figures~6 and 7
in D13 for $M_{\rm bh} \gsim 10^9 \msun$, but we postpone a
more detailed analysis to a future work.}

\section{Calibration against observables}
\label{sec:cali}
As in B12, we anchor our model to a number of observables, with particular attention to
quantities characterizing the MBH population and their accretion properties in the local $z\sim 0$  
universe, where all the currently available MBH spin measurements are performed.

\begin{figure}
\includegraphics[width=8.5cm]{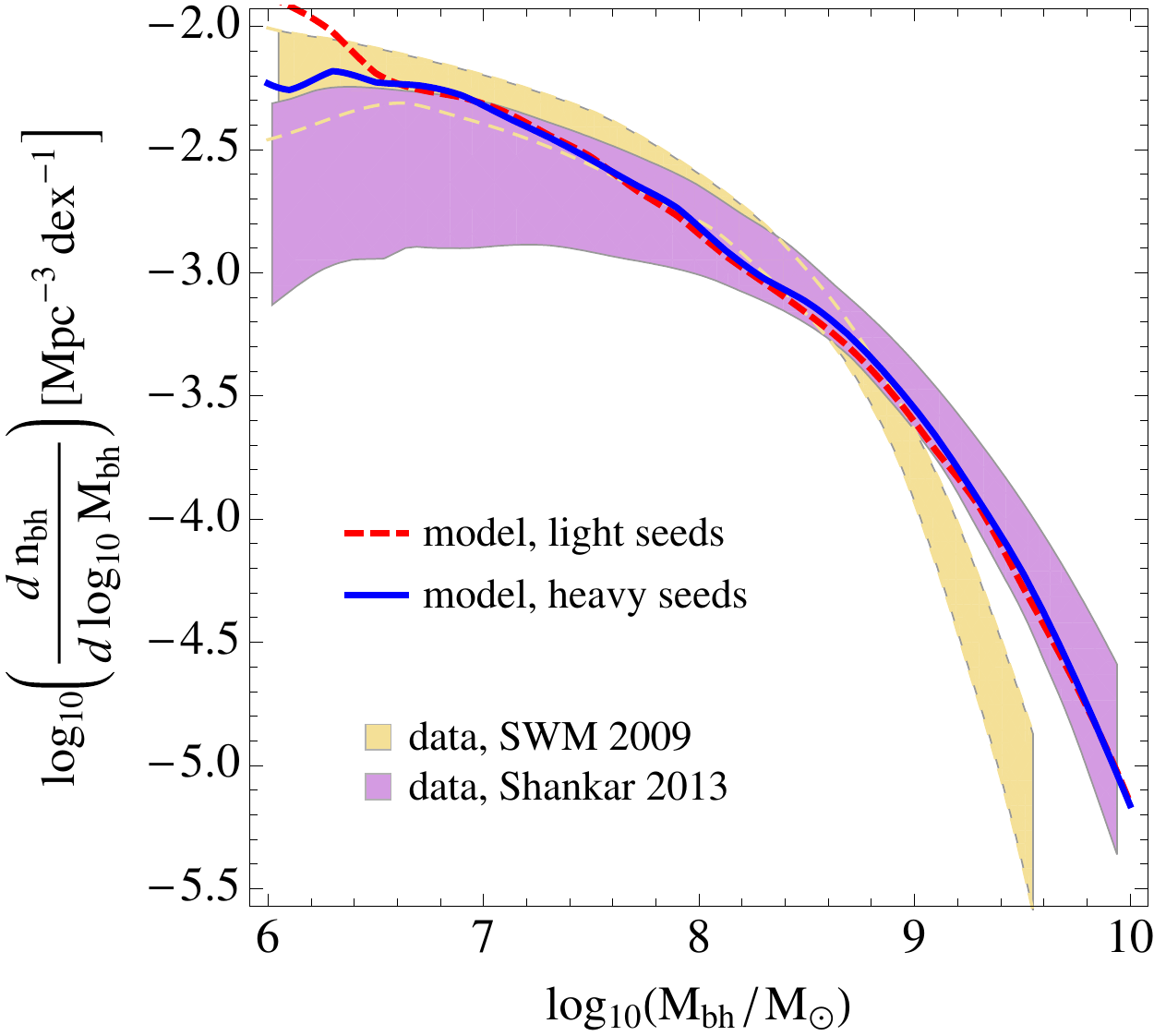}
\caption{
The MBH mass function at $z=0.1$ predicted by our  model vs the ranges, 
reconstructed from observational data, given by \citet{swm09} and \citet{shankar13}. 
Note that although the results of \citet{shankar13} are still 
preliminary and are based only on bulge-dominated galaxies, it is clear that
systematic errors still dominate the uncertainties in the high-mass end.
\label{fig:phiBH}}
\end{figure}
\begin{figure}
\includegraphics[width=8.5cm]{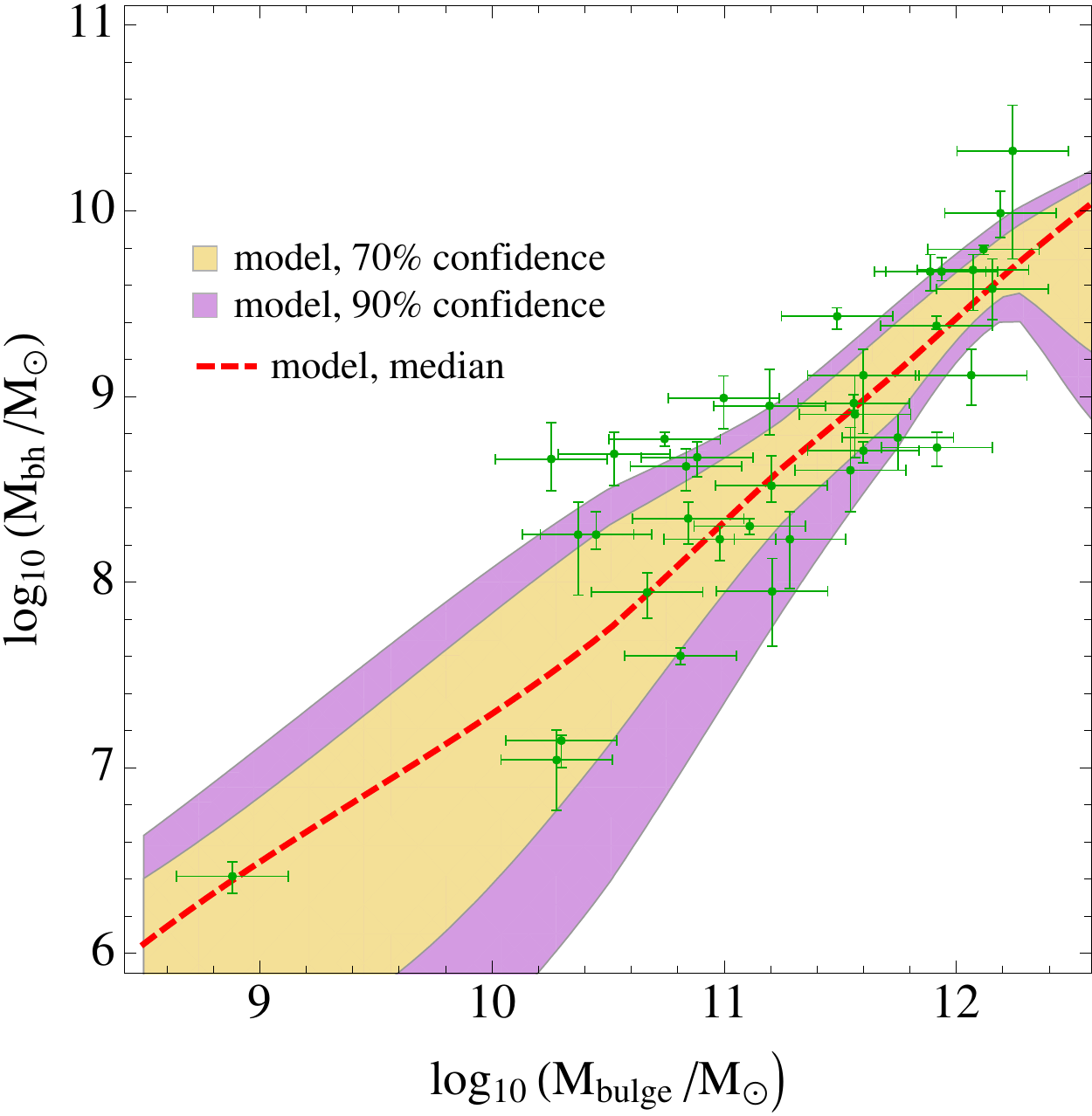}
\caption{
The relation between the MBH mass and the dynamical mass of the bulge predicted by our  model 
at $z=0$, compared to the data compilation of \citet{mcconnell_ma}. This figure adopts a light-seed model, but the heavy-seed model 
leads to similar results.
\label{fig:magorrian}}
\end{figure}
\begin{figure}
\includegraphics[width=8.5cm]{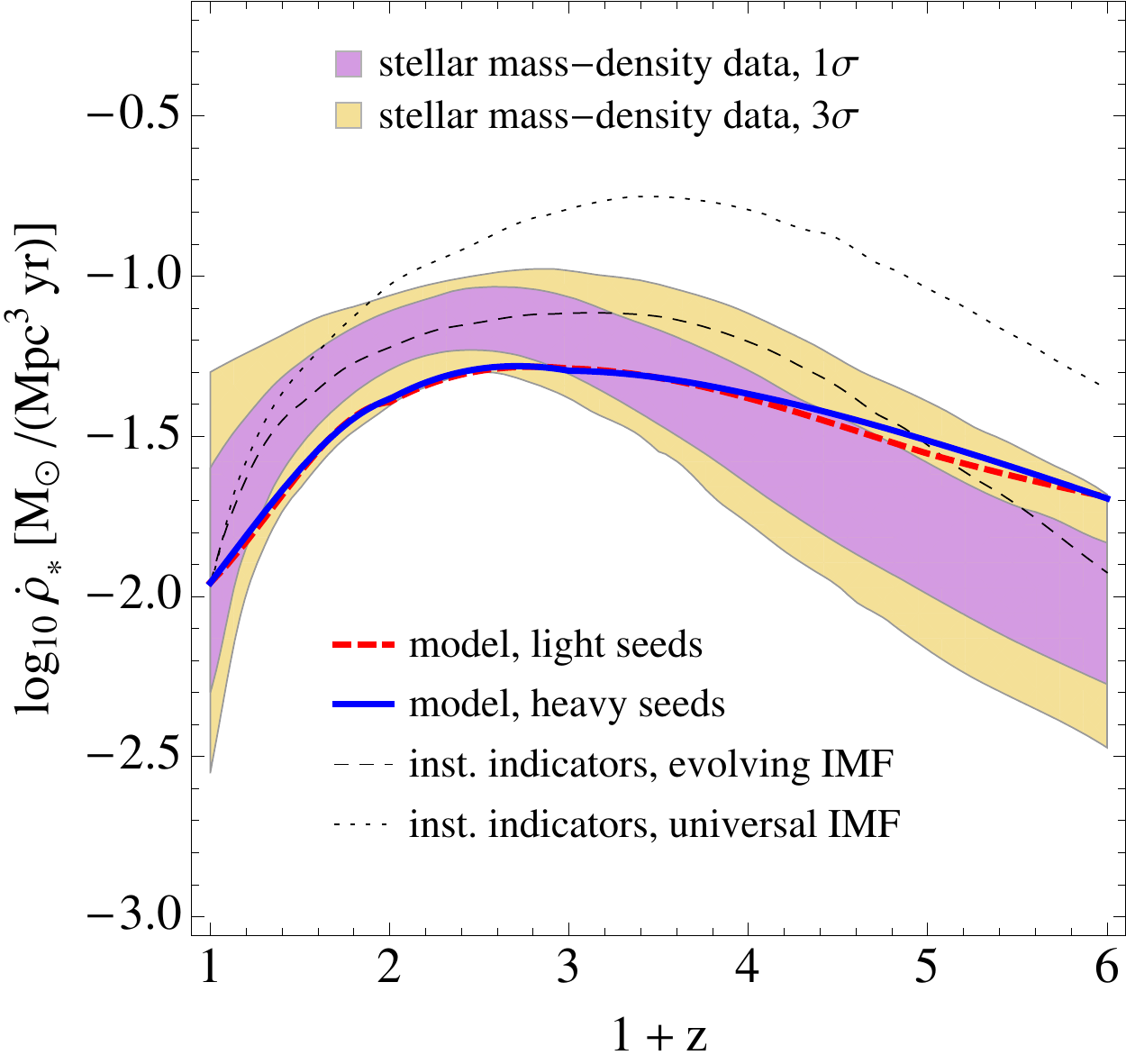}
\caption{The star formation density as function of redshift, as predicted by our model vs 
the one derived by \cite{sfr_history} from observations of the stellar mass density and assuming an evolving IMF (shaded regions,
corresponding to $1\sigma$ and $3\sigma$ confidence regions). We also show fits to instantaneous star-formation density indicators,
assuming either a universal or an evolving IMF~\citep{sfr_history}.
\label{fig:sfr}}
\end{figure}
\begin{figure}
\includegraphics[width=8.5cm]{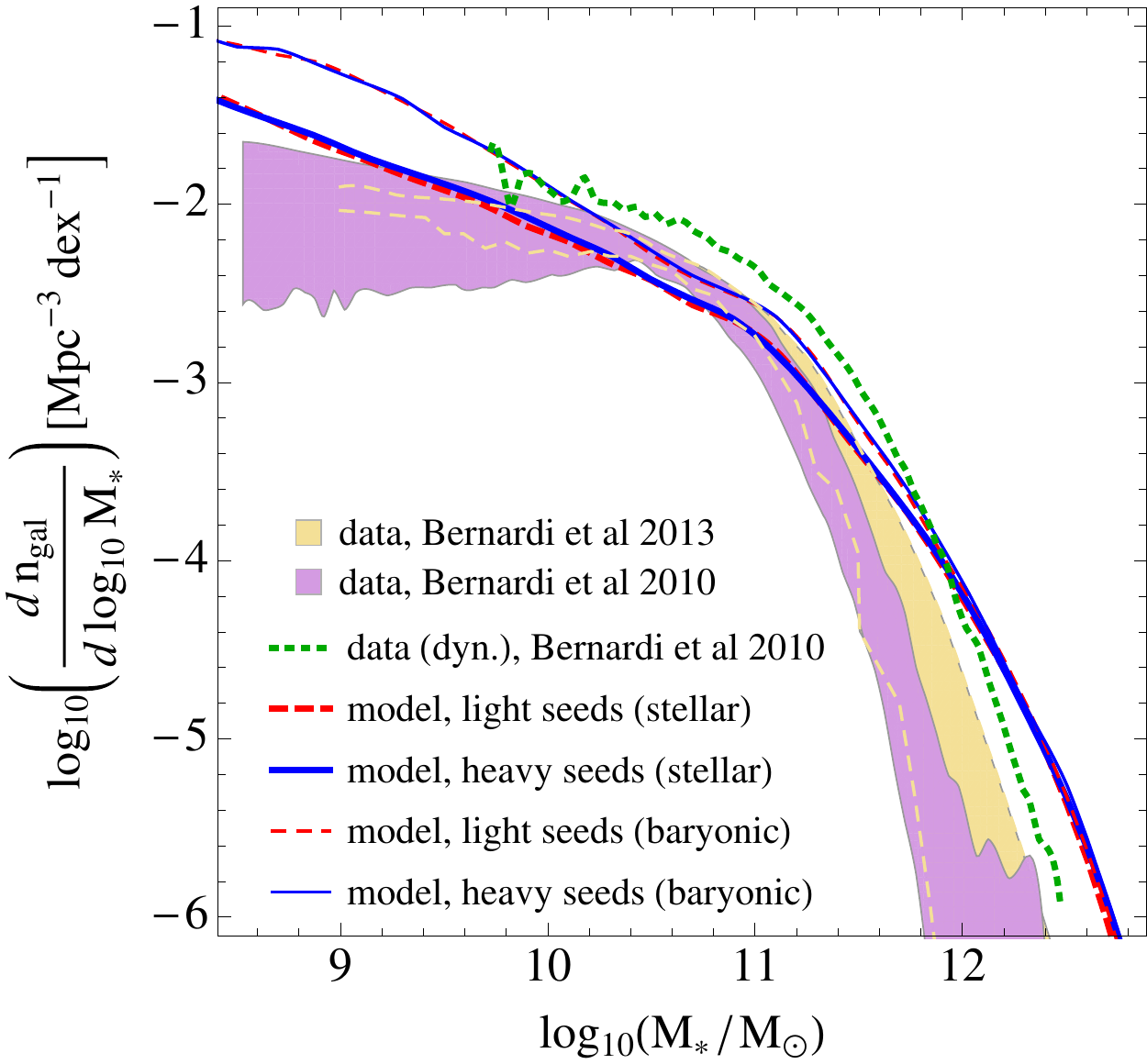}
\caption{
The mass function of galaxies at $z=0$ predicted by our model (including the stellar mass alone, or the total
baryonic mass in gas and stars) vs the estimates of \citet{bernardi10} and \citet{bernardi13} for the stellar-mass function from luminosity observations, and the estimate of \citet{bernardi10} for the dynamical mass function (obtained by reconstructing the dynamical masses through the galaxy's size and velocity dispersion).
\label{fig:phiGal}}
\end{figure}
\begin{figure}
\includegraphics[width=6.2cm,angle=-90]{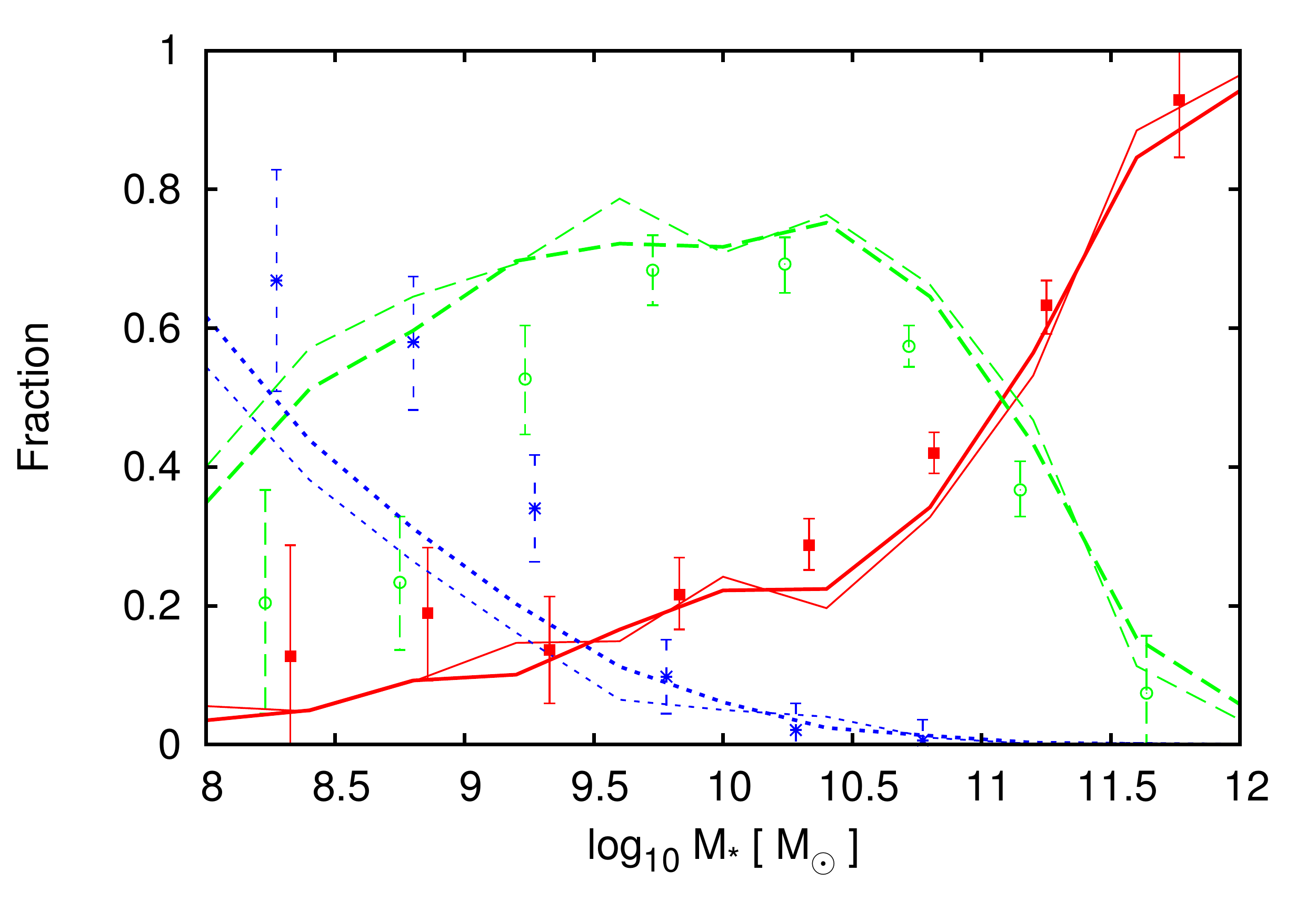}
\caption{
Morphologies predicted by our model at $z=0$ as a function of the stellar mass
(thick lines: heavy seeds; thin lines: light seeds), vs data from \citet{conselice}
(squares: ellipticals; circles: spirals; stars: irregulars).
\label{fig:morphology}}
\end{figure}

More precisely, we calibrate our free parameters (c.f. Table~\ref{table:parameters}) against
observational data for the QSO luminosity function at $z=0.1$ (Figure~\ref{fig:phiL}), 
the MBH mass function at $z=0$ (Figure~\ref{fig:phiBH}), the local ``Magorrian'' relation between the MBH mass and the
bulge dynamical mass in ellipticals (Figure~\ref{fig:magorrian}), the star formation density for $0\leq z \lesssim 5$ (Figure~\ref{fig:sfr}), the stellar mass function at $z=0$ (Figure~\ref{fig:phiGal}), and the observed fraction of galaxies with
given morphology at $z=0$ (Figure~\ref{fig:morphology}). 
We stress, however, that the results of our model (and the overall conclusions of this paper) 
do not depend strongly on the values of these free parameters. 
Also, the results shown in this section are for the {\it pseudobulge}
model (c.f. Section \ref{sec:implementation}), but the {\it disk} and {\it hybrid} models produce similar results.

As can be seen from Figure~\ref{fig:phiL}, our model gives reasonable predictions 
for the $z=0.1$ QSO luminosity function when compared
to the compilation of observational data by \citet{hopkins}, despite slightly underestimating it at low luminosities (the disagreement
at the high-luminosity end is less statistically significant). The vertical error bars on the data are from \citet{hopkins},
while for the horizontal ones we have assumed $-0.5$ dex and $+0.1$ dex to account for possible overestimation of the bolometric correction \citep{lusso}.  
Similarly, Figure~\ref{fig:phiBH} shows that our predictions for the $z=0$ black-hole mass function are in agreement with recent estimates
by \citet{shankar13}. While the reconstruction of the MBH mass function performed by \citet{shankar13}
is still preliminary and based only on bulge-dominated galaxies, from a comparison to the observational estimates of
\citet{swm09} (also shown in  Figure~\ref{fig:phiBH}), it is clear that systematic errors still dominate the determination of the mass
function at the high-mass end, and that our model's prediction is clearly within the observational uncertainties. 
Figure~\ref{fig:magorrian} shows the agreement between a compilation of dynamical bulge masses and black-hole
masses in elliptical galaxies form \cite{mcconnell_ma} \citep[see also][]{haring,magorrian} and 
our predictions for ellipticals.

\begin{figure}
  \centering
   \includegraphics[width=0.44\textwidth]{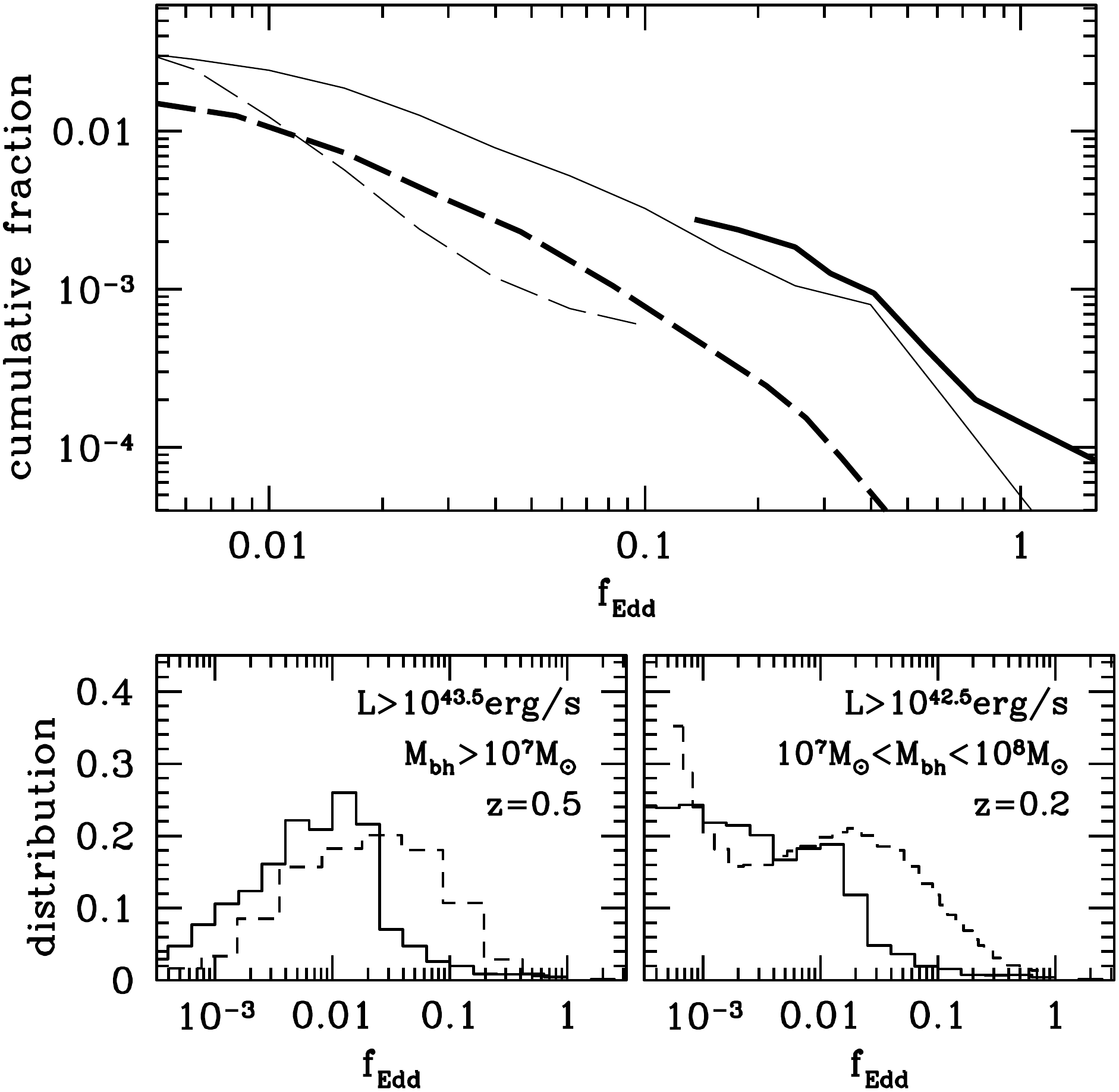}
\caption{Distribution of Eddington ratios $f_{\rm Edd}$. In the top panel the
cumulative fraction of MBHs accreting above a certain Eddington ratio is shown 
for two different MBH mass bins. Thin lines show the predictions 
of our model at $z=0.1$ for the mass bins $2\times10^6\msun<M<5\times10^6\msun$ (solid) 
and $5\times10^7\msun<M<2\times10^8\msun$ (long--dashed). 
Those are compared with observational estimates from \cite{heckman04} 
(thick lines) evaluated at $M_{\rm bh}=3\times10^6\msun$ (solid)
and $M_{\rm bh}=10^8\msun$ (long--dashed). In the bottom panels, the $f_{\rm Edd}$ distribution for selected systems (as labeled in each panel) are compared to data from \cite{hickox09} (left) and \cite{heckman09} (right). Solid histogram are the predictions of our model, while dashed histograms are the data.}
\label{eddf}
\vskip 0.5cm
\end{figure}

\begin{figure*}
\begin{center}
\includegraphics[width=1.4\columnwidth,angle=-90]{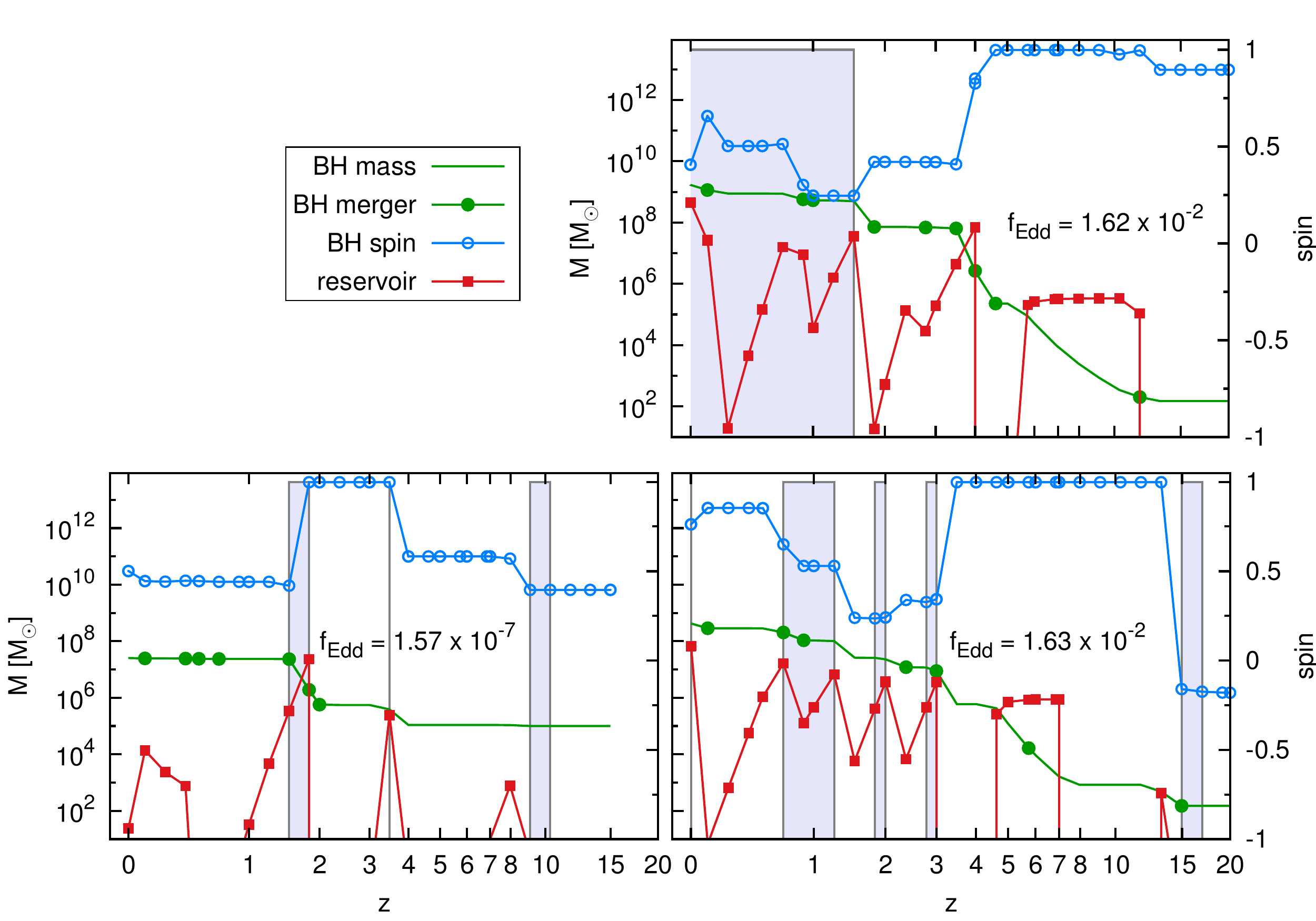}
\caption{Examples of main-progenitor evolutionary tracks of three MBHs selected in the {\it pseudobulge} model. The MBH spin is shown
with a blue solid line with empty circles; the mass of the MBH with a solid green line (with filled circles representing the MBH mergers); the reservoir mass with
a solid lines with filled squares. The lavender shaded areas indicates the redshifts at which the host is bulge dominated ($B/T>0.7$).}
\label{tracks_bulge}
\end{center}
\end{figure*}
\begin{figure*}
\begin{center}
\includegraphics[width=1.4\columnwidth,angle=-90]{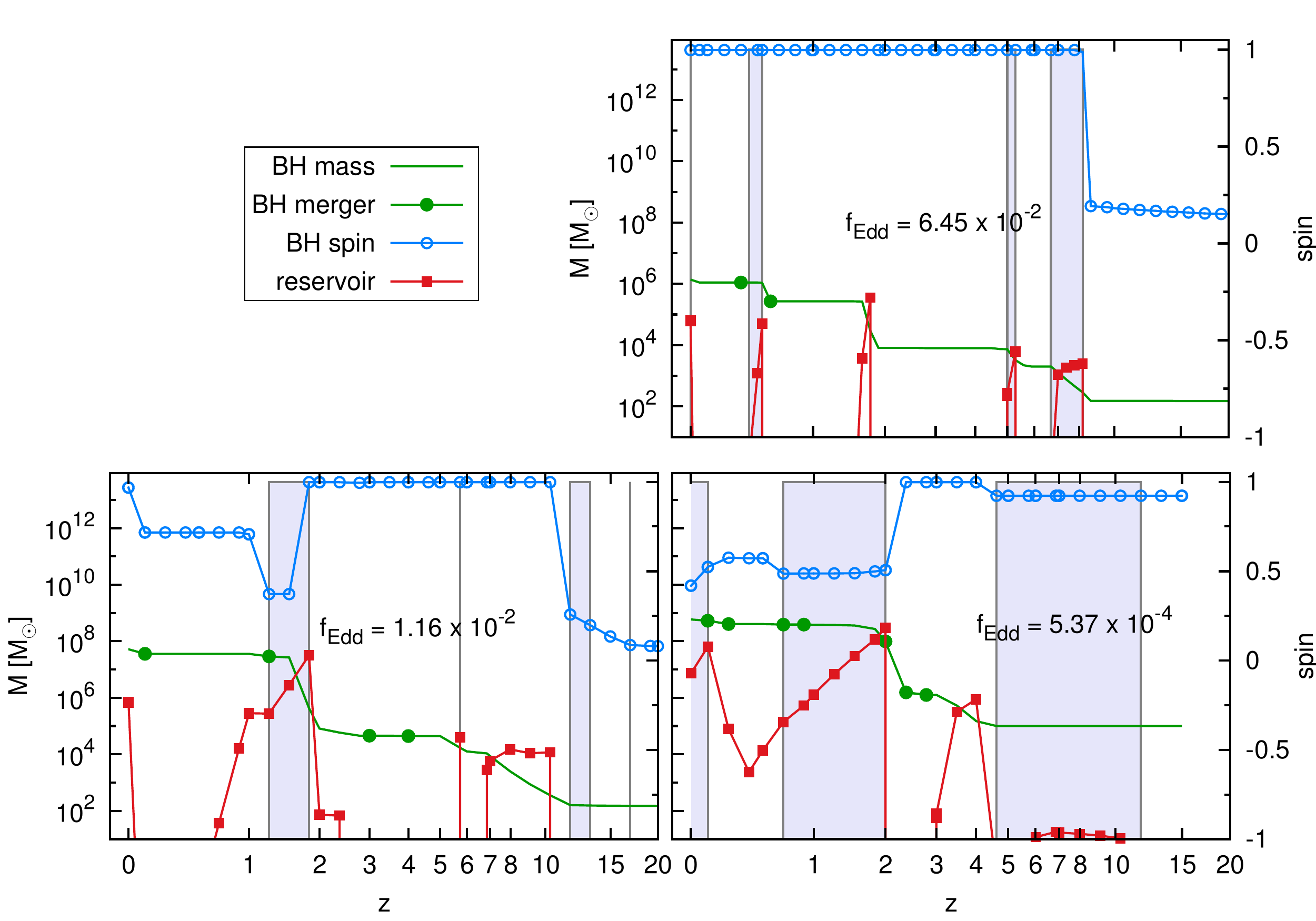}
\caption{Same as Figure \ref{tracks_bulge}, but for the {\it disk} model.}
\label{tracks_kassin}
\end{center}
\end{figure*}
 
The improved star-formation implementation described in Section~\ref{sec:SF} allows our model to better 
reproduce data for the star-formation history
than the original prescription of B12. Figure~\ref{fig:sfr} compares our predictions to reconstructions based either
on observations of the stellar mass density and assuming an evolving IMF \citep{sfr_history}, or to fits to instantaneous 
star-formation density observations, assuming either a universal or an evolving IMF~\citep{sfr_history}.
Our predictions for the local galaxy mass function (Figure~\ref{fig:phiGal}) 
are instead slightly higher than the observational determinations by \citet{bernardi10} 
and \citet{bernardi13} at the low- and high-mass ends. The same problem is often seen in galaxy 
formation models, which tend to over-produce stars \citep[see, e.g.][]{delucia2007,guo,dimatteo2014}, possibly
because of overly-simplified treatments of AGN and supernova
feedback. More sophisticated feedback schemes might improve the match with observations, and we plan to 
investigate this issue further in future work. However, one should also keep in mind that
determinations of the mass function are typically obtained by estimating galaxy masses through their luminosity,
which results in large systematic uncertainties (especially at the high-mass end), due to 
the assumptions on the stellar mass-to-light ratio, as well as the different possible ways in which one can 
fit the light profiles \citep{bernardi13}.
Finally, Figure~\ref{fig:morphology} shows the predicted fraction of ellipticals, spirals and irregulars as a function of stellar mass,
together with the observational estimates of \cite{conselice}. More specifically, following \citet{guo}, 
we use the fraction of total baryonic mass in the bulge, $B/T=M_{\rm b}/(M_{\rm b}+M_{\rm d})$ (with $M_{\rm b}$ and $M_{\rm d}$ the bulge and disk masses), to
discriminate the various morphologies, i.e. we classify a galaxy as an elliptical when $B/T>0.7$, as a 
spiral when $0.03<B/T<0.7$, and as an irregular when $B/T<0.03$.\footnote{Qualitatively similar results
are obtained using stellar rather than total baryonic masses to define $B/T$, but it seems preferable to use
total baryonic masses if one wants to apply the same classification to high redshifts, where gas may dominate
over stars.} In spite of this simplistic classification, 
our model reproduces, at least qualitatively, the observational results.
Since in section \ref{sec:obs} we will be dealing with accreting systems, in Figure~\ref{eddf} we compare the Eddington ratio, $f_{\rm Edd}$, distribution predicted by our model to observations. In the top panel, we compare the local cumulative distribution of $f_{\rm Edd}$ as measured by \cite{heckman04}, who inspected 23.000 active galaxies in a complete sample of 123.000 galaxies from the SDSS in the redshift range $0.05<z<0.2$, to the predictions of our {\it pseudobulge} model at $z=0.1$ (other models yield similar results). The model reproduces the $f_{\rm Edd}$ observed trend satisfactorily, even though it slightly underpredict the fraction of accreting systems at $f_{\rm Edd}>0.01$. The lower panels highlight the difference between the model and the data at $f_{\rm Edd}\approx0.1$. The figure show that our model catches the main trends in the low redshift Eddington ratio distributions, although not perfectly. A better calibration of the model against this quantity may be considered in future work. 

\section{MBH spin evolution}

\label{sec:results}
The spin evolution of each MBH is dictated by a complex interplay of $i)$ the variable supply of mass available for accretion, $ii)$ the time-changing host galaxy properties and morphology, and $iii)$ MBH binary coalescence events. We will first inspect some examples of individual MBH-galaxy evolutionary tracks, to illustrate the effect of those ingredients (Section \ref{individualspin}). 
 We stress that the few examples we consider do not represent the whole variety of evolutionary paths predicted by our model. Predictions on 
statistically significant samples of objects will be presented in Section \ref{spinglobal}.

\subsection{Spin evolution of individual MBHs}
\label{individualspin}

Figure \ref{tracks_bulge} gives three examples of evolution of $z=0$
spirals for the {\it pseudobulge} model. More specifically, we show the 
evolution of the MBH mass (with MBH mergers shown with filled circles) along the
main MBH-progenitor history (i.e. at each MBH merger we follow the more massive progenitor back in time).
Also shown are the MBH spin, the mass of the reservoir available for accretion onto the MBH, as well
as the host galaxy morphology (with the redshift intervals in which the galaxy is elliptical, i.e. $B/T>0.7$,
marked by a shaded lavender area). 

The two right panels show two MBHs accreting at $f_{\rm Edd} \approx 1.6\times 10^{-2}$
at $z\approx0$. Despite the similarity of the accretion rates and galaxy morphologies at $z\approx0$, the final MBH
spins are quite different.
In both cases, the spin grows rapidly early on,
and remains almost maximal as long as the MBH mass
is $\lesssim 10^6 M_\odot$. This is
because, as discussed earlier, for such low masses the
MBH spin is smaller than the typical angular momentum of
the accretion disk arising from the capture of a cloud, and as a result the Bardeen-Petterson
effect aligns the MBH spin to the disk's angular momentum, thus making accretion
effectively coherent.
However, the later evolution of the two cases is quite different and reflects
the wide kinematic range of bulges and pseudobulges shown in Figure
\ref{bulges}, with $v/\sigma$ above (below) 1 resulting in 
high (moderate) MBH spins. For instance, in the bottom right panel
the spin first drops to $\approx 0.25$ and then rises again to $\approx 0.8$ following a series of MBH mergers and accretion events (triggered by disk instabilities and major mergers).
A major galactic merger at $z\approx0$ then triggers a MBH merger and an accretion event that result in a final spin $\approx 0.75$.
In the top right panel, instead, accretion and mergers also cause the
spin to drop to $\approx 0.4$ at $z\approx 4$, but the degree of anisotropy
of the MBH fueling (i.e., in the pseudobulge model, the value of $v/\sigma$ 
of the host galaxy's stellar population) is never large enough to allow for very large spins in the subsequent evolution.
In the left panel, we show a highly sub-Eddington
system, which has a final MBH spin of
$\approx0.5$. In this case, a major merger at $z\approx2$ changes the
morphology of the galaxy from spiral to elliptical, and the following highly
``incoherent'' (i.e. with $v/\sigma \ll 1$) accretion causes a dramatic
spin down of the MBH. After that episode, the galaxy re-acquires a
substantial disk because of minor mergers and accretion/cooling of gas, thus becoming a spiral again, but
the accretion rate onto the central MBH never becomes high enough to
substantially change its spin.

These trends can be compared to those found in the {\it disk} model, shown in Figure \ref{tracks_kassin}.
In the top right panel we consider a galaxy containing an accreting MBH with $M_{\rm bh}\sim 10^6 M_{\odot}$
at $z=0$. For such a small mass, the distinction between ellipticals and spirals
and between the various accretion models (e.g. disk vs pseudobulge) is unimportant because the Bardeen-Petterson
effect always aligns the MBH spin to the angular momentum of the incoming cloud. This makes accretion
essentially coherent and results in almost maximal spins. The bottom left panel shows again an accreting MBH in a spiral galaxy, but with larger final mass ($M_{\rm bh}\sim 10^8 M_{\odot}$
at $z=0$). Again, accretion is effectively coherent (and the spin close to maximal) until the MBH grows 
larger than $\sim 10^6 M_{\odot}$. After that, the spin evolution during accretion events 
depends on the galaxy morphology. In particular, 
at $z\approx 2$ a major galactic merger (\textit{not} accompanied by a MBH merger) triggers a morphology change to
elliptical, as well as a QSO event that pushes the MBH spin to $a_{\rm bh}\approx 0.3$.
The galactic disk then re-grows via minor mergers and accretion/cooling of gas, and
accretion becomes ``more coherent''   (because the gas in spirals 
typically has $v/\sigma>1$ at low redshift, c.f. Figure \ref{gas}). As a result, the MBH spin starts growing again, and 
at $z\approx 0$ an ongoing accretion event pushes it up to almost maximal values.
The bottom right panel shows instead a non-accreting (at $z\approx 0$) MBH hosted in an elliptical galaxy.
Again, the spin remains close to maximal until $M_{\rm bh}\sim 10^6 M_{\odot}$, after which
the spin evolution is dominated by accretion, which tends to spin
the MBH down when the host is elliptical (c.f. the QSO event at $z\approx 2$).
When the host is a spiral, instead, the spin tends to mildly grow during accretion events.

From the above examples we can draw a couple of conclusions. Non-accreting systems (which we arbitrarily define as MBHs with $f_{\rm Edd}<0.01$) 
typically retain memory of the last ``violent'' event that changed the central MBH spin. 
Since this is likely to be a major merger/disk instability event that temporarily changes the galaxy morphology to bulge-dominated, 
the MBH spin will likely be small (non-maximal, in any case), whether the galaxy later retains a bulge-like morphology or acquires a disk again. 
Highly accreting systems (i.e. with $f_{\rm Edd}>0.01$) will likely have their spin affected by the ongoing accretion episode. 
In this case, the spin at $z=0$ depends on the typical $v/\sigma$ ratio of the host galaxy. 
Spirals in the {\it pseudobulge} model have a range of possible spins mirroring 
the wide spread in $v/\sigma$. Conversely, spirals in the {\it disk} model are almost guaranteed to have almost maximally spinning MBHs. 
On the other hand, MBHs in accreting ellipticals tend to have low spins, because of their typically low $v/\sigma$ ratios, as shown in Figure  \ref{bulges}. 

\subsection{Spin evolution of the global MBH population}
\label{spinglobal}

\begin{figure}
\begin{center}
\begin{tabular}{c}
\includegraphics[scale=0.38,clip=true]{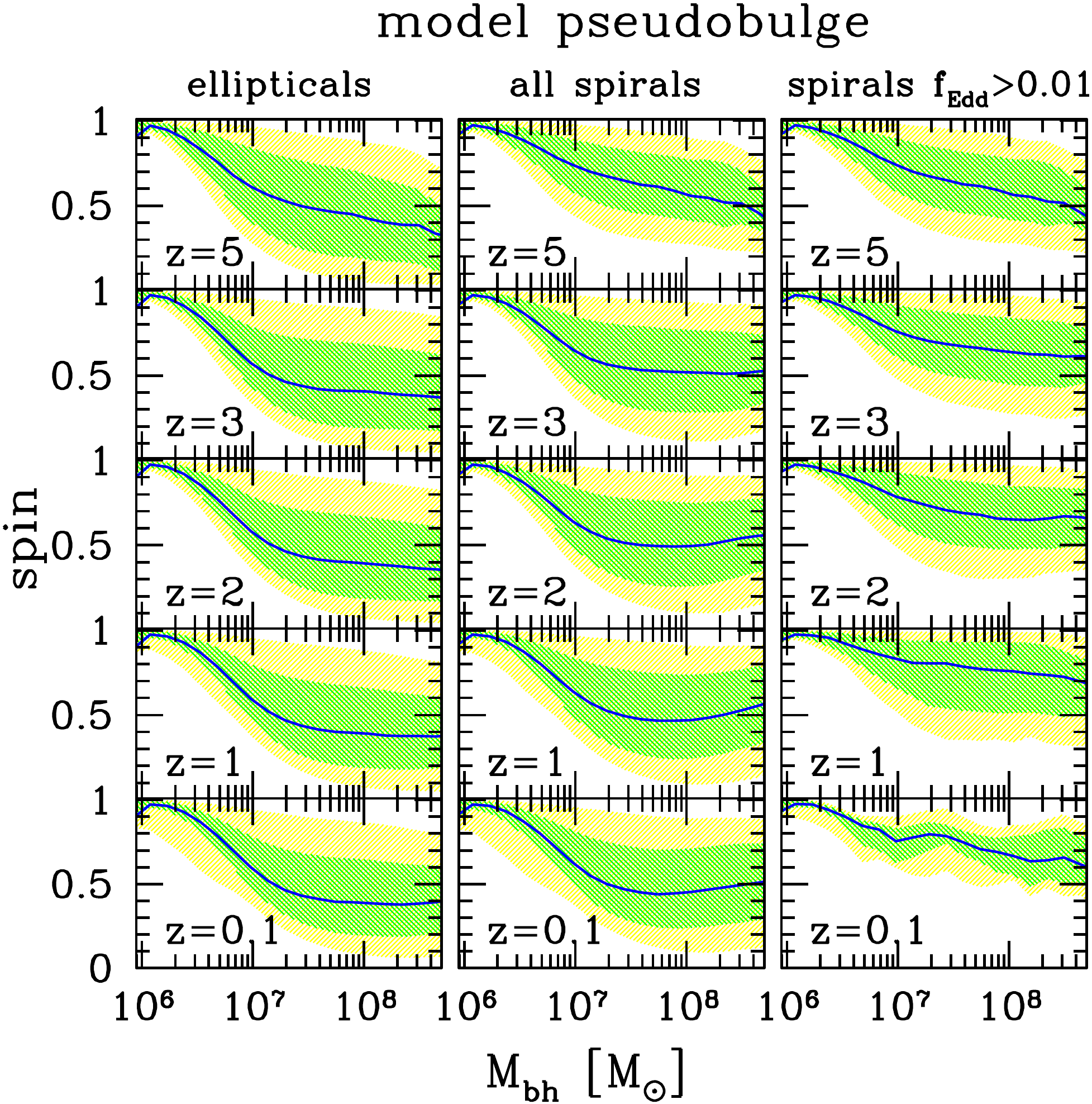}\\
\includegraphics[scale=0.38,clip=true]{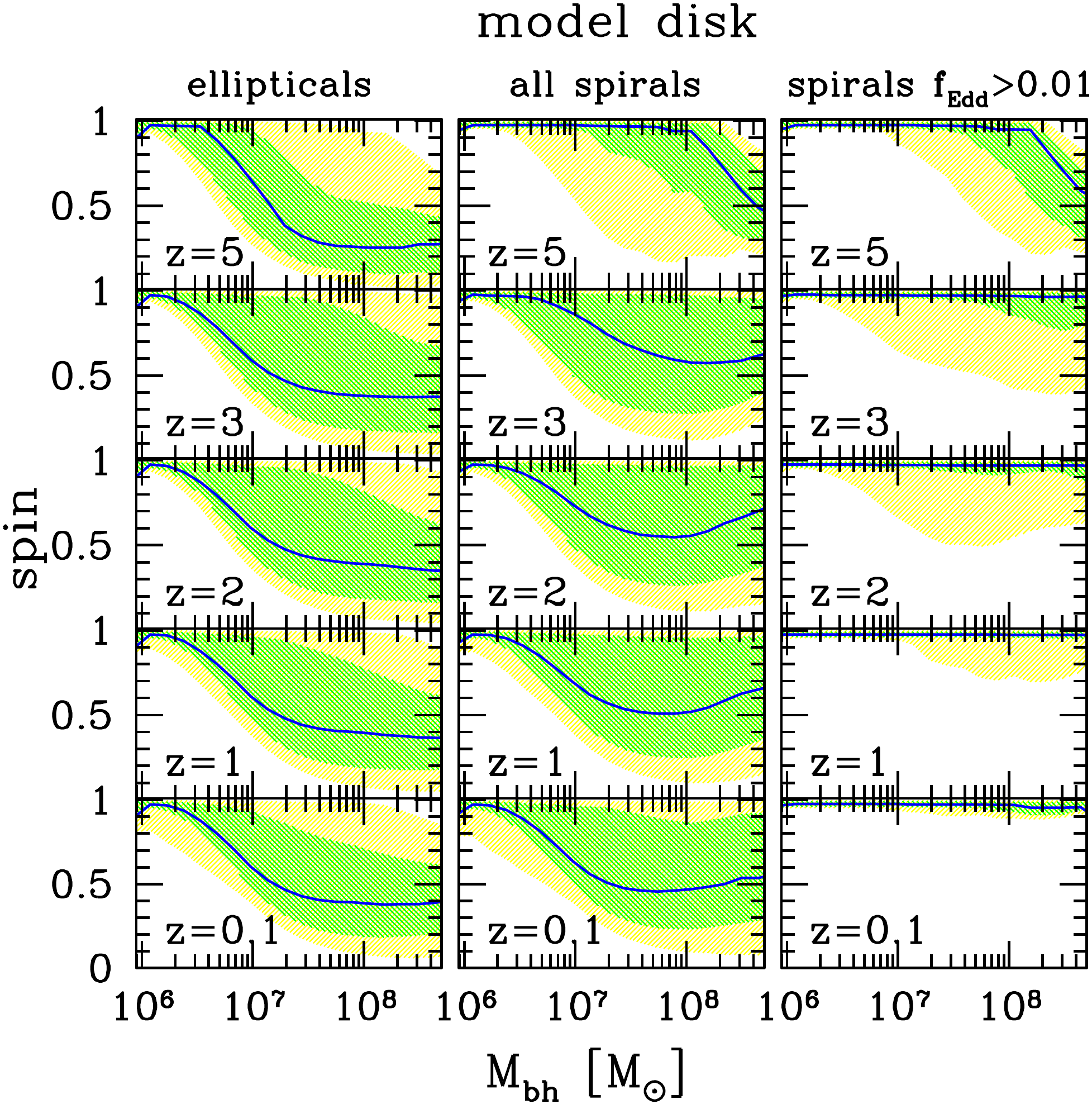}\\
\includegraphics[scale=0.38,clip=true]{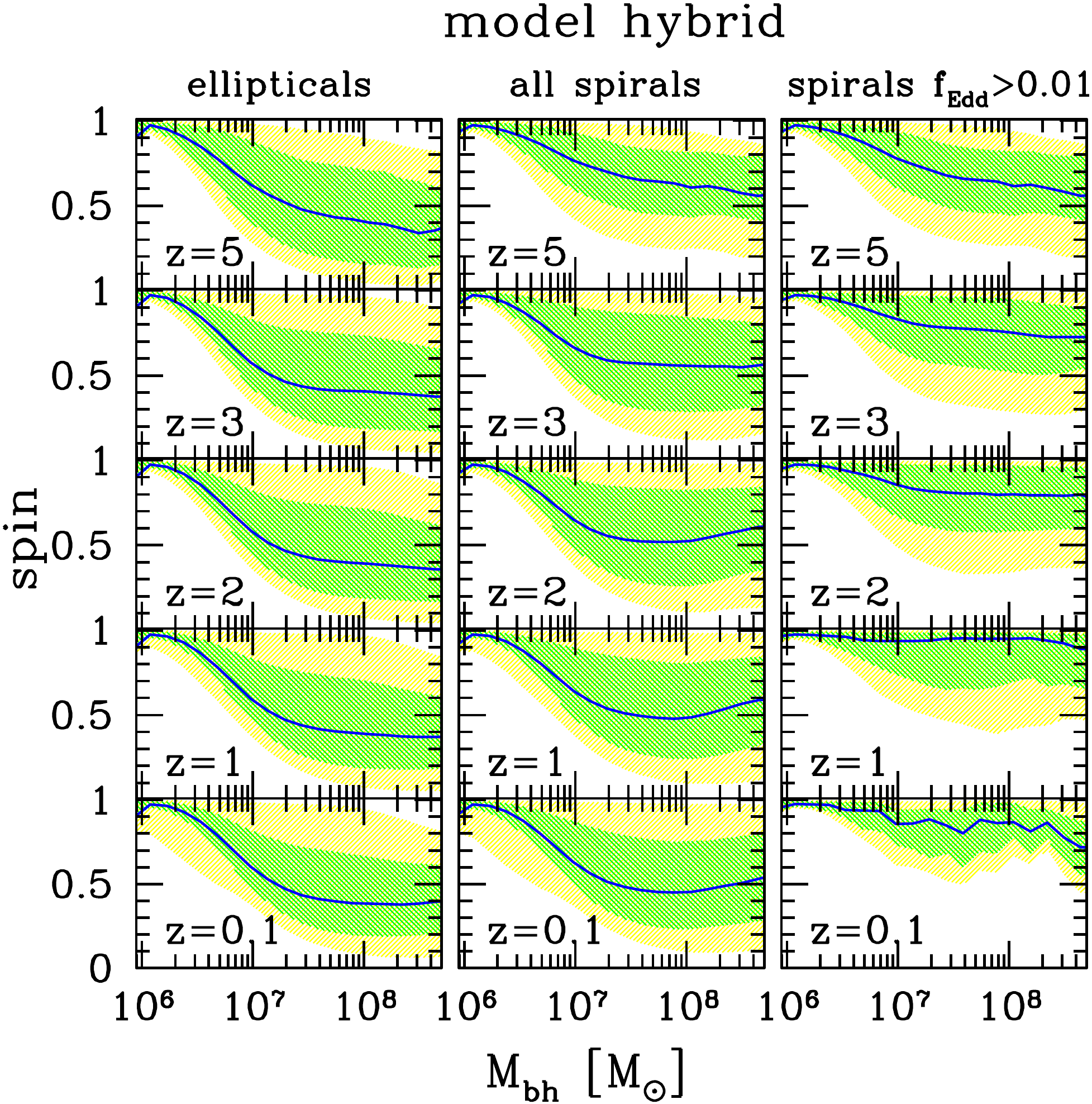}\\
\end{tabular}
\caption{Spin evolution as a function of redshift for different subsample of galaxies. The considered model is specified at the top of each panel. In each plot, the blue line is the median of the spin distribution as a function of MBH mass and as predicted by the model, while green and yellow shaded areas represent the spin ranges enclosing $68\%$ and $95\%$ of the distribution.}
\label{spinzev}
\end{center}
\end{figure}

Having discussed how accretion, mergers and galaxy morphology contribute to the MBH spin build-up in individual galaxies, we now turn to the investigation of the cosmic spin evolution 
of the whole MBH population. In the three panels of Figure \ref{spinzev} we show the spin distributions predicted by the {\it pseudobulge} (top plot), {\it disk} (central plot) 
and {\it hybrid} (bottom plot) models, for selected subsamples of galaxies at different redshifts. 

As mentioned earlier, all three models adopt the same prescription for the spin evolution in elliptical galaxies (right column in each panel), 
which results in very similar distributions that do not evolve much from $z=5$ to the present epoch. 
For $M_{\rm bh}\lesssim 10^6\msun$,  $J_{\rm disk}/2J_{\rm bh} > 1$ and MBHs tend to have maximal spins. 
At large masses,  $J_{\rm disk}/2J_{\rm bh}\ll 1$, and $a_{\rm bh}$ tends to the equilibrium value given by Figure \ref{fig:a_eq}. 
The median $v/\sigma$ of elliptical galaxies is $\approx0.2$, corresponding to $F\approx0.6$, which translates in a median equilibrium spin of $0.3$, in good agreement with Figure \ref{spinzev}. 
The evolution of the overall population of spiral galaxies (central column) also shows similarities in all three models. 
This class of systems experiences more pronounced redshift evolution, and the median of the spin distribution shows a dip between  $10^{7}\msun$ and $10^{8}\msun$, 
appearing at $z\sim3$. The overall distributions are not very different from those of the ellipticals, 
but spins are on average larger by virtue of the typically higher $v/\sigma$, as shown in Figure \ref{bulges}.

\begin{table*}
\begin{center}
\begin{tabular}{cccccccc}
\hline
Object name  & Galaxy type & z & $L_X$[erg s$^{-1}$] & $f_{\rm Edd}$ & log$(M_{\rm bh} [\msun])$ & spin & adopted PDF\\
\hline
1H0707-495        & --       & 0.0411 & $3.7\times10^{43}$ &  1.0  & $6.70\pm{0.4}$      & $>0.97            $ & flat   [0.97,0.998]\\
Mrk1018           & S0       & 0.043  & $9.0\times10^{43}$ &  0.01 & 8.15                & $0.58^{+0.36}_{-0.74}$ & flat   [0,0.94]\\
NGC4051           & SAB(rs)bc& 0.0023 & $3.0\times10^{42}$ &  0.03 & 6.28                & $>0.99            $ & flat   [0.99,0.998]\\
NGC3783           & SB(r)ab  & 0.0097 & $1.8\times10^{44}$ &  0.06 & $7.47\pm{0.08}$     & $>0.88            $ & flat   [0.88,0.998]\\
1H0419-577        & --       & 0.104  & $1.8\times10^{44}$ &  0.04 & $8.18\pm{0.05}$     & $>0.89            $ & flat   [0.85,0.998]\\
3C120             & S0       & 0.033  & $2.0\times10^{44}$ &  0.31 & $7.74^{+0.20}_{-0.22}$ & $>0.95            $ & flat   [0.95,0.998] \\
MCG-6-30-15       & E/S0     & 0.008  & $1.0\times10^{43}$ &  0.4  & $6.65\pm{0.17}$     & $>0.98            $ & hGauss [0.998,0.01] \\
Ark564            & SB       & 0.0247 & $1.4\times10^{44}$ &  0.11 & $<6.90$             & $0.96^{+0.01}_{-0.06}$ & hGauss [0.96,0.04]  \\
TonS180           & --       & 0.062  & $3.0\times10^{44}$ &  2.15 & $7.30^{+0.60}_{-0.40}$ & $0.91^{+0.02}_{-0.09}$ & hGauss [0.94,0.067] \\
RBS1124           & --       & 0.208  & $1.0\times10^{45}$ &  0.15 & 8.26                & $>0.97            $ & hGauss [0.998,0.02]\\
Mrk110            & --       & 0.0355 & $1.8\times10^{44}$ &  0.16 & $7.40\pm{0.09}$     & $>0.89            $ & Gauss  [0.945,0.033]\\
Mrk841            & E        & 0.0365 & $8.0\times10^{43}$ &  0.44 & 7.90                & $>0.52            $ & Gauss  [0.80,0.17] \\
Fairall9          & Sc       & 0.047  & $3.0\times10^{44}$ &  0.05 & $8.41\pm{0.11}$     & $0.52^{+0.19}_{-0.15}$ & Gauss  [0.6,0.1]  \\
SWIFTJ2127.4+5654 & SB0/a(s) & 0.0147 & $1.2\times10^{43}$ &  0.18 & $7.18\pm{0.07}$     & $0.6\pm{0.2}      $ & Gauss  [0.6,0.1]  \\
Mrk79             & SBb      & 0.0022 & $4.7\times10^{43}$ &  0.05 & $7.72\pm{0.14}$     & $0.7\pm{0.1}      $ & Gauss  [0.7,0.1]\\
Mrk335            & S0a      & 0.026  & $5.0\times10^{43}$ &  0.25 & $7.15\pm{0.13}$     & $0.83^{+0.09}_{-0.13}$ & Gauss  [0.81,0.067,$<0.92$]\\ 
Ark120            & Sb/pec   & 0.0327 & $3.0\times10^{45}$ &  1.27 & $8.18\pm{0.12}$     & $0.64^{+0.19}_{-0.11}$ & Gauss  [0.68,0.093]\\
Mrk359            & pec      & 0.0174 & $6.0\times10^{42}$ &  0.25 & 6.04                & $0.66^{+0.30}_{-0.54}$ & Gauss  [0.66,0.33,$<0.96$]\\
IRAS13224-3809    & --       & 0.0667 & $7.0\times10^{43}$ &  0.71 & 7.00                & $>0.987           $ & Gauss  [0.989,0.002]\\
NGC1365           & SB(s)b   & 0.0054 & $2.7\times10^{42}$ &  0.06 & $6.60^{+1.40}_{-0.30}$ & $0.97^{+0.01}_{-0.04}$ & Gauss  [0.97,0.03,$<0.98$]\\
\hline
\end{tabular}
\end{center}
\caption{Sample of MBHs with spin measurements from K$\alpha$ reflection line. The table is compiled using objects from \cite{reynolds13,brenneman13}; galaxy redshifts are taken from the NED database \protect\footnote{http://ned.ipac.caltech.edu/}. Quoted errors in the mass and spin measurements correspond to 68\% and 90\% confidence level respectively. The spin PDF we use in our statistical analysis is given in the last column, which identifies three different functional forms: flat (min and max value given in []), half Gaussian (maximum and $\sigma$ given in []), and Gaussian (maximum and $\sigma$ given in []). In the latter case, a third number in [], when present, defines a sharp cutoff in the PDF.}
\label{tab1}
\end{table*}

The subset of {\it accreting} MBHs (i.e., with $f_{\rm Edd}>0.01$) in spirals (right column) shows instead a remarkably different evolution between the {\it pseudobulge} and {\it disk} models. 
In the {\it pseudobulge} model, these galaxies maintain, on average, higher spins at all redshifts compared to other systems 
(c.f. discussion in Section \ref{individualspin}), but the trend of decreasing spin with increasing mass is preserved. 
Conversely, in the {\it disk} model, MBH spins show a stronger redshift evolution, and at $z<2$ they tend to be maximal independently of the MBH mass. 
This is because MBHs are efficiently accreting gas with $v/\sigma>1$ (c.f. Figure \ref{gas}). 
As expected, the {\it hybrid} model shows an evolution which is half-way between the {\it disk} and {\it pseudobulge} models. 

Our predicted spin distributions are different for different classes of galaxies, which is a testable prediction. 
In general, MBHs in {\it ellipticals} have lower spins than their spiral counterparts, but still with an average value of $a_{\rm bh}\approx0.4$ and a 
long tail extending to much higher values. 
We note that our findings do not rule out spin-powered jet models \citep[see, e.g.,][]{sikora2007}. Indeed, in such models the power of the jet is usually $\propto a_{\rm bh}^2$ \citep{blandford1977,Tchekhovskoy2010,narayan2013}, 
implying a difference of just a factor of a few in luminosity between maximally and mildly spinning MBHs. 
It is also clear that in our model, accreting MBHs tend to be biased toward higher spins than non-accreting ones, especially if located in spiral galaxies. 
In addition, current measurements of the MBH spins require large X-ray fluxes, which are only possible if the MBH shines at a significant fraction of the Eddington luminosity. 
Therefore, the current observed sample of MBH spins may not be an unbiased indicator of the overall cosmic MBH population. 
In the following we will perfom a statistical analysis comparing our predictions to the MBH spin measurements available today at  $z<0.1$,
selecting the appropriate sample in our simulations in order to account for both these selection effects.

\section{Comparison with observations}
\label{sec:obs}

We carry out here a quantitative comparison between MBH spin measurements and the output of our theoretical models, which link the MBH spin evolution to the properties of the galaxy host. However, existing spin measurements are sparse and are affected by several statistical and systematic uncertainties that are often difficult to quantify. The results of the following analysis should therefore be taken as indicative. Nonetheless, we will show that even with the current data, we can rule out some mass and spin growth scenarios, and gain some qualitative insights about the connection between spin evolution and the properties of the galaxy host supplying gas to the accretion flow. 

\subsection{The observed sample}
\label{obssample}

MBH spins have been measured via K$\alpha$ reflection broad line modeling for about 20 objects. We took data from the sample compiled by \cite{reynolds13}, integrated with objects from \cite{brenneman13}. We noted some (minor) discrepancies in the numbers reported by the two Authors; when necessary we inspected the original measurement papers (see references in \cite{reynolds13} and \cite{brenneman13}) and obtained the values directly from there. All the relevant properties of the sample are shown in Table \ref{tab1}. A meaningful comparison to a theoretical model is possible only if $i)$ observations have meaningful errors and $ii)$ the general properties of the observed subsample can be isolated and selection effects are understood. Both issues are somewhat tricky here. 

Errors quoted in Table \ref{tab1} represent 68\% confidence level in the mass and 90\% confidence level in the spin measurements. A complete knowledge of the probability density functions (PDF) of those quantities would be desirable, but such information is often missing in the literature, and we can at best put forward educated guesses. For the (log of the) MBH mass, we consider $i)$ a Gaussian PDF with $\sigma$ equal to the quoted error when the latter is symmetric with respect to the best measured value, $ii)$ a flat PDF within the given errors when those are asymmetric and $iii)$ a Gaussian PDF with an arbitrary $\sigma=0.3$ when errors are absent. Spin PDFs are instead derived by visually inspecting the function $\chi^2(a_{\rm bh})$, which is always given in the relevant papers. We identify three families of measurements: $i)$ sometimes $\chi^2(a_{\rm bh})$ has an approximately flat minimum, and in this case we take a flat PDF in the corresponding range; $ii)$ for few objects $\chi^2(a_{\rm bh})$ is extremely 
skewed 
on the left of the minimum, and in such cases we take as PDF the left half of a Gaussian distribution; $iii)$ more often, $\chi^2(a_{\rm bh})$ is quite symmetric (at least in the 99\% confidence region), calling for a Gaussian model of the PDF. Details of the PDFs are given in the last column of Table \ref{tab1}. Being aware of the arbitrary nature of this procedure, we also consider an alternative model in which we take all errors in masses and spins to have a flat PDF within the range quoted in column 7 of Table \ref{tab1}. We show in Appendix \ref{app:sanity} that our results are largely independent of the adopted shape of the PDFs.         
 
The obvious selection effect for a K$\alpha$ line measurement is that the source has to be bright in hard X \citep[see also discussion in][]{brenneman11} in fact most of the MBHs in the sample resides in Seyfert 1 or narrow line Seyfert 1 (NLS1) galaxies. Hosts are usually spirals or lenticular galaxies, with the exception of Mrk841. The observed sample is neither flux nor volume limited. X-ray luminosities and associated MBH masses have distributions consistent with being log-flat in the ranges $10^{42}-10^{45}$erg s$^{-1}$ and $10^6-3\times10^{8}\msun$ respectively. Eddington ratios $f_{\rm Edd}$ are also evenly distributed in the range $0.01$-1. The redshift distribution is also approximately flat in the range $0<z<0.07$, with two outliers at $z>0.1$. Given these facts, the best matching sample in our model is provided by spiral galaxies (defined by $B/T<0.7$) with $f_{\rm Edd}>0.01$ at low redshift. For the purpose of the analysis, we will therefore exclude Mrk841 (which is an elliptical) and include all galaxies in the observed sample with 
unknown morphology (thus implicitly assuming they are spirals/lenticulars). Discarding the latter would weaken our results (because there would be only 12 objects in the observed sample) but not qualitatively change them.

\subsection{The theoretical sample}
\label{theosample}

Theoretical distributions are computed with our model on a grid in the $M_{\rm bh}-a_{\rm bh}$ parameter space. The MBH mass range $10^6\msun<M_{\rm bh}<5\times10^8\msun$ is divided in five equally log-spaced bins, and for each bin, spin distributions are computed on 20 linear bins covering the range $0<a_{\rm bh}<1$. We applied Gaussian kernel smoothing to each measured spin, with $\sigma=0.05$, to get smoother distributions{\footnote{The Gaussian kernel smoothing has mostly the effect of making the distribution visually smoother, and we checked that the results of our analysis are independent of it.}}. To check the robustness of our results against spin binning, we also constructed two theoretical distributions using 10 and 30 linear $a_{\rm bh}$ bins, and one distribution considering 20 bins both in log mass and spin.  As a sanity check, a comparison among the results obtained with different binning choices is performed in Appendix \ref{app:sanity}. 

\begin{figure}
\begin{center}
\begin{tabular}{c}
\includegraphics[scale=0.38,clip=true]{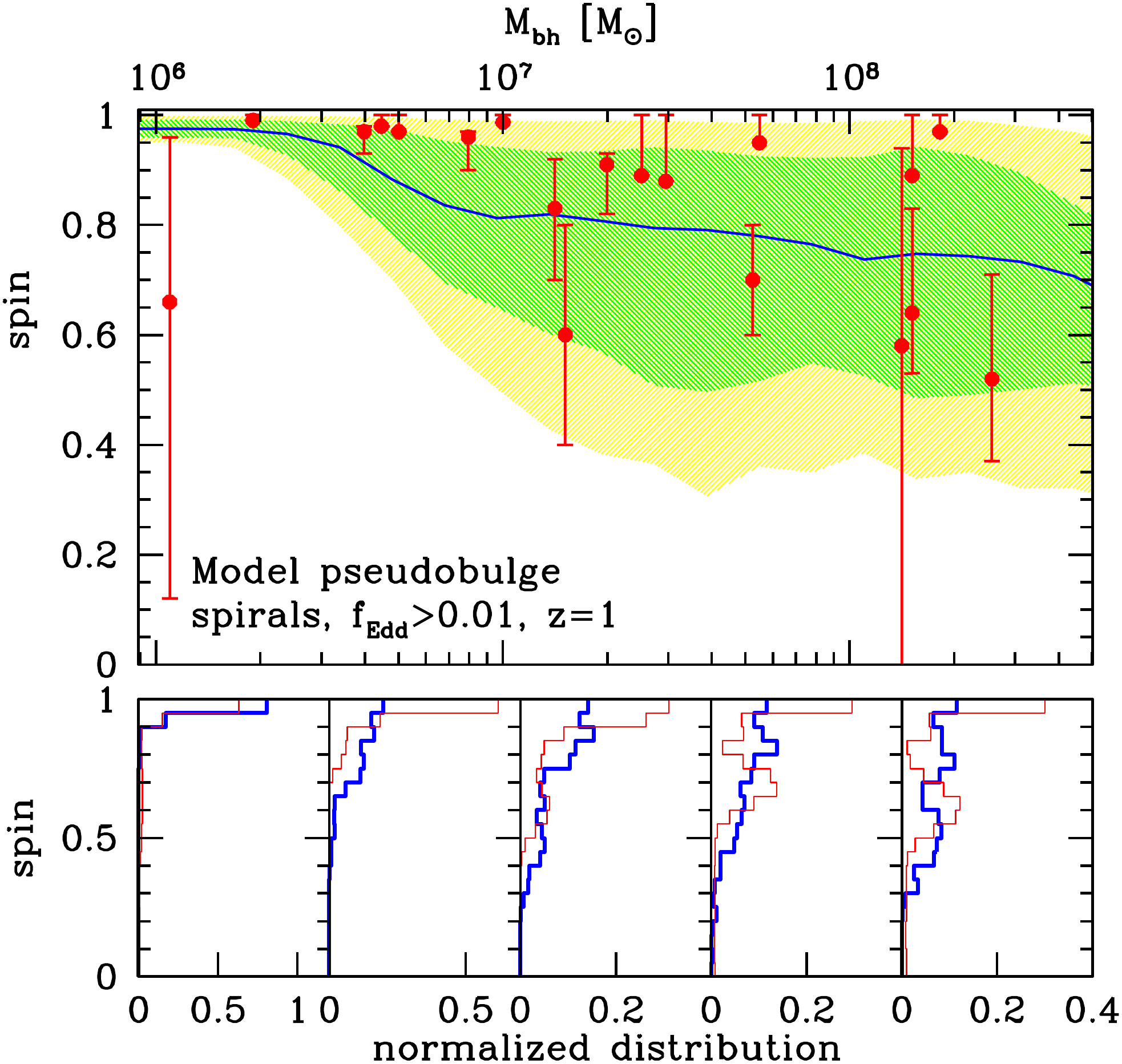}\\
\includegraphics[scale=0.38,clip=true]{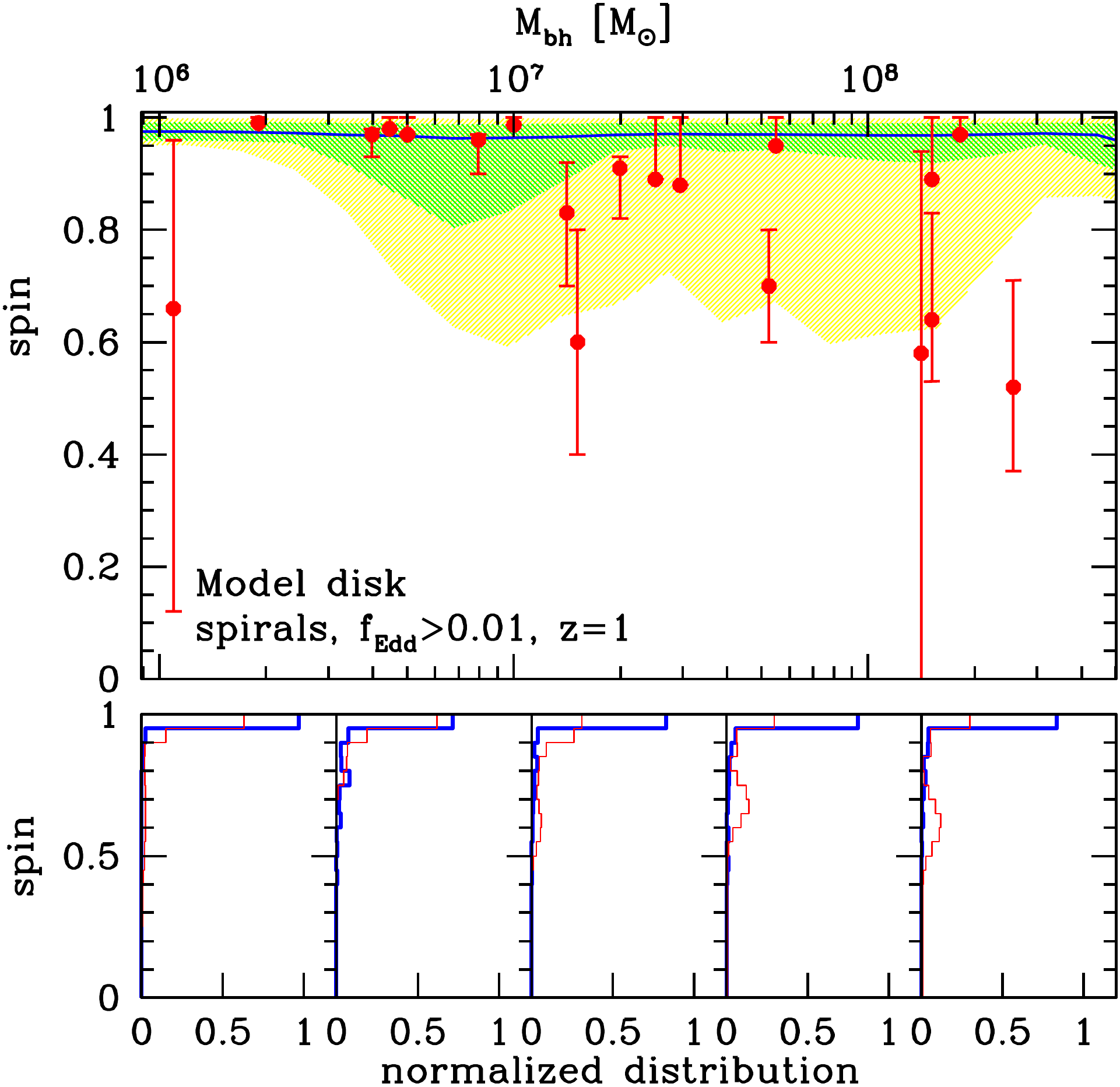}\\
\includegraphics[scale=0.38,clip=true]{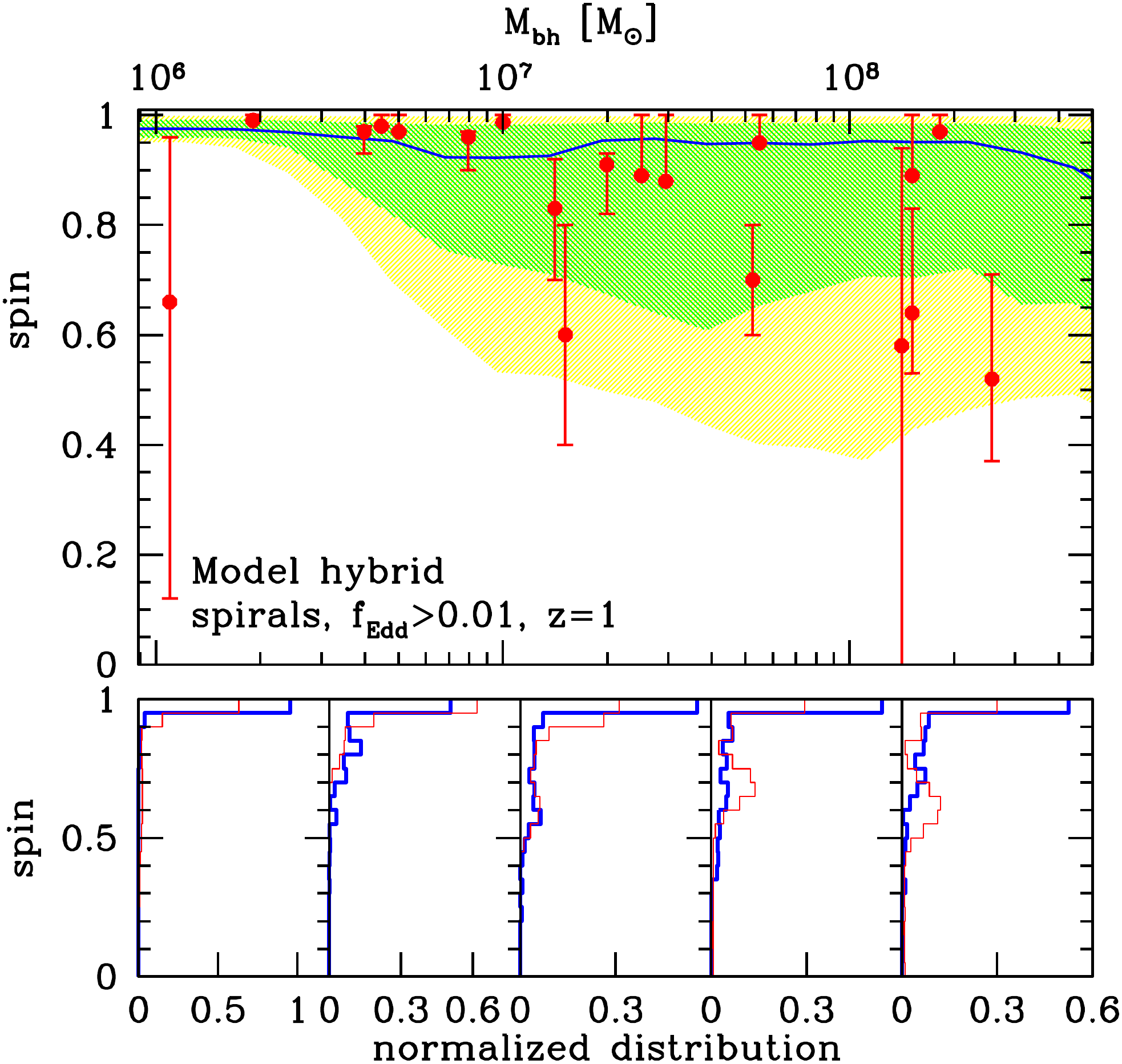}\\
\end{tabular}
\caption{{ Comparison between measured MBH spins and predictions of our models for the subsample of accreting MBHs ($f_{\rm Edd}>0.01$) in spirals. In all three plots, the top panel shows the outcome of the model with the same line and color code as in Figure \ref{spinzev}, and red dots are measurements with errors as described in Table \ref{tab1}. The five small bottom panels show the spin distributions predicted by our models in different mass bins (thick--blue histograms) and the PDF of the measured spins in the same mass bin, including observation errors (thin--red histograms). 
Theoretical distributions are taken at $z=1$.}}
\label{spinobs}
\end{center}
\end{figure}

As mentioned above, all (but one) MBHs with spin measurements are accreting at $f_{\rm Edd}>0.01$ and reside in spiral or lenticular galaxies at $z<0.1$. Those systems, although present in fair number, are necessarily rare in our runs (because accreting systems becomes much rarer than non-accreting ones at low redshift, c.f. Figure~\ref{eddf}), which makes it difficult
to construct a smooth 2D mass-spin distribution to compare to observations, even by adding together our runs for light and heavy seeds. Figure \ref{spinzev}, however, shows that the overall spin evolution of the galaxy population is small at $z<1$, as a consequence of the reduced galactic activity both in terms of gas cooling/star formation and mergers. Therefore, to improve the statistics of our analysis, we compare observations to $z=0.1$ to a {\it fiducial theoretical sample taken at $z=1$}. For the sake of completeness, effects due to the evolution of the galaxy population at $z<1$ are investigated in Appendix \ref{app:sanity}, 
where we show that they do not significantly affect the conclusions of our analysis.

The predicted spin distributions of accreting MBH hosted in spirals at low redshift are shown in Figure \ref{spinobs}, together with the measurements of Table \ref{tab1}. The {\it pseudobulge} model qualitatively reproduces the observed trends, with most of the measurements and relative errorbars falling in the 95\% confidence region predicted by the models. The {\it disk} model displays a slightly poorer match, with many events outside the 68\% confidence region predicted by the model, and just a minor hint of the  general spindown observed with increasing MBH mass. The {\it hybrid} model nicely matches the data, reproducing both the observed high spin sub-sample, and the objects with moderate spin.

\subsection{2D-Kolmogorov-Smirnov test} 
\label{2dks}
Although the qualitative comparison between theoretical models and measurements is encouraging (in the sense that we find the right trends in the right subset of galaxies to reproduce observations), and seems to favor the {\it hybrid} model over the others, we want to corroborate our findings with some more quantitative indicators. We first performed two dimensional, one sample Kolmogorov-Smirnov tests \citep[2D-KS,][]{press92} to compare the observed samples to the different MBH subsamples at different redshifts. Although not as rigorous as its one dimensional counterpart, the test returns an approximate probability, $p_{\rm KS}$, that the data are drawn from the model distribution \citep[see,][for details about the 2D-KS statistics]{press92}.  

To account for observational errors, we computed the average $p_{\rm KS}$ over $10^4$ realizations of the observed samples, picking the mass and spin of each individual object from the corresponding PDF. Looking at the results shown in Table \ref{tab2}, a number of conclusions can be drawn:
\begin{itemize}
\item the selected subsample matters. In particular ellipticals and spirals hosting non-accreting MBHs (i.e., those with $f_{\rm Edd}<0.01$) have a spin distribution that poorly matches observations, yielding $p_{\rm KS}\approx10^{-3}$ and $p_{\rm KS}\approx10^{-2}$ respectively for the {\it pseudobulge} and {\it hybrid} models; 
\item in the {\it hybrid} model, the subsample of accreting MBHs (i.e., ones with $f_{\rm Edd}>0.01$) in spirals is perfectly consistent with the observed sample;
\item the {\it hybrid} model is favored, yielding $p_{\rm KS}>0.5$ (we show in Appendix \ref{app:sanity} that this result is independent of binning, redshift and observational error PDF);
\item the {\it pseudobulge} model is also consistent with the data, yielding $p_{\rm KS}>0.2$;
\item in the {\it disk} model, the subsample of accreting MBHs in spirals provides a a poor match with observations, disfavoring the model.
\end{itemize}
Some of the above results, in particular those given by the {\it disk} model, are somewhat bin-dependent. We show in Appendix \ref{app:sanity} that this issue prevents us from ruling out the {\it disk} model on the basis of the relatively sparse observed sample, but it does not affect our general conclusions.

\begin{table}
\begin{center}
\begin{tabular}{c|ccc}
\hline
Model & Ellipticals & Spirals, $f_{\rm Edd}<0.01$ & Spirals, $f_{\rm Edd}>0.01$\\
\hline
{\it pseudobulge}  & 0.0020 & 0.0271 & \red{\bf 0.3614}\\ 
{\it disk}   & 0.0015 & 0.0969 & \red{\bf 0.0521}\\   
{\it hybrid} & 0.0016 & 0.0589 & \red{\bf 0.5328}\\
\hline
\end{tabular}
\end{center}
\caption{2D-KS test results on different samples of galaxies for our three fiducial spin evolution models. Reported is the (approximate) probability with which the spin measurements can be reproduced by a given model and a given sample.
The results for the sample that matches the observations (spirals hosting MBHs with $f_{\rm Edd}>0.01$) are highlighted in bold red.
The theoretical sample is taken at $z=1$, as explained in the main text.}
\label{tab2}
\end{table}

\subsection{Bayesian model selection} 
\label{secbayes}
The 2D-KS test provides useful indications about the consistency of the data with the predictions of a given model. It is, however, not very practical in assessing which one is the best model, among a set of different options. This latter issue can be tackled within the framework of Bayesian model selection, by computing the posterior probability of the parameters of a certain model, given a set of data. Because our models are non-parametric, this boils down to the evaluation of the odds ratio for different pairs of models, which we now describe.

Following \cite{sesana11}, we assume that the observations are uncorrelated, so that the number of objects, $n_i$, measured in a particular $\Delta{M}_{\rm bh}\Delta{a}_{\rm bh}$ bin
in parameter space can be drawn from a Poisson probability distribution with parameter $r_i$ equal to the bin-integrated rate:
\begin{equation}
p(n_i) = \frac{(r_i)^{n_i} {\rm e}^{-r_i}}{n_i!}\,.
\label{poiss}
\end{equation}
If we divide the parameter space up into a certain number $K$ of bins, then the information that comes from the {\it data} ($D$) is the number of events in each bin ($n_i$). The overall {\it likelihood} $p(D|X)$ of seeing this data under the model $X$ is the product over all $K$ bins of the Poisson probabilities to see $n_i$ events in the $i$-th bin given the rate $r_i(X)$ predicted by model $X$:
\begin{equation}
p(D|X) = \prod_{i=1}^K \frac{(r_i(X))^{n_i} {\rm e}^{-r_i(X)}}{n_i!}\,.
\label{like}
\end{equation}
It is straightforward to take the limit of this expression as the bin sizes tend to zero to derive a continuum version of this equation~\citep{gair10}, but in this analysis we will stick to binned distributions for simplicity. In the presence of measurement errors, the $j$-th event can then be assigned to different bins, with a probability given by its PDF $\rho_j(M_{\rm bh},a_{\rm bh})$. We can therefore construct the entire set of possible arrays of events falling in each bin, $n_i$, with their relative probabilities. Each array is then analyzed and weighted according to its probability to get the overall likelihood of the sample given the model.

If we want to compare two competitive models $A$ and $B$, Bayes' theorem allows us to assign to model $A$ a probability 
\begin{equation}
p(A|D) = \frac{p(D|A)P(A)}{p(D|A)P(A)+p(D|B)P(B)},
\label{bayes}
\end{equation}
whereas probability of model $B$ is just obtained by swapping $A$ and $B$ in equation (\ref{bayes}). Here, $p(D|X)$ is the likelihood given by equation (\ref{like}) and $P(X)$ is the prior probability assigned to model $X$. The odds ratio of model $A$ over model $B$ is 
\begin{equation}
\Lambda_{AB} = \frac{p(D|A)}{p(D|B)} \frac{P(A)}{P(B)} .
\label{oddratio}
\end{equation}
If we do not have any reason to prefer a priori model $A$ to model $B$, then $P(A)=P(B)=0.5$, and the odds ratio becomes the likelihood ratio. Moreover, equation (\ref{bayes}) implies that, in a two model comparison, we can simply assign a 'relative probability' $p_A=p(D|A)/(p(D|A)+p(D|B))$ to model $A$, and $p_B=1-p_A$ to model $B$.

\begin{table}
\begin{center}
\begin{tabular}{c|ccc}
\hline
\multicolumn{1}{c|}{model pairs} & \multicolumn{3}{c|}{}\\
\hline
{\it hybrid/pseudobulge} & log$\Lambda_{hp}=$1.0804 & $p_h=$0.9233 & $p_p=$0.0767\\ 
{\it hybrid/disk} & log$\Lambda_{hd}=$2.4749 & $p_h=$0.9966 & $p_d=$0.0034\\ 
{\it pseudobulge/disk}  & log$\Lambda_{pd}=$1.3944 & $p_p=$0.9612 & $p_d=$0.0388\\
\hline
\end{tabular}
\end{center}
\caption{Model selection results: pair comparisons between our three fiducial spin evolution models  {\it pseudobulge (p)}, {\it disk (d)} and {\it hybrid (h)}. For each two model comparison, we report the log of the likelihood ratio $\Lambda_{AB}$, i.e. the
ratio between the probability of model $A$ ($p_A$) and model $B$ ($p_B$).}
\label{tab3}
\end{table}

We compute the odds ratios of our three fiducial models: {\it pseudobulge} ({\it p}), {\it disk} ({\it d}) and {\it hybrid}  ({\it h}) according to equation (\ref{oddratio}), where we assume $P(A)=P(B)=0.5$. Likelihoods are computed according to equation (\ref{like}), folding-in measurement errors. The shape of the error PDFs of each single observation are described in Section \ref{obssample}, and we factorize $\rho_j(M_{\rm bh},a_{\rm bh})=g(M_{\rm bh})h(a_{\rm bh})$, assuming uncorrelated mass and spin measurements. 

Results are shown in Table \ref{tab3}. We get log$\,\Lambda_{hd}\approx2.5$, implying that the {\it hybrid} model provides a better description of the data than the disk model, at $>99.5$\% confidence level. This strengthen the results of the 2D-KS test, providing compelling evidence in favor of the {\it hybrid} model. Model {\it pseudobulge} sits somewhat in the middle; it is preferred at $\approx95$\% level over model {\it disk}, but it is disfavored at about the same level with respect to model {\it hybrid}. To summarize, the odds ratio analysis provides moderate (decisive) evidence that the {\it hybrid} model is a better description of the data than the {\it pseudobulge} ({\it disk}) model.

\section{Discussion and Conclusions}
\label{sec:conclusions}

In this paper, we presented the results of a semianalytical model for
the evolution of galaxies and MBHs. The model keeps track (although in
simplified ways) of the morphology of the galaxies, as well as the MBH
masses and spins. For the first time we link the dynamical properties
of the gas fueling the MBHs to the host galaxy properties, through
different observationally based prescriptions, and derive predictions testable with existing observations. We stress that all the
other similar investigations predicting MBH spin distributions either
assumed ``chaotic'' accretion (MBHs accreting small gas clouds with
isotropically oriented angular momenta) or ``coherent'' accretion
(MBHs accreting all the time on a fixed plane).

Our model predicts different spin distributions for different types of
galaxies. To date, only $\approx 20$ MBH spins have been directly measured
through K$\alpha$ iron line fitting \citep{reynolds13,
  brenneman13}. All these MBHs (but one) are hosted in low-redshift late
type galaxies, preventing us from testing our results for different
galaxy types. We therefore tested our model by selecting low-redshift
accreting MBHs hosted in spirals. { For this galaxy class, we have
  assumed three different gas dynamics prescriptions. In the {\it
    pseudobulge} model we assumed that the dynamics of gas fueling the
  central MBHs is similar to that of the stars both in galaxy  bulges
    or pseudobulges, according to the nature of the host's spheroidal
    component. In the {\it disk} model, instead, the fueling gas has
  the more coherent dynamics of the large scale gaseous disk. The {\it
    hybrid} model shares the same gas dynamics as the {\it disk} model
  for isolated galaxies, but assumes a gas dynamics similar to
    that of stars in spheroids during merger driven accretion events (see
  Section~\ref{sec:link} for details)}.

Our models predict interesting features in the MBH spin distribution, and a statistical comparison to observations yields a number of interesting results:
\begin{enumerate}
\item different galaxy morphologies result in different MBH spin distributions. Because the geometrical properties of the accretion flow are related to the large scale kinematics of the galaxy, MBHs hosted in ellipticals tend to have lower spins than those hosted in spirals; 
\item in general, MBH spins are a decreasing function of the MBH mass: $\sim 10^{6}\msun$ MBHs tend to be maximally spinning, whereas for masses $>10^{8}\msun$ a wide range of values is possible for the spin, depending on the host morphology and MBH accretion rate; 
\item accreting MBHs in spirals, i.e. those matching the observational sample, tend to spin fast, and do {\it not} provide an unbiased indicator of the underlying spin distribution of the overall MBH population;
\item 2D-KS tests show that, in general, the {\it hybrid} and {\it pseudobulge} models are consistent with observations, while the {\it disk} model is 
disfavored. A quantitative assessment of the compatibility of the {\it disk} model with the data, however, strongly depends on the maximum value to which MBHs can be spun up by accretion. This issue is discussed in detail in Appendix \ref{app:sanity}, where we show that if this value is $a_{\rm bh}=0.998$~\citep{thorne}, the {\it disk} 
model can be safely ruled out at $>99\%$ confidence level;
\item a likelihood odds ratio study shows that the {\it hybrid} model -- which takes into account the possible angular momentum reshuffling due to  merger-driven phases of nuclear activity --  provides the best match to the observed data. In a two model comparison: $i)$ it is strongly favored (at $>99\%$ confidence level) over the {\it disk} model; $ii)$ it is moderately favored (only at about $90\%$ confidence level) over the {\it pseudobulge} model; 
\item For the sake of comparison, in Appendix \ref{app:co-cha} we have also
run three separate sets of simulations adopting the standard
coherent-accretion model, as well as two flavors of the
chaotic-accretion model. These models provide a much worse description of the data; the {\it coherent} model shows features similar to the {\it disk} model, but with a poorer agreement to the data, while the chaotic models are unambiguously ruled out by observations.   
\end{enumerate}

We emphasize that the study of the evolution of the MBH spin
distribution in a cosmological framework is still in its infancy.
Because our model is idealized from many points of view, here
we highlight some still-needed theoretical and observational improvements.
$i)$
  The treatment of the Bardeen-Petterson effect, which plays a major role in determining
  the fraction of prograde vs retrograde accretion events, is
  based on simple isotropic-$\alpha$-viscosity driven accretion
  disks, and this assumption has been criticized by \cite{sorathia13a,
    sorathia13b}. The impact of this strong and debated assumption is
  unknown, since no working models of the MBH spin evolution relaxing
  it are available to date. $ii)$ We assumed that every
  gas cloud fueling the MBHs weighs $M_{\rm cloud}\sim 3\times10^4 \msun$. Giant molecular
  clouds are observed to have a broad mass spectrum. Observational
  studies of local galaxies reported typical molecular cloud mass
  functions with cut-offs at about $10^{5 - 6} \msun$
  \citep[e.g.][]{williams97, engargiola03, fukui08}, making the average cloud
  mass comparable to the mass we that assume in this paper. We note however that the
  cut-off mass seems to depend on the host galaxy properties
  \citep[e.g.][and references therein]{hopkins12b}. A simple
  implementation of a ``characteristic'' cloud mass function is not
  trivial, and beyond the scope of this paper. The effect of
  different average cloud masses can anyway be easily understood: the
  cloud mass contributes to setting the maximum MBH mass at which a
  complete alignment of the spin occurs at every accretion event,
  effectively making the fueling dynamics unimportant. To first
  approximation (see D13 for a more detailed discussion),
  an order of magnitude variation of the average cloud mass results in
  an order of magnitude shift along the MBH mass axis of all the spin
  distributions. The results not showing any significant feature in the
  spin evolution (e.g. the $f_{\rm Edd}>0.01$
  {\it disk} model at low redshift, but see also the {\it coherent} and {\it chaotic I} models discussed in Appendix B) would thus remain mostly unchanged under a
  variation of the cloud mass. On the other hand, the {\it pseudobulge}
  and {\it hybrid}
  models would predict higher spins for higher masses (lower spins for
  lower masses) if the cloud mass were higher (lower). 
  We checked that a shift
of the  spin distributions
by one order of magnitude toward larger masses does not affect the compatibility of those two models with
  the data, yielding a KS probability $p_{\rm KS}>0.3$. A similar shift of the 
  distribution toward lower masses results in a slightly worse description of 
  the data, but still with a KS
  probability $\gsim 0.1$. $iii)$ Our dynamical prescriptions for the
  nuclear gas dynamics are mostly based on gas and/or stellar dynamics
  observations on scales much larger than the accretion disk or the MBH
  sphere of influence. Our model would greatly benefit from a
  description of the gas dynamics on scales $\lsim 100$ pc,
  already achievable in some cases with high resolution ALMA
  spectroscopy \citep[e.g.][]{combes13, combes14}. Finally, $iv)$ in
  our model we completely neglected the effect of rotational energy
  extraction from the MBH on the spin. This could in
  principle power relativistic jets through the Blandford-Znajek
  mechanism, limiting the maximum spin to lower values in radio
  galaxies.

From an observational point of view, the main improvement would
consists in enlarging the currently small sample of measured MBH spins, and 
to extend it to higher redshift and different galaxy types{\footnote{{ While finalizing this manuscript, \cite{reis14} reported a spin measurement of the $z=0.658$ lensed quasar 1RXS J113151.6-12315 of $a=0.87^{+0.08}_{-0.15}$ (3$\sigma$ errors). The estimated MBH mass is $2\times10^{8}\msun$, the accretion rate computed from its bolometric luminosity is nearly Eddington, and the host is a Seyfert 1 spiral galaxy \citep{claesken06}. Thus, this system falls in our sample of accreting spirals at $z=1$ shown in figure \ref{spinobs}, and its inferred mass and spins are perfectly consistent with all the models shown there.}}}.
In this regard, the forthcoming Astro-H satellite will have exquisite 
combined spectral resolution and sensitivity, making K$\alpha$-based 
spin measurements possible perhaps up to $z\sim1$ \citep{astroH2012}. 
Along the same lines, Athena+, 
to be launched in 2028, will push such 
measurements to $z\sim2$ for extremely X-ray bright systems,
significantly expanding the observed spin sample \citep{dovciak2013}. 
On a shorter timescale, the eROSITA satellite will observe $>10^5$ AGNs.
Image stacking in luminosity and redshift bins will reveal the
'average' shape of the K$\alpha$ line of these systems, making
possible to study trends in the typical spin of AGNs across 
a large mass range  and  for redshifts perhaps up to $z\sim1$~\citep{merloni2012}, thus providing
an important benchmark for comparison to theoretical models. 
On a different note, the eLISA mission, now selected by ESA for
the L3 launch slot, will measure the spins of merging MBH binaries 
across the cosmic history to high precision, providing a sample
that will not be biased toward high X-ray luminosities. 

Even though the future of spin measurements looks literally bright, we
should also keep in mind that the uncertainties on the value of
measured spins depend on the technique used. Different methods have
been used to constrain the MBH spin distribution at different
redshifts and for different masses. Some methods estimate the spin
value from the radio properties of the AGNs \citep[e.g.][and
  references therein]{dali13}, from the ionizing flux required to
produce the observed broad emission lines \citep{netzer13}, or fitting
the continuum from the accretion disk \citep{laordavis11}. Some of these
methods are still matter of debate \citep[e.g.][]{gallo12, raimundo12,
  laor11}, and all of them require an independent measure of the MBH
mass to estimate its spin. In this paper we compared our predictions
with the mass independent measurements obtained through
relativistically broadened iron K$\alpha$ line fitting, to prevent any
possible spurious mass spin correlation. We refer to \cite{reynolds13}
and \cite{brenneman13} for a detailed discussion of the uncertainties
of these measurements. Since our models predict a significant dependence of the
MBH spins on the MBH masses, the uncertainties on the 
masses are fundamental as well. For
{ example, Mrk 359} has an estimated mass of $\approx 10^6 \msun$,
obtained through a single epoch measurement of the continuum and broad
H$\beta$ FWHM \citep{2005ApJ...618L..83Z}. The width of the line is only 480
km s$^{-1}$ significantly smaller than the typical value for type 1
AGN as well as than the threshold for being classified as a
NLS1. Since the MBH is hosted in a spiral galaxy the line is
peculiarly large to be associated with a single narrow line, and a
broad component is expected because of the point like emission clearly
visible in the HST imaging \citep{1998ApJS..117...25M}. The observed line
width could be so narrow because of orientation effects, already
suggested to be significant for NLS1s \citep[][]{decarli08}, causing an
underestimation of the MBH mass. (To be on the safe side, we checked that the
    contribution of Mrk 359 to our analysis is negligible. Its MBH
    spin is so poorly constrained that removing it from the
    observational sample has no effect on our results.) 
More independent measurements of the
masses of MBHs with spin estimates would help better constrain
their values. The detection of gravitational waves will also greatly
improve the situation, giving high precision measurements of both the MBH
masses and spins, which could be compared with predictions for merging
systems.

{ Regardless of the large uncertainties on which we commented above, it is
remarkable that two of the models ({\it pseudobulge}
and {\it hybrid}) that we investigated generate samples of MBH masses and
spins consistent with observations. Since these models assume that
the accreting gas has a less coherent dynamics than the larger-scale
gas structure (although, in the hybrid case, only in recent merger remnants)}, our
analysis seems to suggest the existence of a physical process that,
while decreasing the gas angular momentum magnitude (triggering
accretion in the first place), reshuffles the gas angular momentum
direction as well.  Such a reshuffling could be caused by local
processes, i.e. star formation and the torques exerted by the gas
self-gravity \citep[e.g.][]{maio13}, as well as larger-scale
gravitational instabilities and violent gas inflows
\citep[e.g.][]{hopkins12, dubois14}.  On the other hand, in order to achieve
such a good agreement between the predictions of the model and the
observed spins, we have to require the accreting gas to have, on average, a
non-zero angular momentum, i.e. the gas must \textit{not} accrete
isotropically onto the MBH. We have in fact shown that a perfectly
isotropic accretion flow would result in significantly lower spins 
for intermediate to large MBH masses, which is inconsistent with
observations (c.f. Appendix \ref{app:co-cha}).

{ We further note that the {\it pseudobulge} and the {\it hybrid}
  models, while providing the best descriptions of the data, hint at
  different (although possibly coexisting) evolutionary scenarios.
The hybrid scenario has the highest statistical significance and a
simple physical interpretation, with a clear candidate for the source of
turbulence.  The gas dynamics remains quite coherent for most of the time, which is reasonable because 
gas can efficiently dissipate turbulent motions on few
local orbital times \citep[see the discussion in][and references
  therein]{maio13}. Only during galaxy mergers the violent reshuffling of the angular 
momentum allows for more isotropic gas
inflows. We note that, strictly speaking, in the {\it hybrid} model we adopt a ``bulge-like''
dynamics for the accreting gas during the {\it whole} of the merger driven accretion events, even though the
AGN activity lasts for more than the galaxy merger itself. Thus, during 
the tails of the nuclear activity, accretion may be more coherent than in our model's assumptions. 
However, during a merger-driven AGN phase most of the gas is accreted
in a short almost-Eddington burst, lasting about
a Salpeter time $\lesssim 0.2$ Gyr.  The long lived accretion tails, still observable
at low redshifts,   present instead a lower Eddington ratio $f_{\rm Edd}\sim 0.01$ 
and cannot therefore significantly change the MBH spin magnitude.
 
Although the pseudobulge model is disfavored by our statistical
analysis, it still provides a good description of the observational
data and is consistent with the evidence that low redshift AGNs with
moderate to high Eddington ratios are preferentially hosted in
pseudobulges \citep{heckman14}. Pseudobulges have recently been
associated with stellar bars, i.e., structures able to drive gas
inflows toward the galaxy nucleus \citep[see, e.g.][for up to date
  reviews]{sellwood13, kormendy13}. After the initial instability
responsible for the bar formation, the bar itself generally undergoes
a thickening that can be either caused by a second ``buckling''
instability or by a vertical resonance~\citep{sellwood13}, possibly responsible for the pseudobulge
formation~\citep{kormendy13}. In this scenario, the gas could accrete with a
more disk like dynamics until the bar thickens, and fall toward the
MBH with a higher velocity dispersion after the pseudobulge forms if
the buckling instability and/or any vertical resonances significantly
affect the gas motion.

The paucity of data and the many uncertainties about the different
fueling processes do not allow us to firmly reject either of these
models.  If either model is confirmed by more data, it could
point toward a physically motivated galaxy-MBH coevolution scenario
for low redshift disk galaxies, in which the dynamics of gas in
nuclear bars and/or mergers drive the MBH mass and spin evolution.}

\section*{Acknowledgments}
We are indebted to  M. R. Krumholz and D. Calzetti for invaluable insights into the 
metallicity dependency of star formation. We also thank  M. Colpi, J. Gair, M. Kesden, 
A. Merloni, C. Reynolds, A. Tchekhovskoy and M. Volonteri  for insightful comments and discussions.
We acknowledge support from the European Union's Seventh Framework Program (FP7/PEOPLE-2011-CIG) through the Marie Curie Career Integration Grant GALFORMBHS PCIG11-GA-2012-321608 (to E.B.), and support from the DLR (Deutsches Zentrum fur Luft- und Raumfahrt) through the DFG grant SFB/TR 7 Gravitational Wave Astronomy and (to A.S.).
Computations were performed on the gpc supercomputer at the SciNet HPC Consortium, as well as on the ``Projet Horizon Cluster'' at the Institut d'astrophysique de Paris.

\bibliographystyle{apj.bst}

\bibliography{references}

\appendix
\section{A: The star formation}
\label{ap:SF}
{ 
}

Observations of our and nearby galaxies have established that star formation is ultimately associated with Giant Molecular Clouds (GMCs), {\it giant} because they can be very massive ($M\sim 10^6 M_{\sun}$) and extended ($\sim 100$ pc).
There, the star formation rate is determined by the mass fraction $f_{\rm c}$ in cold gas (generally, but not only, in molecular form) and the timescale $t_{\rm SF}$ needed to convert it into stars.
This latter depends both on the cloud properties, such as its density (the higher the faster is the process) and on the presence of feedback-driven turbulence, which slows the collapse. 
One of the main (observationally supported) assumptions  is that the GMC properties are environment independent, as long as the surrounding interstellar  
medium has lower pressure than the GMC itself. In particular, the  cloud surface density $\Sigma_{\rm cl}$ is set by internal processes to be always around $\Sigma_{\rm cl} \approx  85 ~ M_{\sun}~{\rm pc^{-2}}$ in more tenuous environments, where the gas surface density is $\Sigma_{\rm g} \le 85 ~ M_{\sun}~{\rm yr^{-1}} \equiv \Sigma_{\rm th}$ \citep[e.g.][]{bolatto11}.
Above this threshold, pressure equilibrium between the interstellar medium and the GMC sets  $\Sigma_{\rm cl} \simeq \Sigma_{\rm g}$.

\subsection{Star formation in the galactic disk}

In this framework, the local star formation in the galactic disk can be described by 
\be
\dot{\Sigma}_* =  \frac{f_{\rm c} \Sigma_{\rm g}}{t_{\rm SF}}, 
\label{eq:SF_disc}
\ee
 where the timescale $t_{\rm SF}$ has two regimes, according to whether the cloud density $\Sigma_{\rm cl}$ is set by the interstellar pressure, 
\be
t_{\rm SF}^{-1} = \frac{M_6^{-0.33}}{0.8 ~{\rm Gyr}}  \max \left[1,\left(\frac{\Sigma_{\rm g}}{\Sigma_{\rm th}}\right)^{0.67}\right],
\label{eq:k7}
\ee
 \citep{krumholz09}, where $M_6 = M/(10^6 M_{\sun})$ is the GMC mass. The local Jeans mass $M_{\rm j} \approx \sigma_{\rm g}^2/G^2 \Sigma_{\rm g}$
 gives a good estimate of the cloud mass, where $\sigma_{\rm g}$ is the gas velocity dispersion.  In a galactic {\it disk} environment, we may assume a condition of marginal gravitational stability, and obtain a function of the background gas surface density only,
\be
 t_{\rm SF}^{-1} =  (\rm 2.6 ~Gyr)^{-1} \times  \left\{ \begin{array}{l l}
 \left(\frac{\Sigma_{\rm g}}{\Sigma_{\rm th}}\right)^{-0.33}, &  \Sigma_{\rm g} <  \Sigma_{\rm th},\\
  \left(\frac{\Sigma_{\rm g}}{\Sigma_{\rm th}}\right)^{0.34}, &  \Sigma_{\rm g} > \Sigma_{\rm th}.
\end{array} \right.
\label{eq:tsf_disc}
\ee

The cold-gas mass fraction $f_{\rm c}$ is equivalent to the molecular mass fraction at metallicity greater than $\sim 1\%$ solar, when H$_2$ has time to form before collapsing and forming stars.
At lower metallicities, instead, star formation will occur in a cold atomic gas phase rather than a molecular phase \citep{krumholz12}. In general, as the metallicity decreases, the cold gas available for forming stars decreases as well.  However, recent observations of nearby  spirals and dwarfs \citep{bigiel10} and of the Small Magellanic Cloud \citep{bolatto11} suggest that $f_{\rm c}$ levels off, around $2\%$ rather than dropping to zero. These considerations led us to adopt  the following prescription,
\be
f_{\rm c} =  \left\{ \begin{array}{l l}
 1-\left[1+\left(\frac{3}{4} \frac{s}{1+\delta}\right)^{-5}\right]^{-1/5},  & ~{\rm ~if}~f_{\rm c} > 2\%,\\
 2\%, &  ~{\rm otherwise}
 \end{array} \right.
\label{eq:fh2}
\ee
(Krumholtz private communication), with
$$ s = \ln{(1+0.6 \chi)}/(0.04 \Sigma_{\rm 1} Z^{'}),$$
$$ \chi = 0.77 (1+ 3.1 Z^{'0.365}),$$
$$\delta = 0.0712 \left(0.1 s^{-1} + 0.675\right)^{-2.8},$$
where $\Sigma_{\rm 1} = \Sigma_{\rm g}/ (M_{\sun}~ \rm{pc}^{-2})$ and $Z^{'}$ is the metallicity in Solar units.

Integrating equation~\eqref{eq:SF_disc}  over the entire disk surface, we obtain the total star formation rate in the disk.

\subsection{Star formation in the galactic bulges}
The very same prescription can be applied to the bulge, during the periods of ``quiescent star-formation'' in the galaxy, i.e. when violent star bursting events are not triggered.
Practically, we use a volumetric star formation law, 
\be
\dot{\rho}_* = \frac{f_{\rm c}  \rho_{\rm g}}{t_{\rm SF}},
\label{eq:SF_bulge}
\ee
which we derive directly from equation (\ref{eq:SF_disc}), by expressing $\Sigma_{\rm g}$ and $M$ as a function of the volumetric gas density $\rho_{\rm g}$ and  the local isothermal sound speed $c_{\rm s}$. 

Let us start by considering the star formation timescale (equation (\ref{eq:k7})). In regions with densities below threshold, $t_{\rm SF} \propto M_{\rm j}^{0.33}$, where 
we can simply write the Jeans mass in the more familiar way
\be
M_{\rm j} = \frac{\pi}{6} \frac{c_{\rm s}^{3}}{G^{3/2} \rho_{\rm g}^{1/2}} \approx 10^6~M_{\sun} \left(\frac{c_{\rm s}}{8.3 ~\rm km/s}\right)^{3} \rho_1^{-1/2},
\label{eq:jeans_mass}
\ee
with $\rho_1 = \rho_{\rm g} /(M_{\sun}/\rm pc^3)$.
In denser regions of the bulge, one has an additional dependence on  $\Sigma_{\rm cl}$, i.e. $t_{\rm SF} \propto M_{\rm j}^{0.33} \Sigma_{\rm cl}^{-0.67}$ (c.f. 
equation (\ref{eq:k7})).
To relate the surface density to the volume density, let us note
that for a spherical cloud of mass $M$ and characteristic size $L$, $\rho_{\rm cl} \approx M/L^3$ and
 $L \approx \Sigma_{\rm cl}/\rho_{\rm cl}$. Eliminating $L$ in these two expressions,
we  obtain
$\rho_{\rm cl} \approx  M_{\sun}{\rm pc}^{-3} \left(\Sigma_{\rm cl}/\Sigma_{\rm th}\right)^{3/2} M_6^{-1/2}$.
Inserting equation (\ref{eq:jeans_mass}) and recalling the pressure equilibrium condition, $\rho_{\rm g} \approx \rho_{\rm cl}$ \citep{krumholz_mckee},
we finally get
\be
\left(\frac{\Sigma_{\rm g}}{\Sigma_{\rm th}}\right) = \rho_1^{1/2}  \left(\frac{c_{\rm s}}{8.3 ~\rm km/s}\right).
\label{eq:conversion}
\ee
Thus, the overall timescale expression becomes
\be
 t_{\rm SF}^{-1} =  (\rm 0.8 ~Gyr)^{-1} \times  \left\{ \begin{array}{l l}
  \rho_1^{0.165}  \left(\frac{c_{\rm s}}{8.3 \rm km/s}\right)^{-0.99} &  \Sigma_{\rm g} <  \Sigma_{\rm th},\\
  \rho_1^{1/2}  \left(\frac{c_{\rm s}}{8.3 \rm km/s}\right)^{-0.32} &  \Sigma_{\rm g} > \Sigma_{\rm th}.
\end{array} \right.
\label{eq:tsf_bulge}
\ee

We now turn our attention to the fraction $f_{\rm c}$ of mass in cold gas available for star formation (equation (\ref{eq:fh2})). Unlike 
$ t_{\rm SF}$, $f_{\rm c}$ depends on the  {\it gas} density and \textit{not} on the GMC density. In more tenuous regions, out of pressure equilibrium,
the cloud can be denser than the background gas, i.e. $\eta \equiv \rho_{\rm cl}/\rho_{\rm g} \geq 1$. 
Therefore, in principle, in equation (\ref{eq:fh2}) we should use a modified version of equation \eqref{eq:conversion}, e.g. an expression 
$\Sigma_{\rm g} \propto \eta^{-1/3} \rho_{\rm g}^{1/2} c_{\rm s}$, 
which could be used both above ($\eta \approx 1$) and {\it below} threshold. However, since $f_{\rm c}$ has a floor of 2\% (c.f. equation \eqref{eq:fh2}), which
kicks in when  $\eta \approx $ a few to several, the modification factor ($\eta^{-1/3} \approx 1-2$) is negligible relative to the other uncertainties in the derivation of equation \eqref{eq:conversion}. 
We will therefore always use directly equation \eqref{eq:conversion} into equation \eqref{eq:fh2}, whether above or below threshold.

\subsubsection{Starburst in Merging Galaxies}
Observations suggest a link between starbursts and mergers. The main evidence is that the strongest starbursts (Ultra- and Hyper-Luminous Infrared Galaxies) are predominantly merging systems at all redshifts \citep[e.g.][]{elbaz_cesarsky}. More recently, CO observations of galaxy populations hinted that starburst/merging galaxies have a different, more efficient, star formation law \citep{daddi10, genzel10}. 
We therefore assume that in merging systems the star formation is driven by different dynamical processes in the bulge, which induce star formation over a dynamical time. In practice,  we use 
the same procedure as in B12: in the gaseous bulge that forms after the merger, star formation is regulated by
\begin{equation}\label{burst_SF}
   \dot{\rho}_{*}=  \frac{\rho_{\rm g}}{t_{\rm ff}}\,,
\end{equation}
where $t_{\rm ff}=\sqrt{3\pi/(32 G \rho_{\rm g})}$ is the local dynamical time for the gas. Again, this expression
can be integrated over the bulge volume to yield the total star formation rate.

\section{B: Coherent and chaotic accretion scenarios}
\label{app:co-cha}
Alongside our fiducial models, for the sake of comparison we also investigate three models implementing the coherent and chaotic scenarios often used in the literature:
\renewcommand{\labelenumi}{\roman{enumi}}
\begin{enumerate}
\setcounter{enumi}{4}
\item {\it coherent} model. Each accretion event takes place in a well defined plane, persisting for the duration of the episode and efficiently spinning--up the hole \citep{thorne}; c.f., for instance, the coherent model of \cite{berti08};
\item {\it chaotic I} model. For each accretion event, we always take $F=w=0.5$, independently of the accreted mass \citep[c.f. for instance the chaotic model of][]{berti08}{\footnote{Note that in our framework, once $F$ is given, $w$ cannot be arbitrarily set to 0.5, because it is a function of the MBH and cloud angular momenta through equation (\ref{eq:w}). In this respect, model {\it chaotic I} corresponds to the limiting case where $F=0.5$ and ${M_{\rm disk}}\rightarrow0$ (which implies ${J_{\rm disk}}\rightarrow0$).}};
\item {\it chaotic II} model. The model of this paper, but with $F=1/2$, i.e., with isotropic distribution for the angular momentum of the gas clouds \citep{king06}.
\end{enumerate}

\begin{figure*}
\begin{center}
\begin{tabular}{ccc}
\includegraphics[scale=0.28,clip=true]{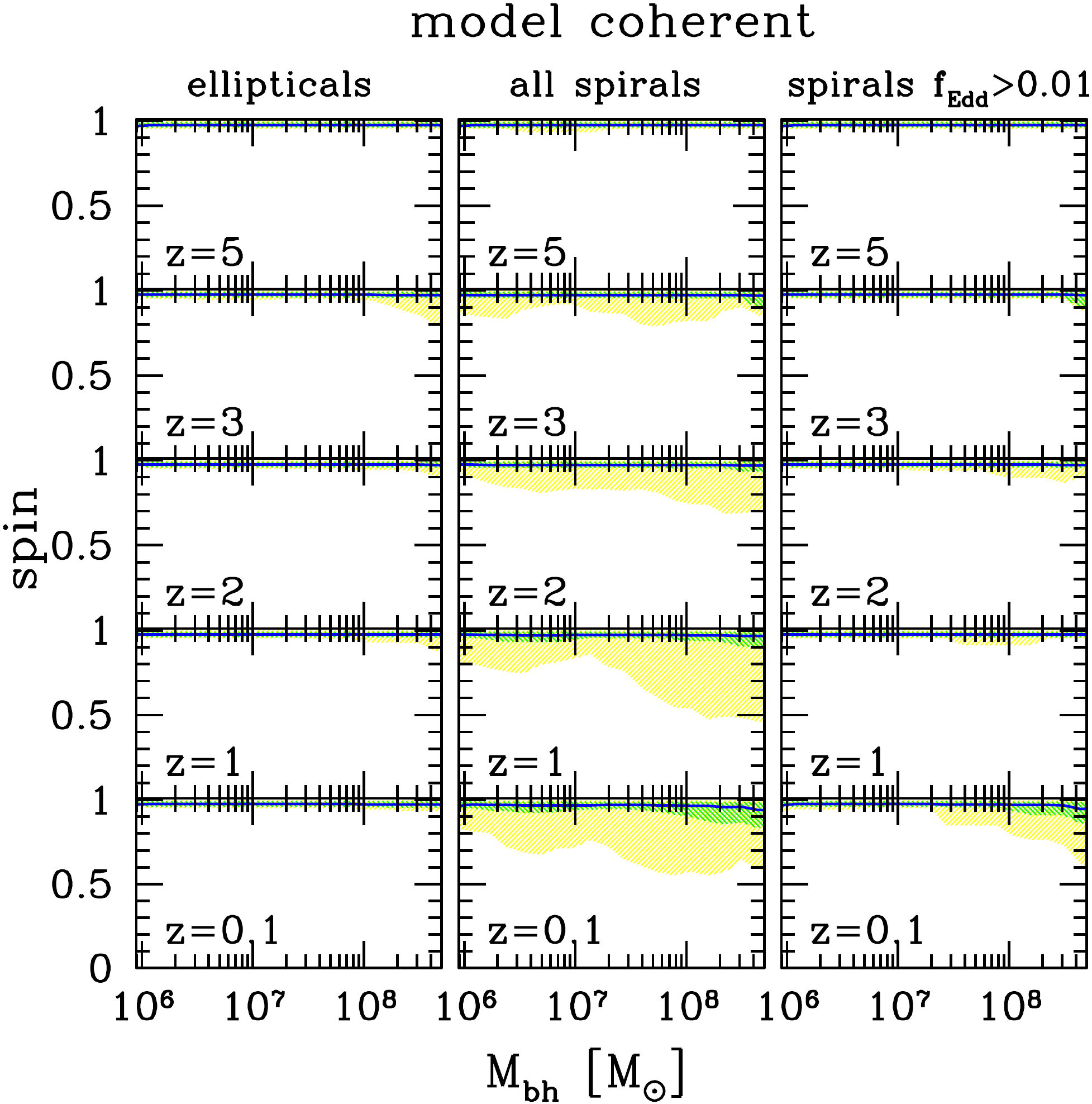}&
\includegraphics[scale=0.28,clip=true]{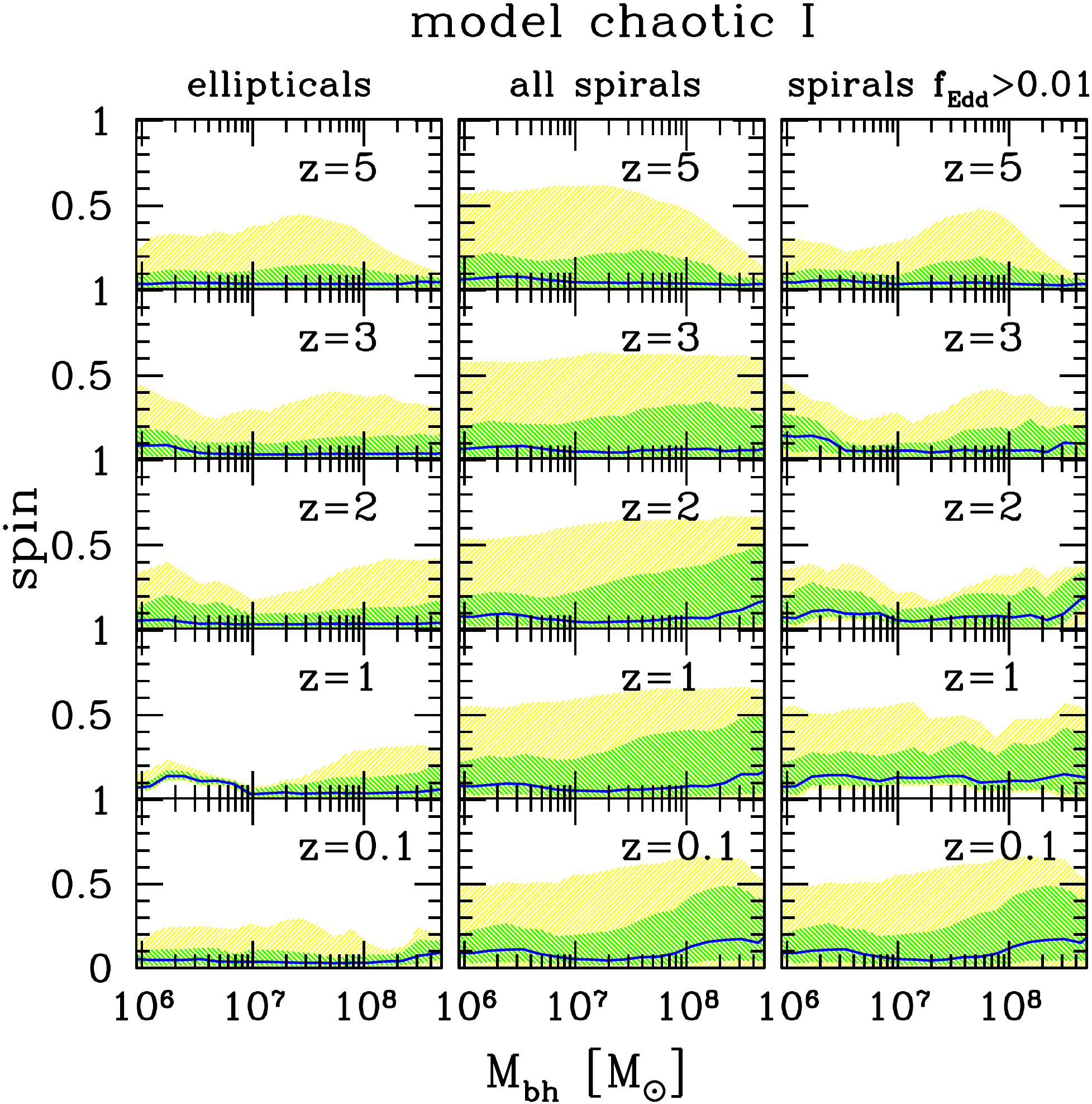}&
\includegraphics[scale=0.28,clip=true]{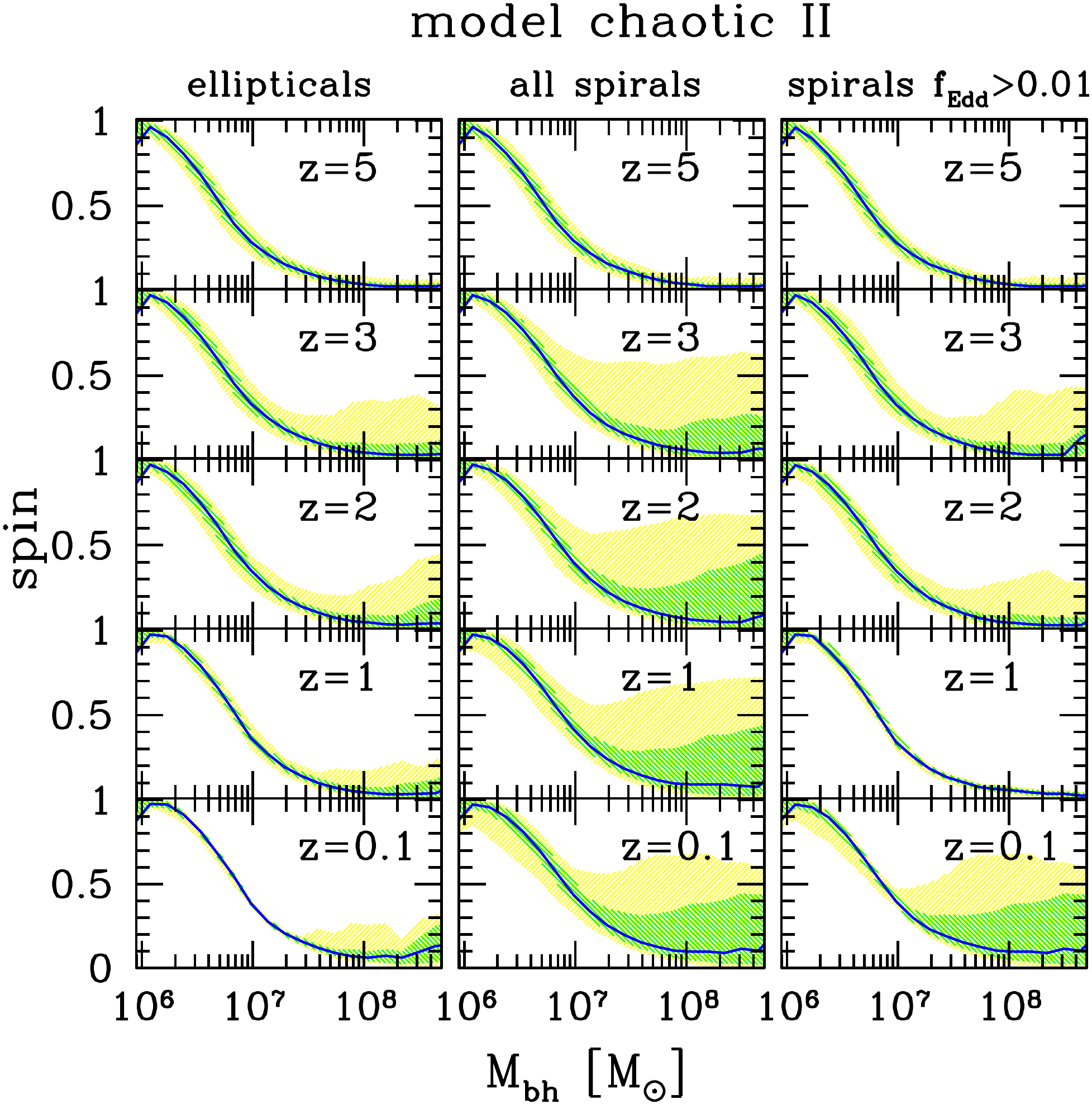}\\
\end{tabular}
\caption{Same as Figure \ref{spinzev} but for the {\it coherent}, {\it chaotic I} and {\it chaotic II} models, as specified at the top of each panel. In each plot, the blue line is the median of the spin distribution as a function of MBH mass and as predicted by the model, while green and yellow shaded areas represent the spin ranges enclosing $68\%$ and $95\%$ of the distribution.}
\label{spinzevcch}
\end{center}
\end{figure*}

The overall redshift evolution of the MBH spin distributions for models {\it coherent}, {\it chaotic I} and {\it chaotic II} is shown in Figure \ref{spinzevcch} (to be compared to Figure \ref{spinzev}). In the {\it coherent} model, MBHs tend to be maximally spinning irrespective of redshift, accretion rate and galaxy host morphology. The opposite is true for the {\it chaotic I} model, where spins tend to be small ($a_{\rm bh}<0.5$) and cluster around zero. In the {\it chaotic II} model, there is a transition between maximally spinning MBHs at low masses and non-spinning MBHs at high masses. The difference from the {\it chaotic I} model is that each accreting lump of matter has a defined angular momentum (although with random direction); light MBHs thus align with it and are efficiently spun up, whereas massive MBHs do not and effectively experience zero-angular momentum accretion averaged over many episodes.

The features in the spin distribution discussed above already hint at the difficulty of reconciling these three models with observations. We do not show here a visual comparison, but we report the results of the same statistical analysis performed in Sections \ref{2dks} and \ref{secbayes}. In this case we summed up accreting MBH in spirals {\it and} in ellipticals, when building spin distributions. The procedure does not affect the results because in these simplistic models the spin evolution is not connected to the nature of the host galaxy, and it allows us to increase the size of the theoretical sample. The {\it coherent} model shows properties similar to the {\it disk} one, but with a stronger clustering toward maximally spinning MBHs, resulting in a poorer match to the data, quantified by a KS probability $p_{\rm KS}<10^{-2}$. Both the {\it chaotic I} and  {\it chaotic II} models result in very different distributions favoring low spins (as shown in the bottom right panels of figure \ref{spinzevcch}), which are impossible to reconcile with observations ($p_{\rm KS}<10^{-4}$ in both cases, see numbers in Table \ref{tab4}). This is confirmed by the computation of the odds ratios against, e.g., the {\it hybrid} model, which completely discard all three scenarios.

\section{C: Sanity checks on the statistical analysis}
\label{app:sanity}

As explained in the main text, to carry out our analysis we used our model to construct discrete theoretical distributions on a specific grid in the $M_{\rm bh}-a_{\rm bh}$ parameter space. We then compared spin measurements at $z<0.1$ to theoretical predictions at $z=1$. In this Appendix, we show that our conclusions do not depend significantly  on these specific choices.

\begin{figure*}
\begin{center}
\begin{tabular}{cc}
\includegraphics[scale=0.4,clip=true]{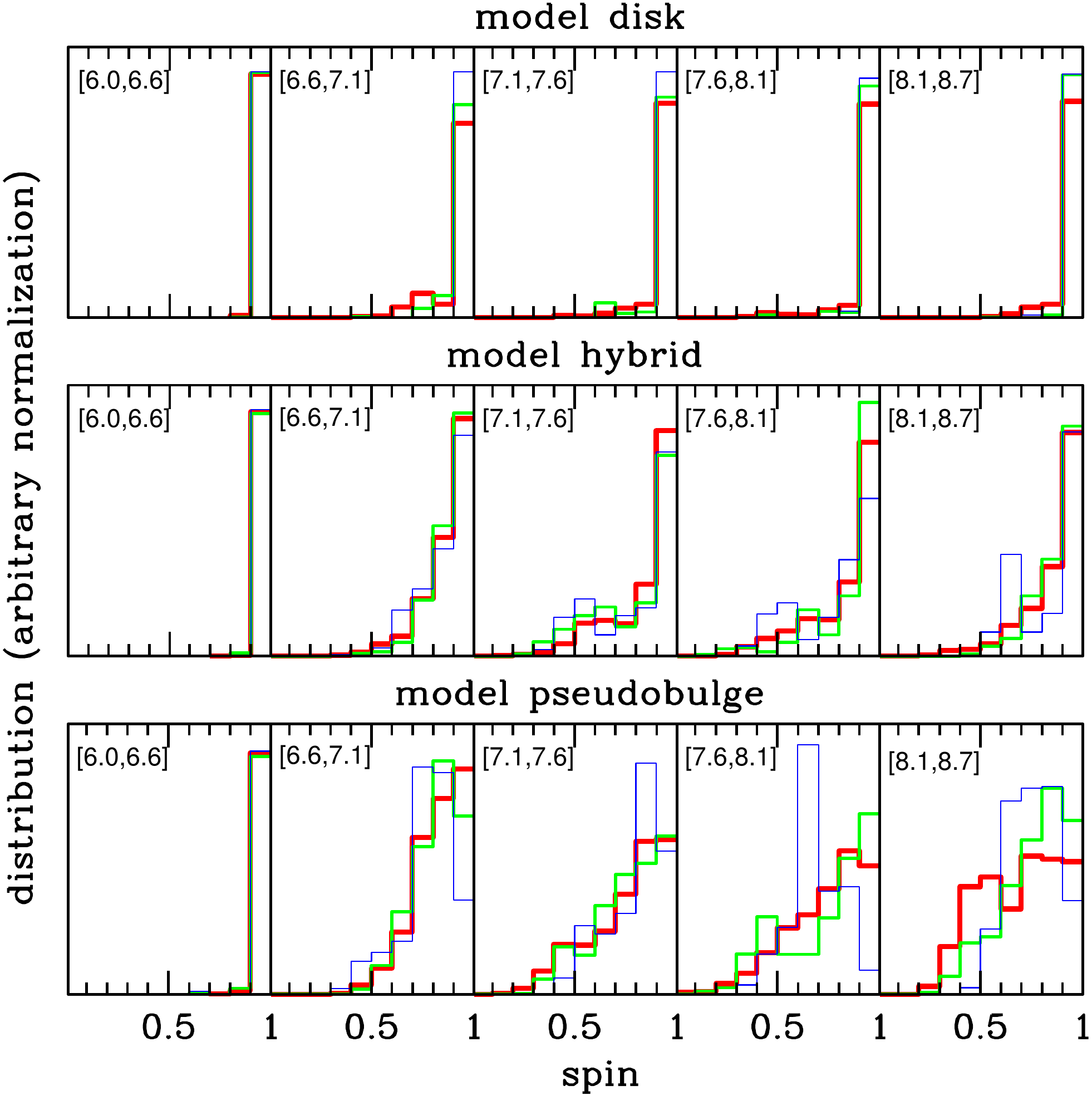}&
\includegraphics[scale=0.4,clip=true]{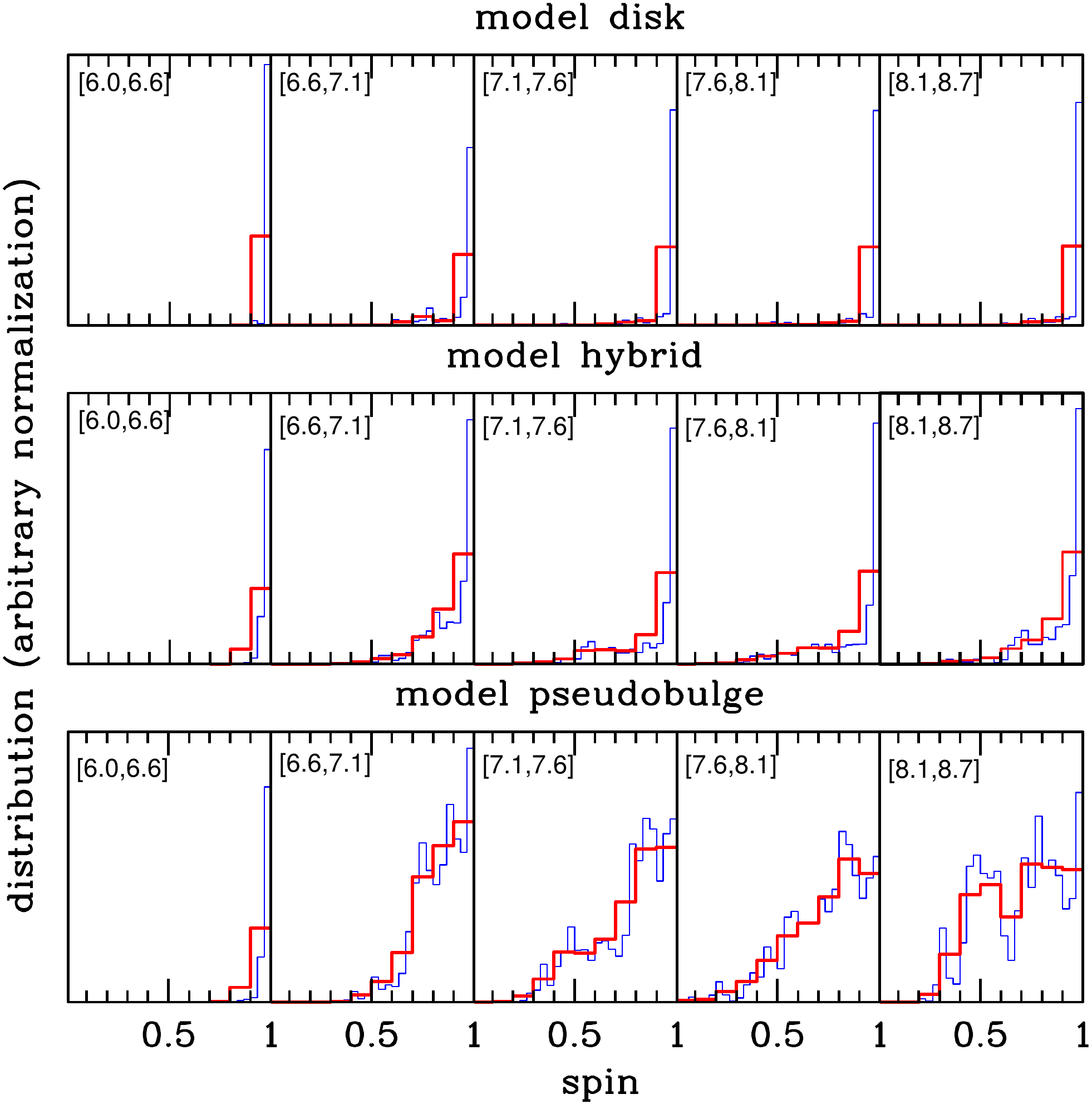}\\
\end{tabular}
\caption{The left panels show spin distributions in different mass bins for accreting MBHs in spirals, for our three fiducial models. In each panel, histograms are for $z=1$ (thick--red), $z=0.5$ (medium--green) and $z=0.1$ (thin--blue). The numbers in square parenthesis in each panel represent the extremes of the considered log$\,M_{\rm bh}$ interval. The right panels show spin distributions in different MBH mass bins and  at $z=1$, but now considering different $a_{\rm bh}$ binnings. Thick-red and thin-blue histograms are computed using 10 and 30 $a_{\rm bh}$ bins respectively. The numbers in square parenthesis in each panel represent the extremes of the considered log$\,M_{\rm bh}$ interval.}
\label{hist}
\end{center}
\end{figure*}

As mentioned, we choose to compare observations to theoretical distributions at $z=1$ rather than at $z=0.1$ in order 
to have a larger statistical sample. The idea behind this choice is that theoretical distributions should not change significantly at $z<1$,
where QSO and AGN activity, as well as star formation are fading. This is shown to be indeed the case in the left panel of Figure \ref{hist}, where we plot the redshift evolution of the population of accreting MBHs in spirals, for different mass bins and models. The figure shows that distributions generated with the {\it pseudobulge} and {\it hybrid} models do not change much at $z\leq1$, but the effect of the small statistical sample at low redshifts is clear, especially at $z=0.1$ (thin blue histograms) for the {\it pseudobulge} model. Distributions generated with the {\it disk} model are slightly more redshift dependent.  At $z=0.1-0.5$ (thin blue and medium green histograms) they show more clustering at almost maximal spins, and the moderate spin tails present at $z=1$ (thick red histograms) are suppressed. (As we will show below, this has an impact when estimating the model's compatibility with the data). 

For the analysis of the main text, we evaluated theoretical distribution  on a grid made of five equally log-spaced mass bins, and 20 linear bins covering the spin range $0<a_{\rm bh}<1$. We construct here distributions using instead 10 and 30 linear $a_{\rm bh}$ bins (by keeping 5 mass bins), and also considering 20 bins both in mass and spin. In the right panel of Figure \ref{hist}, we show the dependence of the distributions on the $a_{\rm bh}$ binning. The distributions for the {\it disk} and {\it hybrid} models both depend on the binning at the high-spin end. This is not surprising because the two models adopt similar accretion prescriptions for MBHs in spirals that recently experienced galactic-disk instabilities. However, as we show below, in the {\it hybrid} model this binning dependence does not have a large impact on the comparison with the data. This is because a significant tail at moderate spins is guaranteed by the fact that the model is also partly {\it pseudobulge} in nature (cf. model description in Section \ref{sec:implementation}). 

Being aware of these issues, we assess here the impact of the theoretical sample's binning and redshift on the results of our statistical analysis. Results for the 2D-KS tests are reported in Table \ref{tab4}. The {\it hybrid} and {\it pseudobulge} models yield probabilities that are largely independent on the choice of binning and redshift. We just notice a significant discrepancy at $z=0.1$ in the subsample of accreting MBHs in spirals in the {\it pseudobulge} case. This is entirely due to the small number of objects found in our simulations, which causes a very noisy distribution. Models {\it chaotic I} and {\it chaotic II} are also unaffected by the specific redshift and binning choice, and are always ruled out at high significance. Conversely, the match between the {\it disk} and {\it coherent} models and observations is binning dependent. Although apparently problematic, this dependence has a physical origin. In our thin accretion disk model, the maximum MBH spin is $a_{\rm bh}=0.998$~\citep{thorne}. Basically all the simulated MBHs falling in the highest $a_{\rm bh}$ bin have indeed this spin value. By binning the spin distribution, we are ``spreading'' those systems on the width of the bin; the larger the bin, the larger the spread, making it easier to reconcile the theoretical distribution with the large number of spins measured in the [0.85,1] range (c.f. Table \ref{tab1}). If the maximum MBH spin is indeed $a_{\rm bh}=0.998$, then the finest spin binning should provide the most trustworthy  results. We caution, however, that such a narrow spin distribution, peaked close to the maximal spin $a_{\rm bh}=1$, is affected by some simplifications that we made in our spin evolution model. As described in Sections~\ref{sec:spin}~and~\ref{sec:Fvsigma}, we assumed  $i)$ that at the beginning of each accretion episode the MBH spin is parallel to the total angular momentum of the reservoir, and $ii)$ that whenever $J_{\rm disk}/2J_{\rm bh} > 1$, the alignment process is very fast and accretion is effectively coherent. Relaxing these two assumptions would affect the highly spinning MBHs, because when $a_{\rm bh}\approx 1$, a small amount of retrograde accretion can efficiently spin the black hole down. The effect of relaxing assumption $i)$ has already been discussed in Section~\ref{sec:Fvsigma}, so here we will simply estimate the effect of relaxing prescription $ii)$. For a broad range of MBH masses, $J_{\rm disk}/2J_{\rm bh}$ can be larger than but still close to 1. In these cases, if the accretion event is initially misaligned with respect to $\boldsymbol{J}_{\rm bh}$ by more than $\pi/2$, accretion would be retrograde during the first part of the spin realignment process. If a fraction of 10$\%$ of the gas mass were accreted on retrograde orbits, the equilibrium spin would be the same as in the case $F\approx 0.9$ when neglecting the spin re-alignment, i.e. $a_{\rm eq} \approx 0.9$ (see Figure~\ref{fig:a_eq}). Such an effect has been observed and discussed in studies where the spin direction  was evolved during each single accretion event \citep[][D13]{dotti10}. Furthermore, different accretion disk models allow MBH spins only up to $a_{\rm bh}=$0.9-0.95~\citep[e.g.][]{gammie04}, and efficient angular momentum extraction via jets might set a maximum equilibrium spin lower than $a_{\rm bh}=$0.9 \citep[see, e.g.,][]{moderski96,moderski98}. It is therefore impossible to rule completely out the {\it disk} and {\it coherent} models. 

\begin{table*}
\begin{center}
\begin{tabular}{c|cccccccccccc}
\hline
\multicolumn{1}{c|}{} & \multicolumn{3}{c|}{pseudobulge} & \multicolumn{3}{c|}{disk} & \multicolumn{3}{c|}{hybrid} & \multicolumn{1}{c|}{coherent} & \multicolumn{1}{c|}{chaotic I} & \multicolumn{1}{c|}{chaotic II}\\
\hline
assumptions & E & S & S acc & E & S & S acc & E & S & S acc & S$+$E acc & S$+$E acc & S$+$E acc\\
\hline
$z=1$ / 10$a_{\rm bh}$ / flat        & 0.0034 & 0.0280 & \red{\bf 0.2336} & 0.0034 & 0.0969 & \red{\bf 0.1969} & 0.0035 & 0.0577 & \red{\bf 0.5831} & 0.0685 & $<10^{-4}$ & $<10^{-4}$\\
$z=1$ / 10$a_{\rm bh}$ / Gauss       & 0.0015 & 0.0247 & \red{\bf 0.2468} & 0.0015 & 0.0793 & \red{\bf 0.1945} & 0.0015 & 0.0488 & \red{\bf 0.6094} & 0.0719 & $<10^{-4}$ & $<10^{-4}$\\
$z=1$ / 20$a_{\rm bh}$ / Gauss       & 0.0020 & 0.0271 & \red{\bf 0.3614} & 0.0015 & 0.0969 & \red{\bf 0.0521} & 0.0016 & 0.0589 & \red{\bf 0.5328} & 0.0035 & $<10^{-4}$ & $<10^{-4}$\\
$z=1$ / 30$a_{\rm bh}$ / Gauss       & 0.0015 & 0.0302 & \red{\bf 0.3785} & 0.0015 & 0.0986 & \red{\bf 0.0232} & 0.0016 & 0.0598 & \red{\bf 0.3761} & 0.0007 & $<10^{-4}$ & $<10^{-4}$\\
$z=0.5$ / 10$a_{\rm bh}$ / Gauss     & 0.0019 & 0.0197 & \red{\bf 0.2666} & 0.0018 & 0.0615 & \red{\bf 0.1186} & 0.0020 & 0.0379 & \red{\bf 0.5636} & 0.0869 & $<10^{-4}$ & $<10^{-4}$\\
$z=0.5$ / 20$a_{\rm bh}$ / Gauss     & 0.0020 & 0.0223 & \red{\bf 0.3816} & 0.0018 & 0.0743 & \red{\bf 0.0170} & 0.0019 & 0.0451 & \red{\bf 0.4380} & 0.0063 & $<10^{-4}$ & $<10^{-4}$\\
$z=0.5$ / 30$a_{\rm bh}$ / Gauss     & 0.0018 & 0.0258 & \red{\bf 0.4251} & 0.0018 & 0.0735 & \red{\bf 0.0069} & 0.0019 & 0.0447 & \red{\bf 0.2764} & 0.0013 & $<10^{-4}$ & $<10^{-4}$\\
$z=0.1$ / 10$a_{\rm bh}$ / Gauss     & 0.0021 & 0.0164 & \red{\bf 0.0813} & 0.0023 & 0.0454 & \red{\bf 0.0711} & 0.0024 & 0.0278 & \red{\bf 0.6620} & 0.0656 & $<10^{-4}$ & $<10^{-4}$\\
$z=0.1$ / 20$a_{\rm bh}$ / Gauss     & 0.0021 & 0.0181 & \red{\bf 0.0984} & 0.0023 & 0.0526 & \red{\bf 0.0035} & 0.0024 & 0.0322 & \red{\bf 0.6699} & 0.0028 & $<10^{-4}$ & $<10^{-4}$\\
$z=0.1$ / 30$a_{\rm bh}$ / Gauss     & 0.0018 & 0.0217 & \red{\bf 0.0899} & 0.0023 & 0.0534 & \red{\bf 0.0007} & 0.0023 & 0.0324 & \red{\bf 0.4392} & 0.0005 & $<10^{-4}$ & $<10^{-4}$\\
\hline
\end{tabular}
\end{center}
\caption{2D-KS test results on different samples of galaxies for all spin evolution models under different assumptions about the theoretical distribution's computation and the treatment of observational errors (indicated in the first column). For the models {\it pseudobulge}, {\it disk} and {\it hybrid}, we compared existing spin measurements with theoretical distributions (produced with our model) for  ellipticals (E), spirals (S) and  spirals containing accreting MBHs (S acc, these latter highlighted in bold red
as they match the properties of the observational sample), whereas for models {\it coherent}, {\it chaotic I} and {\it chaotic II} we sum all accreting MBHs (S+E acc) to achieve better statistics.}
\label{tab4}
\end{table*}
\begin{table*}
\begin{center}
\begin{tabular}{c|ccccccccc}
\hline
\multicolumn{1}{c|}{} & \multicolumn{3}{c|}{hybrid/pseudobulge} & \multicolumn{3}{c|}{hybrid/disk} & \multicolumn{3}{c|}{pseudobulge/disk}\\
\hline
assumptions & {log$\Lambda_{hp}$} & $p_{\rm hybrid}$ & $p_{\rm pseudobulge}$ & {log$\Lambda_{hd}$} & $p_{\rm hybrid}$ & $p_{\rm disk}$ & {log$\Lambda_{pd}$} & $p_{\rm pseudobulge}$ & $p_{\rm disk}$\\
\hline
$z=1$ / 10$a_{\rm bh}$ / Flat      & 0.7995 & 0.8631 & 0.1369 & 1.9259 & 0.9883 & 0.0117 & 1.1264 & 0.9304 & 0.0696\\
$z=1$ / 10$a_{\rm bh}$ / Gauss     & 1.1391 & 0.9323 & 0.0677 & 2.0634 & 0.9914 & 0.0086 & 0.9242 & 0.8936 & 0.1064\\
$z=1$ / 20$a_{\rm bh}$ / Gauss     & 1.0804 & 0.9233 & 0.0767 & 2.4749 & 0.9966 & 0.0034 & 1.3944 & 0.9612 & 0.0388\\
$z=1$ / 30$a_{\rm bh}$ / Gauss     & 0.9119 & 0.8909 & 0.1091 & 2.8313 & 0.9985 & 0.0015 & 1.9193 & 0.9881 & 0.0119\\
$z=0.5$ / 10$a_{\rm bh}$ / Gauss   & 1.0901 & 0.9248 & 0.0715 & 3.0761 & 0.9992 & 0.0008 & 1.9860 & 0.9898 & 0.0102\\
$z=0.5$ / 20$a_{\rm bh}$ / Gauss   & 0.9016 & 0.8886 & 0.1114 & 4.1259 & 0.9999 & 0.0001 & 3.2243 & 0.9994 & 0.0006\\
$z=0.5$ / 30$a_{\rm bh}$ / Gauss   & 0.0543 & 0.5312 & 0.4687 & 2.9997 & 0.9990 & 0.0010 & 2.9454 & 0.9988 & 0.0012\\
$z=0.1$ / 10$a_{\rm bh}$ / Gauss   & 2.3982 & 0.9960 & 0.0040 & 11.241 & $>$0.9999 & $<$10$^{-11}$ & 8.8435 & $>$0.9999 & $<$10$^{-8}$\\
$z=0.1$ / 20$a_{\rm bh}$ / Gauss   & 2.5955 & 0.9975 & 0.0025 & 22.541 & 1.0 & 0.0 & 19.945 & 1.0 & 0.0\\
$z=0.1$ / 30$a_{\rm bh}$ / Gauss   & 3.7300 & 0.9998 & 0.0002 & 22.597 & 1.0 & 0.0 & 18.867 & 1.0 & 0.0\\
\hline
\end{tabular}
\end{center}
\caption{Model selection results: comparisons between models  {\it pseudobulge (p)}, {\it disk (d)} and {\it hybrid (h)}. For each two model comparison, we report the log of the likelihood ratio $\Lambda_{AB}$, i.e. the ratio between the probability of model $A$ ($p_A$) and model $B$ ($p_{B}$).}
\label{tab5}
\end{table*}

We also checked the outcome of the Bayesian model comparison for the different assumptions; results are shown in Table \ref{tab5}. As expected from the 2D-KS tests, the consistency of the {\it disk} model with observations is highly dependent on the binning size of the distribution. This is confirmed by our model selection exercise, which shows that both log$\,\Lambda_{pd}$ and log$\,\Lambda_{hd}$ generally increase with the number of $a_{\rm bh}$ bins. Note that {\it for any choice of binning and sample redshift},  log$\,\Lambda_{hd}>2$, implying that the {\it hybrid} model always provides a better description of the data than the disk model, at least at 99\% confidence level. This is particularly interesting, because even though the {\it disk} model computed on 10 $a_{\rm bh}$ bins at $z=1$ passes the 2D-KS test with flying colours (see Table \ref{tab4}), the odds ratio test provides compelling evidence in favor of the {\it hybrid} model. Model {\it pseudobulge} sits somewhat in the middle; it is generally preferred at 90-to-99\% level over model {\it disk}, but is disfavored at about the same level with respect to model {\it hybrid} (with the anomaly of the $z=0.5$, 30 $a_{\rm bh}$ bins). To summarize, the odds ratio analysis always provides moderate (decisive) evidence that the {\it hybrid} model is a better description of the data than the {\it pseudobulge} ({\it disk}) model, thus proving that our main results are independent on the particular choice of binning and redshift of the theoretical samples.


\end{document}